\documentclass[a4paper,12pt,english]{article}

\usepackage[english]{babel}
\usepackage{amsmath,amsfonts,mathtools}
\usepackage{amssymb}
\usepackage{graphicx}
\usepackage{soul}
\usepackage{bbold}
\usepackage{color}
\usepackage{comment}
\usepackage{booktabs}
\usepackage{verbatim}
\usepackage{empheq}
\usepackage{lmodern}
\usepackage{epsfig}
\usepackage[pagebackref=true,colorlinks=true,citecolor=blue,filecolor=blue,linkcolor=blue,anchorcolor=blue,urlcolor=blue]{hyperref}
\usepackage{ae}
\usepackage{aecompl}
\usepackage{cancel}
\usepackage{bm}
\usepackage{palatino}
\usepackage[rftl]{floatflt}
\usepackage{colordvi}
\usepackage{fancybox}
\usepackage{caption}
\usepackage{subcaption}
\usepackage[normalem]{ulem}
\usepackage{vwcol} 
\usepackage{multirow,bigdelim}
\usepackage{dcolumn}
\usepackage{cite}
\usepackage{stackrel}
\usepackage{float}
\usepackage[section]{placeins} 
\usepackage[export]{adjustbox}
\usepackage{authblk}

\makeatletter
\def\@eqnnum{{\normalfont \normalsize (\theequation)}}
\makeatother

\newcommand{\be}{\begin{equation}}
\newcommand{\ee}{\end{equation}}
\newcommand{\bdm}{\begin{displaymath}}
\newcommand{\edm}{\end{displaymath}}
\newcommand{\bea}{\begin{eqnarray}}
\newcommand{\eea}{\end{eqnarray}}
\newcommand{\bpm}{\begin{pmatrix}}
	\newcommand{\epm}{\end{pmatrix}}
\newcommand\lsim{\mathrel{\rlap{\lower4pt\hbox{\hskip1pt$\sim$}}
		\raise1pt\hbox{$<$}}}
\newcommand\gsim{\mathrel{\rlap{\lower4pt\hbox{\hskip1pt$\sim$}}
		\raise1pt\hbox{$>$}}}
\DeclareMathOperator{\Tr}{Tr}

\newcommand{\snn}{\sigma_{nn}}

\newcommand*\xbar[1]{\hspace{0.1em}%
	\hbox{%
		\vbox{%
			\hrule height 0.5pt 
			\kern0.4ex
			\hbox{%
				\kern-0.1em
				\ensuremath{#1}%
				\kern-0.1em
			}%
		}%
	}\hspace{0.1em}%
}

\AtBeginDocument{
	\heavyrulewidth=.08em
	\lightrulewidth=.05em
	\cmidrulewidth=.03em
	\belowrulesep=.65ex
	\belowbottomsep=0pt
	\aboverulesep=.4ex
	\abovetopsep=0pt
	\cmidrulesep=\doublerulesep
	\cmidrulekern=.5em
	\defaultaddspace=.5em
}

\begin{document}
	
\title{Analytical model for predicting induced-stress distributions in polycrystalline materials}

\author[1]{T.~Mede\thanks{corresponding author: 
		timon.mede@ijs.si}}

\author[1]{S.~El~Shawish\thanks{samir.elshawish@ijs.si}}

\affil[1]{Jo$\check{z}$ef Stefan Institute, SI-1000, Ljubljana, Slovenia}


\date{\today}


\maketitle


\vspace{-1cm}

\begin{abstract}
A simple micromechanical model of polycrystalline materials is proposed, which enables us to swiftly produce grain-boundary-stress distributions induced by the uniform external loading (in the elastic strain regime). Such statistical knowledge of local stresses is a necessary prerequisite to assess the probability for intergranular cracking initiation. Model predictions are verified through finite element calculations for various loading configurations, material properties, and grain-boundary types specified by the properties of a bicrystal pair of grains enclosing the grain boundary.
\end{abstract}



\section{Introduction\label{sec:Introduction}}

Metallic components exposed to corrosive environment (and/or irradiation effects) 
and subjected to mechanical loading can undergo structural deformations that compromise their structural integrity and eventually lead to cracking and service failures. This represents a severe challenge for numerous industries.
Among various material-degradation modes involved, one of the most significant is the intergranular stress-corrosion cracking (IGSCC), and its subvariant, the irradiation-assisted stress corrosion cracking (IASCC), which have been observed in a wide range of structural materials otherwise known for their corrosion resistance, e.g., austenitic stainless steels~\cite{nishioka2008,lemillier,stephenson2014,gupta,fujii2019}, zirconium alloys~\cite{cox,cox1990}, nickel-based alloys~\cite{rooyen1975,shen1990,panter2006,IASCC_IAEA}, high strength aluminum alloys~\cite{speidel,burleigh}, and ferritic steels~\cite{wang,arafin}.

Typical of these processes is the development and propagation of cracks along the grain boundaries (GB) which separate different grains in a polycrystalline. 
The initiation phase of the process (microcrack formation) strongly depends on the local stress state and the strength of GBs. Since microcracks are undetectable by non-intrusive inspection methods, there have been various attempts to model the cracking process~\cite{west2011,stratulat2014,zhang2019,fujii2019,sfantos1,BENEDETTI2013249,BENEDETTI201336,ELSHAWISH2023104940} or numerically simulate the stresses. Most of them relied either on the crystal-plasticity finite element~\cite{diard2005,diard2002,kanjarla,gonzalez2014,hure2016,elshawish2018} or crystal-plasticity fast Fourier transform~\cite{lebensohn2012,disen2020} numerical methods. Our goal here is different. We want to construct a simple analytical model that would allow us to quickly estimate intergranular stresses for any external loading in the elastic regime%
\footnote{The reason for using purely elastic strains is twofold: it not only greatly simplifies our model assumptions but also reflects the fact that IGSCC initiation is typically induced by small applied stresses at or slightly below the yield stress~\cite{gupta}.}
and deduce some useful phenomenological relations that could be used in material-damage predictions. 
This introduction is thus not meant so much as an overview of past achievements but rather as a discussion about the philosophy of our analytical modelling approach, explaining its key ideas and their background.

It has been shown that GBs subjected to large \emph{normal stresses} exhibit the highest cracking susceptibility~\cite{west2011}. On the other hand, \emph{shear stresses} acting on the GBs have little effect on the IGSCC nucleation~\cite{fujii2019}.
While the induced normal stress $\sigma_{nn}$ (the opening-mode stress) strongly depends on the orientation of GB and the details of external loading, the GB strength should not. Hence, it is sensible to classify GBs into different GB types and then study the corresponding intergranular normal stress (INS) distribution $\text{PDF}(\sigma_{nn}) := dP/d\sigma_{nn}$ for each type. By definition, all GBs of the same type should have equal GB strength, meaning the same threshold value (critical stress $\sigma_c$) applies to them. Its value indicates the formation of a microcrack whenever the local GB-normal stress $\sigma_{nn}$ exceeds it~\cite{johnson2019}. This simplified picture allows us to effectively decouple the influence of environment (affecting the GB strength $\sigma_c$) and focus only on the induced stresses arising due to mechanical loading and depending on the material properties.
 
In regions where the share of GBs with stresses above the threshold ($\sigma_{nn} > \sigma_c$) is large enough, the density of microcracks reaches a critical level and they merge into macroscopic, observable cracks. Since grains, their crystal lattices, and GBs in a certain material are typically randomly oriented and distributed, this process is stochastic and we can only estimate the probability for cracking under a certain mechanical load. For that task, GB-stress distributions are a sufficient input and precise knowledge of stress at each GB is not required.

In principle, each GB should be specified by its exact location within the aggregate and its entire neighbourhood, i.e., by the configuration of all the grains surrounding it, including their shapes, sizes, crystallographic orientations, possible defects, etc.
However, it can be assumed that grains which are closer to the investigated GB have bigger impact on it than those farther away. The mechanism of IGSCC initiation can thus be treated perturbatively, at each successive order taking as relevant a larger neighbourhood of a chosen GB while modelling the rest of the aggregate as homogeneous and isotropic matrix material. What a sufficient order of such perturbative expansion is, depends then on the desired accuracy of our GB-stress estimates and on the anisotropy of the material --- the more anisotropic that is, the wider the neighbourhood of a GB that needs to be considered. 

For perfectly isotropic grains it is enough to take the simplest possible approximation in which only the GB itself is needed and all the surrounding grains are treated as equal. 
Since all GBs are equivalent, they have the same GB strength and there exists only a single GB type. The expected value of local GB stress $\sigma_{nn}$ in this case corresponds precisely to the value of external stress $\mathbf{\Sigma}$ projected onto that GB ($\sigma_{ij} = \Sigma_{ij}$), and the key parameter that determines it, is the orientation of GB with respect to external loading.
Similarly, if we applied this approximation to anisotropic materials, the resulting INS distribution would represent that of normal stresses on random GBs in the aggregate.

To be able to predict the probability for intergranular cracking initiation in an anisotropic material, different GB types must be treated individually because they correspond to different GB strengths. This requires going beyond the isotropic approximation.
Next order of perturbation considers in addition to GB also the two grains adjacent to it. This is called the \emph{bicrystal model}. 
By embedding a pair of grains in homogeneous and isotropic elastic medium with average (bulk) properties, each GB type is determined by just $5$ macroscopic parameters (in the continuum limit), e.g., the GB-normal vectors $(a, b, c)$ and $(d, e, f)$ expressed in crystallographic system of either grain and the difference of their twist angles ($\Delta\omega := \omega_2 - \omega_1$) about the GB normal. 
Even more, it has been shown that a specific, unique combination of these parameters (namely, the effective GB stiffness $E_{12}$) is already sufficient%
\footnote{This does not mean, however, that all GB types associated with the same value of $E_{12}$ necessarily correspond also to the same GB strength $\sigma_c$.}
for characterizing the INS distributions, at least in materials with cubic lattice symmetry~\cite{ELSHAWISH2021104293,ELSHAWISH2023104940}.

With the growing anisotropy of the material, more and more layers of grains around the chosen GB need to be taken into account. The reason for that can be understood as follows. What we are basically doing, is solving the generalized Hooke's law for the two grains on either side of the grain boundary. The solution for all the components of stress and strain tensors in them requires specifying how the corresponding bicrystal pair deforms as a whole. With this information in hand, the obtained solution would be exact (barring the used approximation of constant stress and strain throughout each grain and neglecting the presence of structural defects). 

In bicrystal approximation the strain of any bicrystal pair can be directly related to the external stress tensor $\mathbf{\Sigma}$, based solely on the GB type it belongs to, and the orientation of the associated GB. This is because the neighbourhood of each grain pair is treated as featureless (homogeneous and isotropic).
In other words, GB-normal stress is only a function of GB type (which includes also the material properties), GB orientation and external stress. In general, this is not the case and depending on the exact configuration of all the grains surrounding the bicrystal pair, the stress induced in it can vary significantly. Consequently, the true GB-stress distributions (e.g., those obtained in numerical simulations) should thus be wider than bicrystal-model estimates. 
This discrepancy becomes more and more pronounced with the growing anisotropy of the grains and indicates a breakdown of the bicrystal approximation. 

Fortunately, there is a way to overcome this issue and substantially improve the accuracy of bicrystal-model predictions. This can be achieved by adopting an alternative approach to include the effect of
each GB experiencing a slightly different neighbourhood.
The contribution of this inhomogeneity of surrounding material can be treated as a small ``fluctuation'' $\Delta\snn$ added to the local value of $\sigma_{nn}$. The more anisotropic the material, the larger these fluctuations (while in isotropic case they should vanish completely%
\footnote{In fact, this happens even in anisotropic materials with cubic lattice symmetry, when subjected to \emph{hydrostatic} loading, $\mathbf{\Sigma}_{\text{hyd}} \propto \mathbb{1}_{3\times 3}$; cf.~\cite{ELSHAWISH2023104940}.}%
). To learn the exact form of $\Delta\snn$ at each GB, the configuration and structure of the entire aggregate would need to be resolved. 

Instead, we can treat the problem statistically. 
Many independent degrees of freedom determine how a specific neighbourhood influences the GB stress. Hence, it is reasonable to assume that the fluctuations ($\Delta\snn$) are normally distributed. For randomly oriented and distributed grains the fluctuations should also be independent of the GB orientation.
Therefore we can for each GB type perform a convolution of the bicrystal-model prediction ($\sigma_{nn}$) with the (Gaussian) distribution of neighbourhood effects such that their combined distribution is appropriately broadened and smoothened. Such approach prevents us from accurately predicting stresses on individual GBs (which is a prerequisite for identifying the crack-initiation sites), but enables us to correctly estimate their distributions.

The same ``broadening'' technique can also be used to correct for the fact that in practice GB stresses are not constant over the GBs as postulated by our simple analytical model. With that we are finally equipped for estimating the true INS distributions for any material, GB type and loading configuration using only a handful of phenomenological parameters determined from fits.

Following a similar line of thought, an even simpler analytical model known as the ``buffer-grain'' model was proposed~\cite{ELSHAWISH2023104940}. This model also enables us to predict the induced INS distributions for any chosen uniform loading, GB type, and material properties. However, there are several notable differences compared to the bicrystal model.
First of all, the model is not easily generalized to materials with non-cubic crystal lattices and is only applicable to random twist angles.
Another drawback of the buffer-grain model is its reliance on multiple free parameters associated with the properties of invented buffer grains (such as their sizes and stiffness). These parameters need to be carefully adjusted to correctly reproduce the average normal stress at GBs of a given inclination. 

From a physical standpoint, the bicrystal model offers a more faithful representation of a realistic situation as it is based on a more complete set of boundary conditions at the microscale. This allows the model to address also other stress-tensor components besides those along the GB-normal directions. Consequently, the bicrystal model more closely adheres to the original perturbative approach, striving for increased precision in representing the true state as we approach the GB. This, in turn, should also provide a rationale for certain ad-hoc assumptions made in the buffer-grain model.

The paper is structured as follows. In Section~\ref{sec:perturbative} the perturbative modelling approach is presented. Its lowest order approximation (the isotropic model) is explored and compared to finite element (FE) results in Section~\ref{sec:isotropic}. The main part of the paper (the bicrystal model) appears in Section~\ref{sec:bicrystal}.
First the previous results obtained in numerical analyses are summarized and then the pure bicrystal model is introduced. At last the model is convoluted with a suitable Gaussian distribution to imitate the effect of inhomogeneous neighbourhood
and the results of this procedure are compared to numerical simulations.
All technical details are relegated to a set of appendices.


\section{Perturbative modelling approach\label{sec:perturbative}}

\begin{figure}[htb]
	\centering
	\begin{subfigure}[b]{0.32\textwidth}
		\centering
		\includegraphics[width=\textwidth]{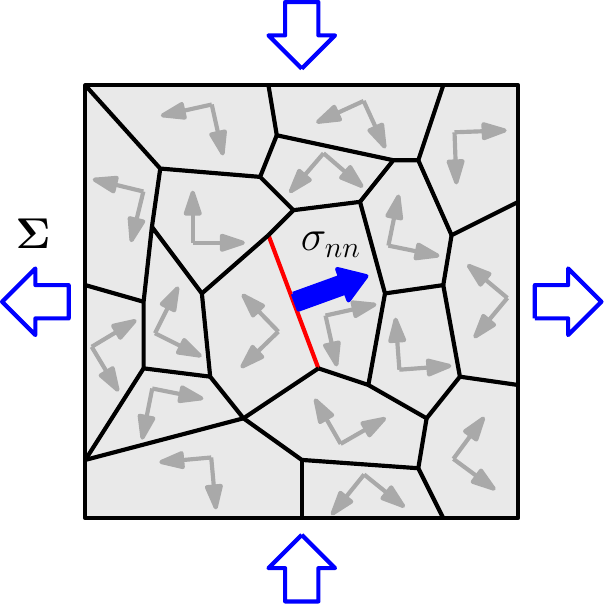}
		\caption{polycrystalline aggregate \\ \centering $(n\to\infty)$}
		\label{fig:polycrystalline-aggregate}
	\end{subfigure}
	\hfill
	\begin{subfigure}[b]{0.32\textwidth}
		\centering
		\includegraphics[width=\textwidth]{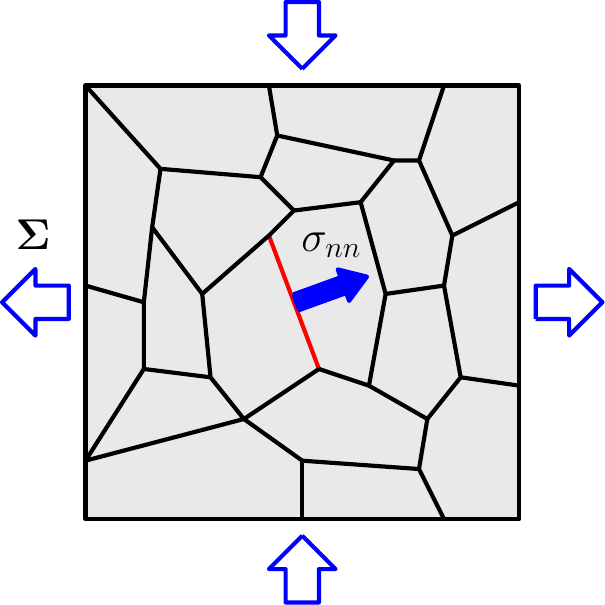}
		\caption{isotropic case \\ \centering $(n = 1)$}
		\label{fig:0th-order-perturbation}
	\end{subfigure}
	\hfill
	\begin{subfigure}[b]{0.32\textwidth}
		\centering
		\includegraphics[width=\textwidth]{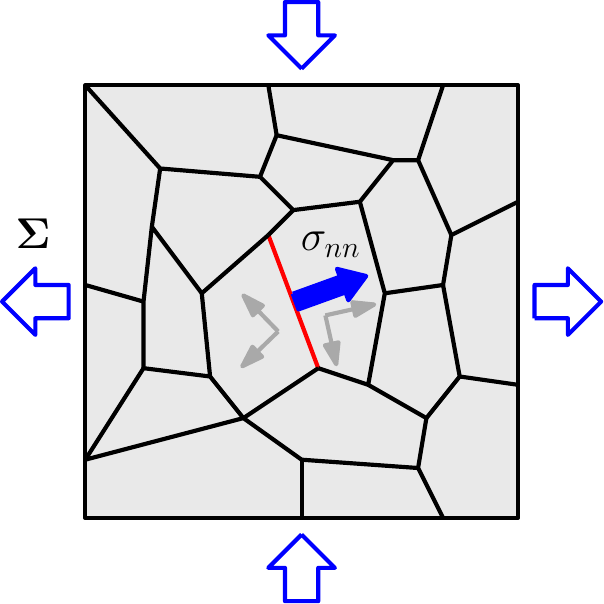}
		\caption{bicrystal approximation \\ \centering $(n = 2)$}
		\label{fig:1st-order-perturbation}
	\end{subfigure}
	\caption{(a) Schematic (2D) depiction of a polycrystalline aggregate with randomly shaped crystal grains (\emph{light gray}) divided by arbitrarily oriented GBs (\emph{black}). The aggregate is exposed to macroscopic external loading $\mathbf{\Sigma}$ which induces a normal stress $\snn$ on the investigated GB (\emph{red}). Crystallographic orientations of grains are indicated by local crystallographic axes (\emph{dark gray}). 
	The lowest two orders of approximation ($n = 1$, $n = 2$) are shown in panels~(b) and~(c). When no crystallographic axes are displayed, the corresponding grain is assumed isotropic (besides being homogeneous as all other grains).}
	\label{fig:schematic}
\end{figure}

Polycrystalline aggregates (Fig.~\ref{fig:polycrystalline-aggregate}) are composed of arbitrary shaped crystal grains in which atoms are arranged in randomly oriented crystal lattices.
Assuming zero crystallographic texture, both the GB orientations and crystallographic orientations of grains are \emph{uniformly} distributed but uncorrelated, i.e., independent of one another.
Computing the induced stress field within the mechanically loaded aggregate requires the knowledge of its entire grain configuration. 
While several experimental techniques exist for determining the distributions of crystallographic orientations in a solid, such as Electron Backscatter Diffraction (EBSD)~\cite{disen2020,johnson2019} in 2D and X-ray Diffraction Contrast Tomography (DCT)~\cite{ludwig2008,johnson2008} in 3D, achieving this goal in practice (i.e., for macroscopic, real-size aggregates) would be impossible.
Besides, the task is also computationally demanding and can only be performed numerically. 

On the other hand, if the objective is to assess the structural integrity of the material under specific mechanical loading conditions, then obtaining precise information about the entire grain configuration and the associated GB stresses would exceed the scope of this analysis (similarly as we typically focus on the collective properties of gas, such as its pressure or temperature, rather than the positions and velocities of individual molecules in it).
Hence, we propose a different approach by treating the problem of induced GB stresses perturbatively, at each order taking into account the less relevant effects.

In the lowest order of perturbation (Fig.~\ref{fig:0th-order-perturbation}), we neglect crystallographic orientations of all the grains and consider just the orientation of GBs for which we want to compute the normal stress components $\snn$. All the grains are thus treated as isotropic.
In the next order approximation (Fig.~\ref{fig:1st-order-perturbation}) we take into account also the crystallographic orientations of the two grains adjacent to the GB.
We can then proceed going to higher orders by adding more and more layers of grains around the bicrystal pair for which their crystallographic orientations are properly taken into account.

Finally, the effect of neglected crystallographic orientations of all the surrounding grains, which were modelled as isotropic (the neighbourhood), is accounted for statistically as a random (Gaussian) fluctuation $\Delta\snn$ on top of the computed $\snn$. While this procedure does not produce the correct values of stresses on individual GBs, their distributions should be much more precise, allowing us to assess the likelihood of cracking initiation (though not also its exact site).

\FloatBarrier


\section{Isotropic approximation\label{sec:isotropic}}

At the lowest order of perturbation, all grains are initially treated as perfectly isotropic (forming a \emph{pure isotropic model}), which enables us to derive analytical expressions for GB stresses. On top of these, we incorporate the statistical influence of the inherent anisotropy of the grains. Finally, we examine the impact of different materials and demonstrate how to generate material-specific estimates, particularly in relation to their elastic anisotropy.
The model's predictions are then scrutinized by comparing them to the results of numerical FE simulations, which serve as a replacement for physical experimentation and are thus meant to represent a realistic response of a perfectly elastic material under mechanical loading.

Since in isotropic approximation all the GBs are equivalent (meaning our analytical model does not distinguish between different GB types), the resulting INS distribution should correspond to that observed on randomly selected GBs in the FE simulation.
For this purpose, we utilized the FE solver Abaqus~\cite{abaqus} in the small strain regime, employing a 3D-periodic Voronoi aggregate~\cite{neper,elshawish2019} with $4000$ grains and $31255$ GBs, which contain a total of $1039562$ quadratic tetrahedral finite elements (C3D10).%
\footnote{For further technical details, please refer to~\cite{ELSHAWISH2021104293} and the references therein.}

To facilitate a comparison with our analytical model, a weighted average of normal stresses on each GB has been produced.
Henceforth, we will refer to this distribution of GB-averaged normal stresses as the \emph{reduced} INS distribution. This term is used to distinguish it from the \emph{full} INS distribution corresponding to normal stresses on all finite elements in the aggregate which are located on the GBs.
Both distributions are weighted, either with respect to the size of GBs or the size of finite elements.%
\footnote{The full INS distribution is comprised of normal stresses $\sigma_{nn,i}^k$ spanning all (pairs of) finite elements $i$ belonging to GBs $k$. These stress values are weighted with respect to the (GB-facet) areas $A_i^k$ of associated finite elements. In the reduced case, a single stress value is first produced for each GB by taking a weighted average over all the finite elements that belong to it.
The mean value of the distribution is the same in both cases, $\langle\snn\rangle = (\sum_k \sum_i A_i^k \sigma_{nn,i}^k)/(\sum_k \sum_i A_i^k)$, but its variance (and consequently the width) is not. For the full INS distribution, it is given by $s^2(\snn) = (\sum_k \sum_i A_i^k (\sigma_{nn,i}^k)^2)/(\sum_k \sum_i A_i^k) - \langle\snn\rangle^2$, whereas in the case of the reduced distribution, it decreases to $s^2(\snn) = (\sum_k (\sum_i A_i^k \sigma_{nn,i}^k)^2/(\sum_j A_j^k))/(\sum_k \sum_i A_i^k) - \langle\snn\rangle^2$.\label{footnote:reduced_vs_full}}

The results are presented in the following way: For the sake of demonstration, $\gamma$-Fe is consistently used in this paper as the reference material.
The validity of the method is then shown for several uniform loadings $\mathbf{\Sigma}$. Only after establishing its efficacy with these loadings, the same technique is extended also to other materials. This extension includes not only cubic materials but also those with more general lattice symmetries (such as orthorhombic).
Relevant properties of the chosen materials required for conducting the FE simulations can be found in Table~\ref{tab:elastic_constants} (Appendix~\ref{app:mat}).

\subsection{Pure isotropic model}

For strictly isotropic grains, the local stress tensor $\bm{\sigma}$ is, 
in every point of the aggregate, equal to the external stress tensor $\mathbf{\Sigma}$.
Since at any chosen GB we are interested in its component along the GB-normal direction ($\sigma_{nn}$), the stress tensor $\mathbf{\Sigma}$ must be transformed from the external loading (\emph{lab}) coordinate system $(X,Y,Z)$ to a local (\emph{GB}) coordinate system $(x,y,z)$:
\begin{align}
	\label{eq:external_load}
	\mathbf{\Sigma}^{\text{lab}}&=\left(
	\begin{array}{ccc}
		\Sigma_{XX} & \Sigma_{XY} & \Sigma_{XZ} \\
		\Sigma_{XY} & \Sigma_{YY} & \Sigma_{YZ} \\
		\Sigma_{XZ} & \Sigma_{YZ} & \Sigma_{ZZ}
	\end{array}
	\right) \quad \to \quad
	\mathbf{\Sigma}^{\text{GB}}=\mathbf{\mathcal{R}} \, \mathbf{\Sigma}^{\text{lab}} \, \mathbf{\mathcal{R}}^T=
	\left(
	\begin{array}{ccc}
		\Sigma_{xx} & \Sigma_{xy} & \Sigma_{xz} \\
		\Sigma_{xy} & \Sigma_{yy} & \Sigma_{yz} \\
		\Sigma_{xz} & \Sigma_{yz} & \Sigma_{zz}
	\end{array}
	\right) \ ,
\end{align}
where
\begin{align}
	\label{eq:Rlab}
	\mathbf{\mathcal{R}}& = \left(
	\begin{array}{ccc}
		\phantom{-}\cos \theta \cos \psi \cos \phi - \sin \psi \sin \phi & \phantom{-}\cos \theta \sin \psi \cos \phi + \cos \psi \sin \phi & -\sin \theta \cos \phi \\
		-\cos \theta \cos \psi \sin \phi - \sin \psi \cos \phi & -\cos \theta \sin \psi \sin \phi + \cos \psi \cos \phi & \phantom{-}\sin \theta \sin \phi \\
		\sin \theta \cos \psi & \sin \theta \sin \psi & \cos \theta \\
	\end{array}
	\right)
\end{align}
is a rotation matrix containing $3$ Euler angles ($\psi$, $\theta$, $\phi$) corresponding to a sequence of rotations ($\mathbf{\mathcal{R}_1}$, $\mathbf{\mathcal{R}_2}$, $\mathbf{\mathcal{R}_3}$) about $\hat{Z}$, $\mathbf{\mathcal{R}}_1 \hat{Y}$ and $\mathbf{\mathcal{R}}_2 \mathbf{\mathcal{R}}_1 \hat{Z} = \hat{z}$, respectively, such that $\mathbf{\mathcal{R}} = \mathbf{\mathcal{R}_3} \mathbf{\mathcal{R}_2} \mathbf{\mathcal{R}_1}$,
and local \mbox{$z$-axis} is directed along the GB normal $\hat{n}$.
Consequently, the GB-normal stress 
\begin{align}
	\begin{split}
		\sigma_{nn} = \Sigma_{zz} & = \left ( \cos^2\theta \right ) \, \Sigma_{ZZ} + \left ( \sin^2\theta \, \cos^2\psi \right ) \, \Sigma_{XX} + \left ( \sin^2\theta \, \sin^2\psi \right ) \, \Sigma_{YY} + \\
		& + \left ( \sin^2\theta \, \sin 2\psi \right ) \, \Sigma_{XY} + \left ( \sin 2\theta \, \cos\psi \right ) \, \Sigma_{XZ} + \left ( \sin 2\theta \, \sin\psi \right ) \, \Sigma_{YZ}
	\end{split}
\label{eq:INS_isotropic}
\end{align}
depends just on the polar angle $\theta$ and azimuthal angle $\psi$ (but not on the twist angle $\phi$, that is only determining the precise orientation of the local $x$ and $y$ axes).
In an aggregate with randomly oriented GBs and no preferential direction, the corresponding INS distribution $\text{PDF}(\sigma_{nn})$ can be either obtained analytically or generated through Monte Carlo sampling over the uniformly distributed Euler angles $\psi$ and $\cos\theta$.%
\footnote{The result is particularly simple for uniaxial loading, when the distribution can be easily expressed in analytical form. Without loss of generality, the lab coordinate system can be chosen so that its $Z$-axis points in the direction of loading. Then the probability density function (PDF) 
is given by 
$\text{PDF}(\tilde{\sigma}_{nn}) := \left (\tfrac{dP}{d\!\cos\theta} \right ) \,  \left (\tfrac{d\tilde{\sigma}_{nn}}{d\!\cos\theta}\right )^{-1} = \left (2 \cos\theta \right )^{-1} = \left (2 \sqrt{\tilde{\sigma}_{nn}}\right )^{-1}$,
with $\tilde{\sigma}_{nn} := \tfrac{\sigma_{nn}}{\Sigma_{ZZ}}$
in the range of $[0,1]$. This distribution is depicted by the \emph{green} curve in the left panel of Fig.~\ref{fig:random_distribution_GammaFe}. Its mean value and standard deviation are $\langle \tfrac{\sigma_{nn}}{\Sigma_{ZZ}}\rangle = 1/3$ and $s(\tfrac{\sigma_{nn}}{\Sigma_{ZZ}}) = 2/(3\sqrt{5}) \approx 0.298$, respectively.}

\FloatBarrier

\subsection{Gaussian modulation}

\begin{figure}[!htb]
	%
	\includegraphics[width = 0.328\textwidth, valign=c]{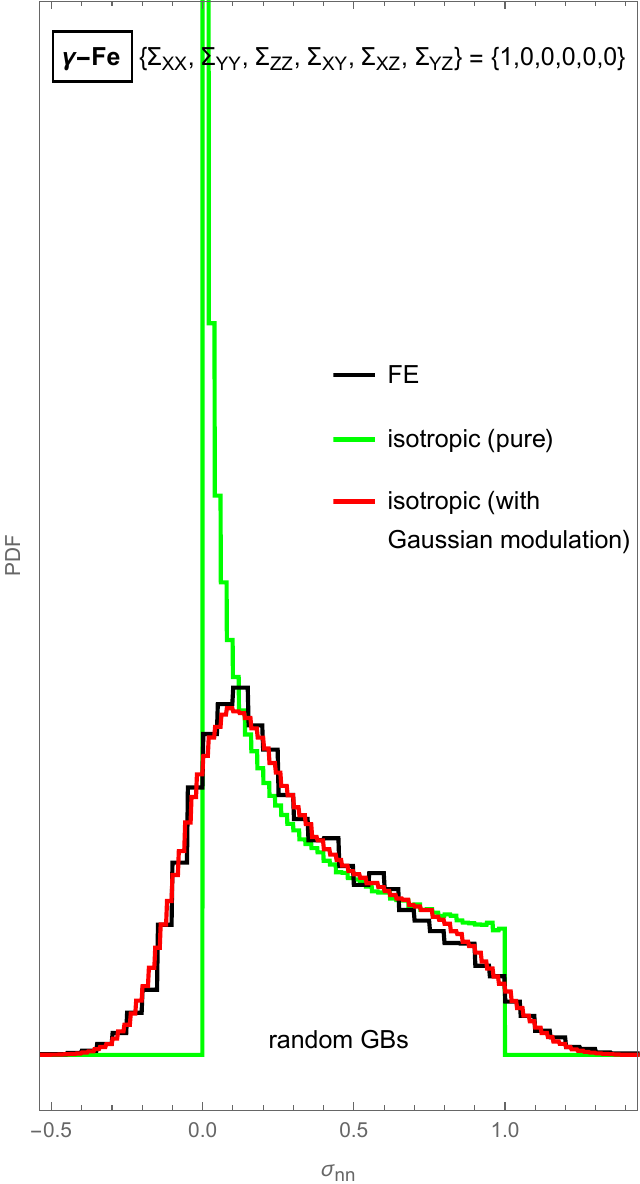}
	\includegraphics[width = 0.328\textwidth, valign=c]{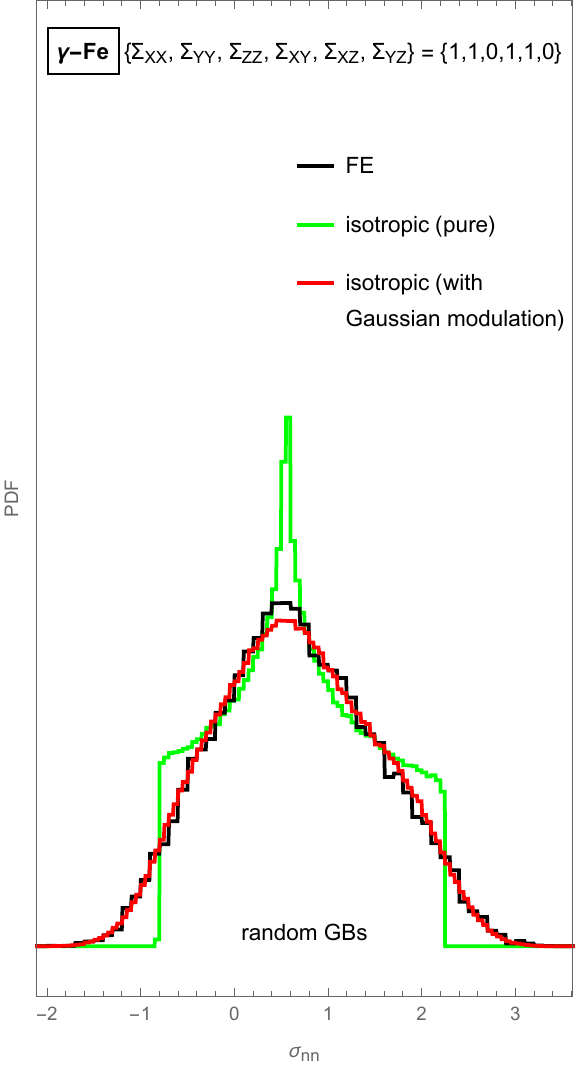}
	\includegraphics[width = 0.328\textwidth, valign=c]{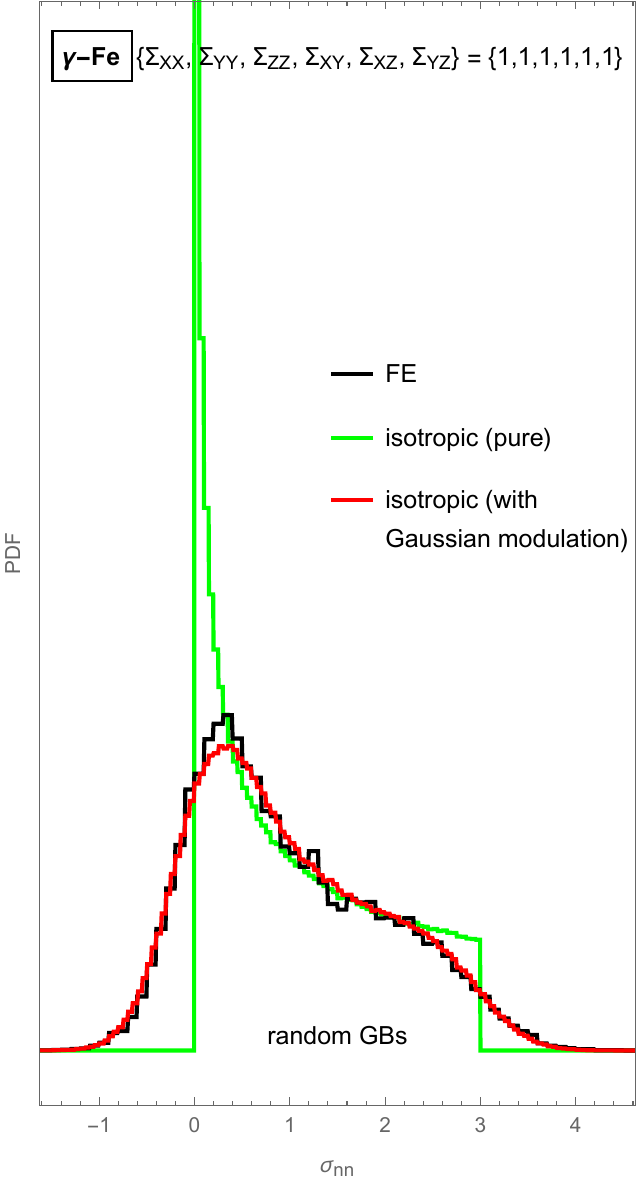}
	\caption{Reduced INS distributions for $\gamma$-Fe and three uniform loadings. Comparison of FE results (\emph{black}; $N=31255$) and estimates from the analytical isotropic model without (\emph{green}) and with Gaussian fluctuations (\emph{red}; $\sigma_G^{(1)} = 0.136 \times \Sigma_{\text{mis}}$).}
	\label{fig:random_distribution_GammaFe}
\end{figure}

The problem with \emph{pure} isotropic model specified in Eq.~\eqref{eq:INS_isotropic} is that its results depend solely on the external loading $\mathbf{\Sigma}$, failing to account for the material properties. This means the model makes identical predictions for all materials, which results in significant flaws when applied to highly anisotropic materials.  That is clearly visible in Fig.~\ref{fig:random_distribution_GammaFe}, where the resulting isotropic INS distributions (\emph{green}) appear notably sharper and narrower than those generated by the FE simulations (\emph{black}). 

The wider and smoother FE distributions can be attributed to the fact that, in realistic aggregates, GBs are not surrounded by an isotropic medium but by specific configurations of anisotropic grains which determine the actual stress experienced at each GB site. 
In other words, it is the non-homogeneous nature of the neighbourhood that dictates the amount of external stress $\mathbf{\Sigma}$ that gets ``transmitted'' to a particular GB. Consequently, the local GB stress is not necessarily equal to $\Sigma_{zz}$, and different GBs can thus generate different responses $\snn$ even when they have the same orientation. 
As a result, realistic INS distributions tend to be wider.
 
This widening effect is determined by numerous independent degrees of freedom associated with the precise arrangement of grains.  
Their contributions to the induced GB stresses $\snn$ act as random fluctuations superimposed onto the estimates provided by the pure isotropic model.
With the central limit theorem in mind, it is reasonable to assume that these fluctuations follow a normal distribution pattern.
Written schematically:
\begin{align} 
	\textbf{isotropic case} + \textbf{Gaussian fluctuations} & \approx \textbf{INS distribution on random GBs} \ . \label{eq:perturbativity_1st_order}
\end{align}

To recap, in order to mimic the effect of inhomogeneous neighbourhood surrounding each GB in anisotropic aggregate, we will model it by a Gaussian distribution~($g$) of width $\sigma_G$ and centered at the origin. 
We will also assume that this distribution has no correlation with the INS distribution generated by the pure isotropic model~($f$). As a result, the final INS distribution~($h$) will be obtained through the convolution of both distributions:
\begin{align}
	f(\snn) & := \text{PDF}(\sigma_{nn}) = \frac{dP}{d\sigma_{nn}} \ , \label{eq:function_f} \\
	g(\Delta\snn) & := \frac{1}{\sqrt{2 \pi} \, \sigma_G} \, e^{-\frac{1}{2} \left ( \frac{\Delta\snn}{\sigma_G} \right )^2} \ , \\
	h(\snn) & = (f\ast g) (\snn) = \int^{\infty}_{-\infty} f(\tau) \, g(\snn-\tau) \, d\tau \ . \label{eq:function_h}
\end{align}
The $n$-th moment of the convolution is computed as
\begin{align}
	\mu_h^{(n)} & = \sum_{i=0}^{n} \binom{n}{i} \mu_f^{(i)} \mu_g^{(n-i)} \ , \label{eq:moments of convolution}
\end{align}
where all the odd central moments of Gaussian distribution $g$ vanish, while for even $k$, they reduce to $\mu_g^{(k)} = (k - 1)!! \cdot \sigma_G^k$.
Since both $f$ and $g$ are also properly normalized, their zero-th moments ($\mu_f^{(0)}$, $\mu_g^{(0)}$) are equal to $1$. Hence,
\begin{align}
	\mu_h^{(1)} & = \langle \snn \rangle_h = \mu_f^{(1)} \ , \label{eq:moment1_conv_isotropic} \\
	\mu_h^{(2)} & = \langle (\snn - \langle \snn \rangle_h)^2 \rangle_h = \mu_f^{(2)} + \sigma_G^2 \ , \label{eq:moment2_conv_isotropic}
\end{align}
where mean value and variance of isotropic INS distribution are given by
\begin{align}
	\mu_f^{(1)} & = \langle \snn \rangle_f = \frac{1}{3} \Tr\mathbf{\Sigma} \ , \label{eq:moment1_isotropic} \\
	\mu_f^{(2)} & = \langle (\snn - \langle \snn \rangle_f)^2 \rangle_f := s^2(\sigma_{nn}) = \left (\frac{2}{3\sqrt{5}} \Sigma_{\text{mis}} \right )^2 
	\ . \label{eq:moment2_isotropic}	 
\end{align}
Both of these expressions depend on the macroscopic loading $\mathbf{\Sigma}$ through rotational invariants ($\Tr{\mathbf{\Sigma}}$ or $\Sigma_{\text{mis}}$, respectively), which retain the same form in all Cartesian coordinate systems.
The von Mises stress $\Sigma_{\text{mis}}$ is defined as:
\begin{align}
	\Sigma_{\text{mis}} & := \frac{\sqrt{3}}{\sqrt{2}} \sqrt{\Tr(\mathbf{\Sigma}^2_{\text{dev}})} \ . \label{eq:von Mises}
\end{align}
It is related to the traceless \emph{deviatoric} stress tensor
\begin{align}
	\mathbf{\Sigma}_{\text{dev}} & := \mathbf{\Sigma} - \tfrac{1}{3}  \Tr(\mathbf{\Sigma}) \, \mathbb{1}_{3\times 3} \ , \label{eq:deviatoric}
\end{align}
that is typically associated with shear deformations that preserve the volume of the aggregate while changing its shape.

The same loading dependence has also been observed in FE simulations~\cite{ELSHAWISH2023104940}. Specifically, it has been found that
the first two (central) statistical moments of both the reduced and the full INS distribution are either equal to $\tfrac{1}{3} \Tr\mathbf{\Sigma}$ or proportional to $\Sigma_{\text{mis}}^2$, respectively.%
\footnote{For non-cubic materials, the latter relation does not hold exactly, although it remains a good approximation for most loading configurations. If it were exact, then under \emph{hydrostatic} loading $\mathbf{\Sigma}_{\text{hyd}} = \Sigma_0 \, \mathbb{1}_{3\times 3}$ (for which $\Sigma_{\text{mis}} = 0$), the true INS distribution would be infinitely narrow: $\text{PDF}(\sigma_{nn}) = \delta(\snn-\Sigma_0)$. However, this is not the case. Therefore, for non-cubic lattice symmetries, the width of fluctuation $g(\Delta\snn)$ does not precisely scale with $\Sigma_{\text{mis}}$ (contrary to what Fig.~\ref{fig:width_Gaussian_random} seems to suggest for CaSO$_4$).} 
Identifying the convoluted distribution $h$ with $\text{PDF}(\sigma_{nn})$ obtained from the FE simulation not only implies that $\sigma_G$ should also scale with $\Sigma_{\text{mis}}$, but it also allows us to use Eq.~\eqref{eq:moment2_conv_isotropic} to determine the value of $\sigma_G$.%
\footnote{In principle, we could take any moment of $h$, obtain its value from the FE distribution, and use it to fit the value of $\sigma_G$. However, from a practical perspective, it is advisable to use the lowest order $(n\geq 2)$ for which $\mu_f^{(n)}\neq 0$, as this yields the most accurate result for an imprecisely known FE distribution (computed only on a finite aggregate).}

Finally, the modelling approach is validated in Fig.~\ref{fig:random_distribution_GammaFe}, where the computed $\sigma_G$ is utilized to generate the convoluted distributions $h$ (\emph{red}) for several different%
\footnote{Two of them ($\{\Sigma_{XX},\Sigma_{YY},\Sigma_{ZZ},\Sigma_{XY},\Sigma_{XZ},\Sigma_{YZ}\} = \{1,0,0,0,0,0\}$ and $\{1,1,1,1,1,1\}$) are rotationally equivalent, i.e., the corresponding tensors are related by rotation $\mathbf{\mathcal{R}}$ for $\psi = 0$, $\cos\theta = \sqrt{2/3}$, and $\phi = -\pi/4$, up to an overall scaling factor of $3$, cf.~Eqs.~\eqref{eq:external_load}--\eqref{eq:Rlab}. In an infinite, untextured (i.e, macroscopically isotropic and thus rotationally invariant) aggregate, these two loading configurations merely represent different choices of lab coordinates. However, in our context, they serve as additional validation that the aggregate used in our numerical simulations is sufficiently large to disregard the finite aggregate-size effects (which can affect the smoothness of the INS distribution and its variation upon aggregate rotation).}
uniform loadings $\mathbf{\Sigma}^{\text{lab}}$ (but here showing only three such examples) and then compare them to the (reduced) distributions of GB stresses from FE simulations (\emph{black}).

\begin{figure}[htb]
	\centering
	\includegraphics[width = 0.5 \textwidth, valign=c]{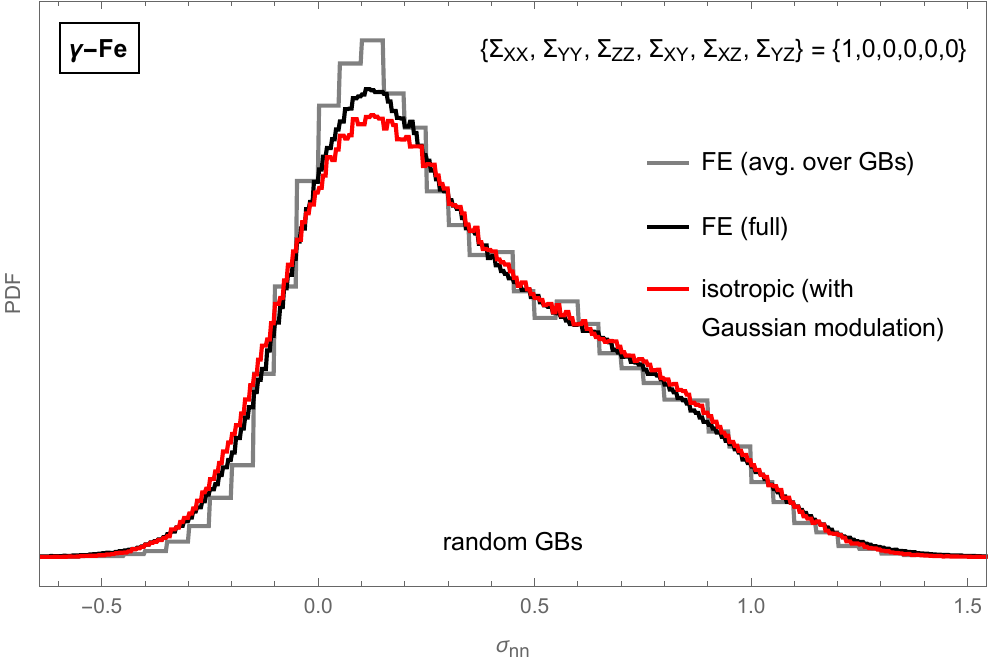}
	\caption{A comparison between the modelled (\emph{red}; $\sigma_G = 0.167 \times \Sigma_{\text{mis}}$) and realistic (\emph{black}; $N = 3105904$) full INS distribution for $\gamma$-Fe and uniaxial tensile loading.
	The reduced INS distribution (\emph{gray}; $N=31255$) is also shown for comparison.
	Both FE distributions are almost identical due to a relatively modest widening needed to correct for the averaging of stresses over the GBs; $(\sigma_G^{(2)})^2 \ll (\sigma_G^{(1)})^2$.}
	\label{fig:random_distribution_NOavg}
\end{figure}

Another factor contributing to the widening of INS distributions in realistic (FE) scenarios is the variation of induced stress across the GB, in contrast to the constant stress assumed in our simplified model.
To account for this behaviour and thus describe the \emph{full} INS distribution (instead of just the \emph{reduced} version), we can perform one more convolution with additional Gaussian (characterized by a different width $\sigma_G^{(2)}$). Since the convolution of two Gaussian distributions results in another Gaussian distribution whose variance is the sum of variances of the original distributions, this procedure is equivalent to a single convolution with a Gaussian of width:
\begin{align}
\sigma_G = \sqrt{(\sigma_G^{(1)})^2+(\sigma_G^{(2)})^2} \ .
\end{align}
Henceforth we will  denote the width of the Gaussian distribution in the reduced case as $\sigma_G^{(1)}$ and the width in the full case as $\sigma_G$. 
The effect of this additional convolution with a normal distribution of width $\sigma_G^{(2)}$ on the shape of the INS distribution is depicted in Fig.~\ref{fig:random_distribution_NOavg}.

\FloatBarrier

\subsection{Effect of the material}

The same procedure can be performed also for other materials.
Because the predictions of the pure isotropic model are material-independent, 
the differences in INS distributions for various materials have to emerge
at the convolution level (owing to distinct, material-specific widths of Gaussian distributions).
In other words, material properties affect only the width of the fluctuations.
Therefore, $\sigma_G$ is also the only parameter that needs adaptation in our analytical description.

\begin{figure}[htb]
	\centering
	\includegraphics[width = \textwidth]{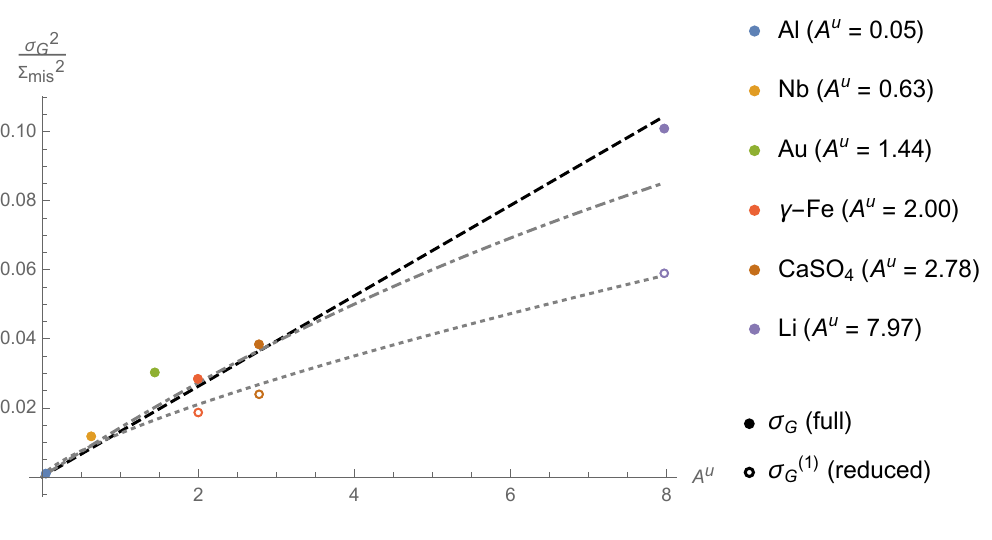}
	\caption{To simulate the response of anisotropic materials to external loading, normally distributed random fluctuations are added to the GB-normal stresses induced in isotropic crystal. The width $\sigma_G$ of the fluctuation distribution is determined by matching the standard deviation of FE distribution. The square of this width is shown as a function of the universal elastic anisotropy index $A^u$ for several materials. Additionally, a \emph{linear} fit (\emph{black dashed line}) is depicted, while the \emph{gray dash-dotted curve} represents the fit from Eq.~\eqref{eq:s_FE}. For comparison, a $\sigma_G^{(1)}$ fit~\protect\cite{ELSHAWISH2023104940} (\emph{gray dotted curve}) is also included, demonstrating that the predominant widening effect arises from the anisotropic neighbourhood of the GB, rather than from the averaging of stresses over each GB (i.e., $\sigma_G^{(1)} > \sigma_G^{(2)}$).}
	\label{fig:width_Gaussian_random}
\end{figure}

Most significant is its correlation with the material's anisotropy.
While there should be no fluctuations in the isotropic case (meaning the Gaussian is infinitely narrow), $\sigma_G$ is expected to increase with the growing anisotropy of the material.
As a measure of material anisotropy, we utilize the \emph{universal elastic anisotropy index} $A^u$~\cite{ranganathan} since it is applicable
to any lattice symmetry (not only cubic).
By matching the width of a (full) INS distribution produced in FE simulation, $\sigma_G$ can be calculated for any material and then shown as a function of its $A^u$, see Fig.~\ref{fig:width_Gaussian_random}.%
\footnote{\looseness-1 Although our primary focus is on 
the actual (full) GB-stress distributions, the same procedure can be applied to the reduced INS distributions, thereby generating the values of $\sigma_G^{(1)}$. However, the statistics in that case are considerably worse, making the fit $\sigma_G^{(1)} = 0.07 \, (A^u)^{0.37} \, \Sigma_{\text{mis}}$, proposed in~\cite{ELSHAWISH2023104940} (including the provision that a $60$\% larger value should be used for random GBs), less reliable (see \emph{gray dotted} curve in Fig.~\ref{fig:width_Gaussian_random}).}

The observed pattern is reasonably well-fitted by the \emph{linear} dependence 
\begin{align}
\sigma_G^2 & = k \, A^u \ , \label{eq:linear_fit}
\end{align} 
which appropriately approaches zero in the isotropic limit ($A^u = 0$).
The current best estimate for the proportionality factor is $k = 0.013$.
In the past, an approximate relation between the standard deviation of the INS distribution on random GBs and the material's anisotropy was proposed for uniaxial loading~\cite{ELSHAWISH2018354}:
\begin{align}
	s_{FE} \left (\frac{\snn}{\Sigma} \right )& \approx \frac{4}{3\sqrt{5}} \, \frac{A^u+6}{A^u+12} \ . 
	\label{eq:s_FE}
\end{align}
However, this relation tends to undershoot the true values of $s_{FE}(\snn)$, especially for large $A^u$. As a result, the approximation $(\frac{2}{3\sqrt{5}})^2+\sigma_G^2 = \left (\frac{4}{3\sqrt{5}} \, \frac{A^u+6}{A^u+12} \right )^2$ also underestimates the size of Gaussian fluctuations $\sigma_G$ for materials with larger anisotropy, as shown by the \emph{gray dash-dotted} curve in Fig.~\ref{fig:width_Gaussian_random}.
 
\begin{figure}[htb]
	\centering
	\includegraphics[width = 0.325\textwidth]{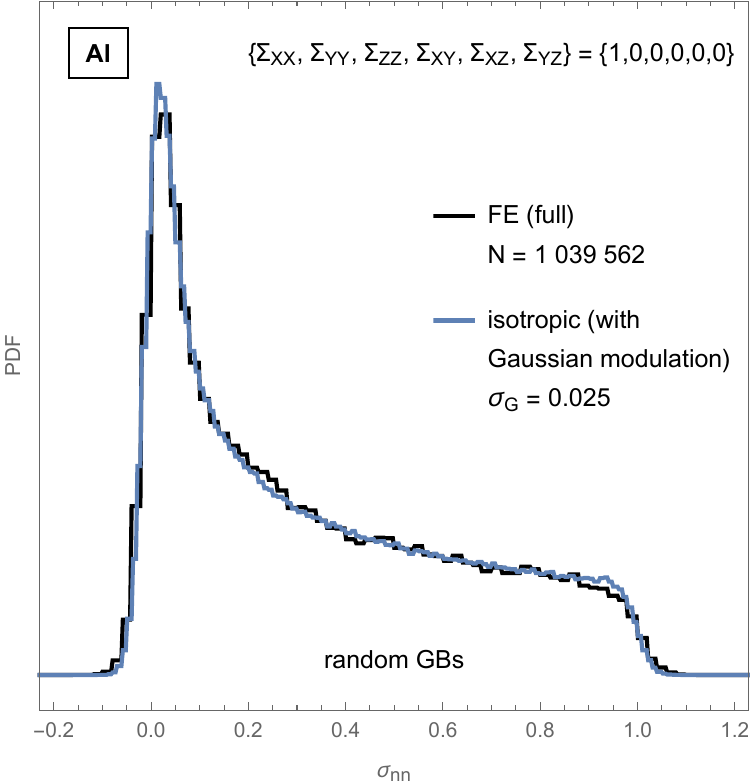}
	\includegraphics[width = 0.325\textwidth]{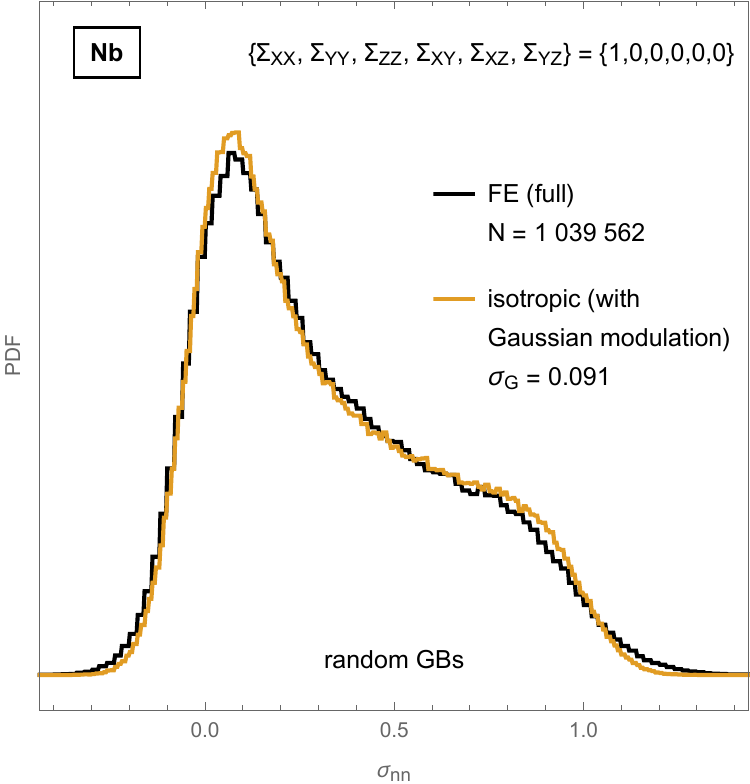}
	\includegraphics[width = 0.325\textwidth]{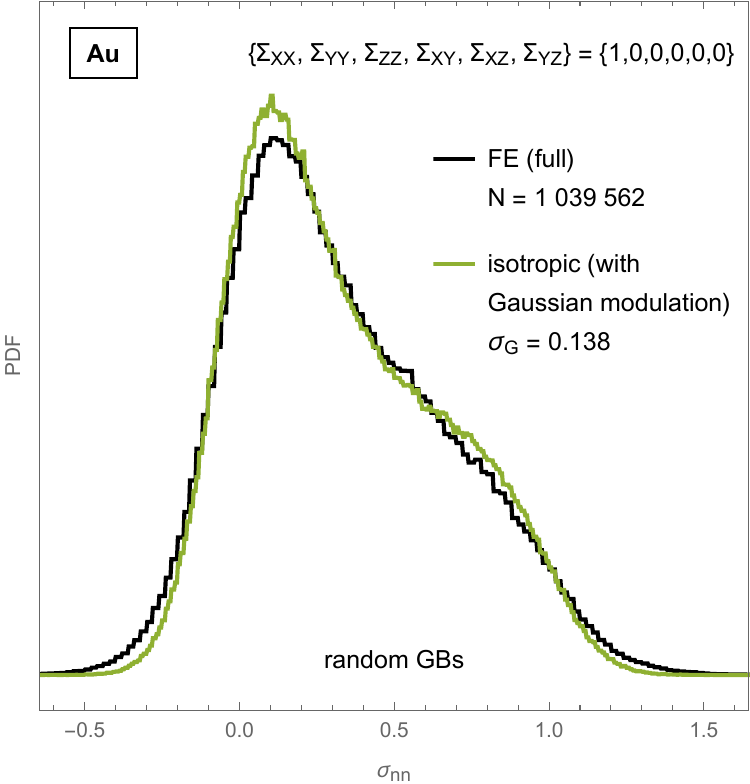}
	\includegraphics[width = 0.325\textwidth]{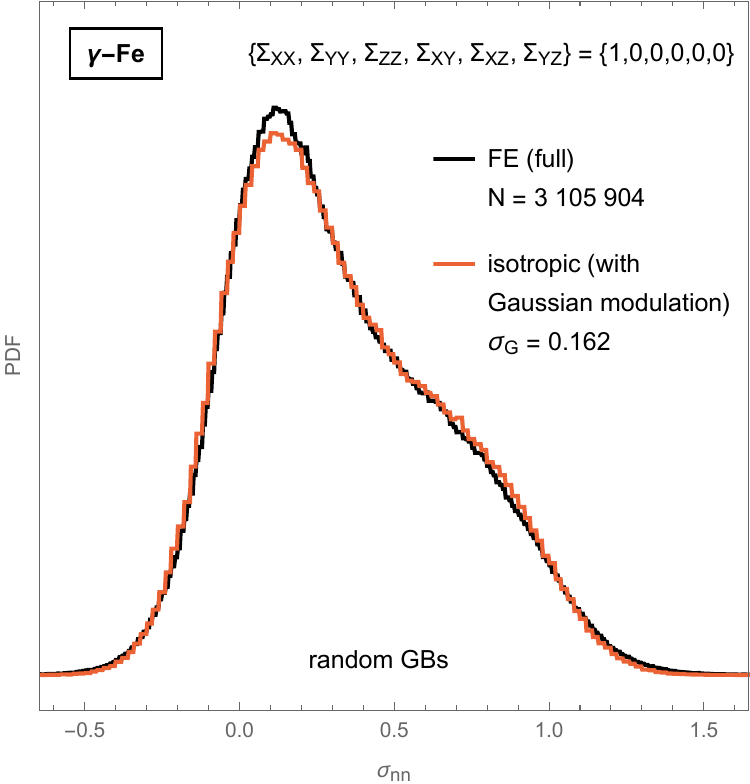}
	\includegraphics[width = 0.325\textwidth]{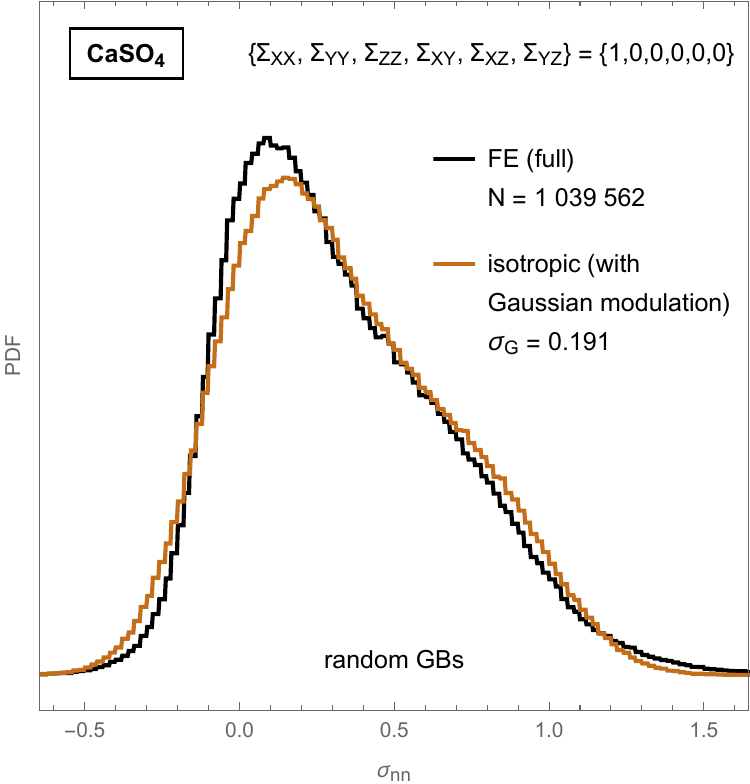}
	\includegraphics[width = 0.325\textwidth]{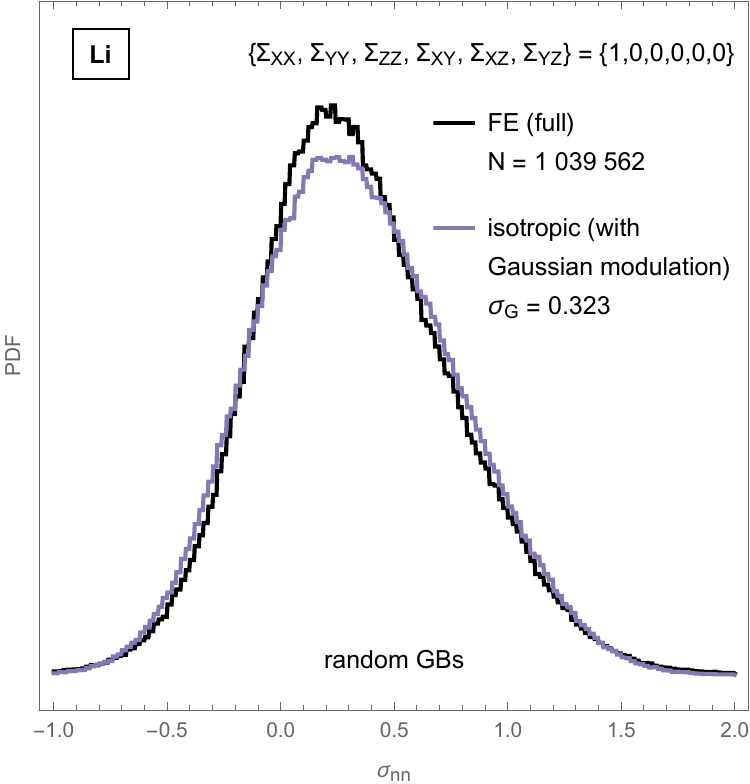}
	\caption{INS distributions induced by uniaxial tensile loading for various selected materials. A comparison is shown between the results of numerical simulation and an isotropic model convoluted with a Gaussian distribution. The value of its width $\sigma_G$ was determined from the linear fit shown in Fig.~\ref{fig:width_Gaussian_random}. The number of finite elements in the aggregate is denoted by $N$. For $\gamma$-Fe, the data from three different Voronoi aggregates has been combined, which effectively corresponds to using a single aggregate with $12000$ grains.}
	\label{fig:distributions_random_100000_materials}
\end{figure}

A comparison between the modelled INS distributions (with $\sigma_G$ determined from the linear fit in Eq.~\eqref{eq:linear_fit}) and those obtained numerically is presented in Fig.~\ref{fig:distributions_random_100000_materials}. 
A (fairly) good agreement between both can be observed.

\begin{figure}[htb]
	\centering
	\includegraphics[width = 0.49\textwidth]{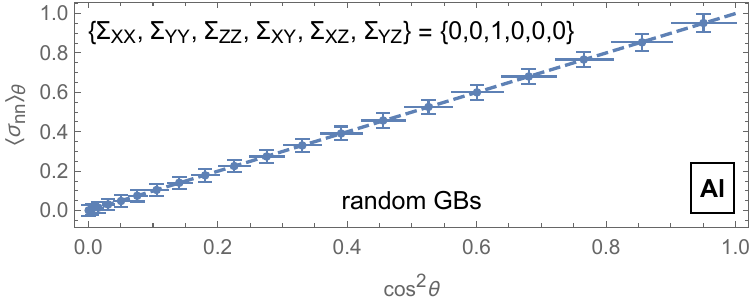}
	\includegraphics[width = 0.49\textwidth]{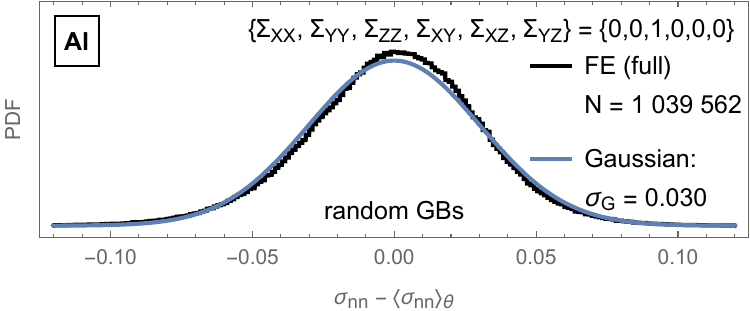}
	\includegraphics[width = 0.49\textwidth]{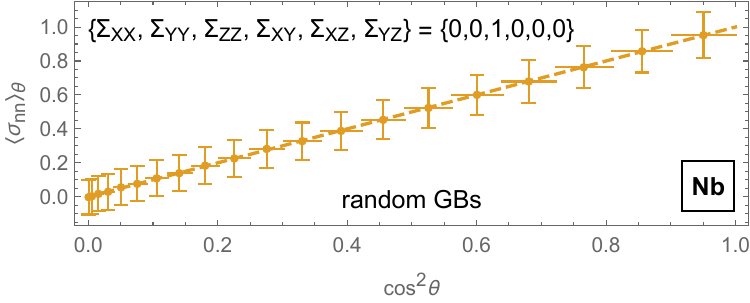}
	\includegraphics[width = 0.49\textwidth]{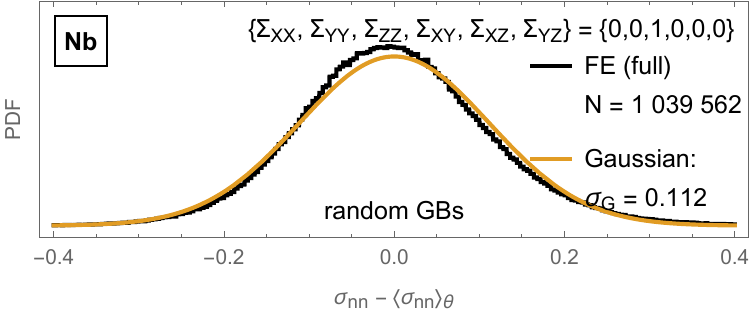}
	\includegraphics[width = 0.49\textwidth]{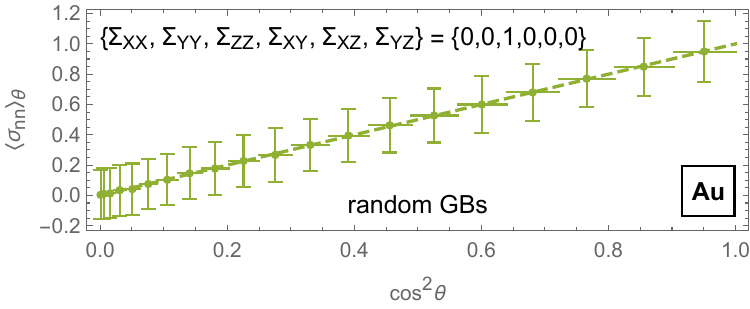}
	\includegraphics[width = 0.49\textwidth]{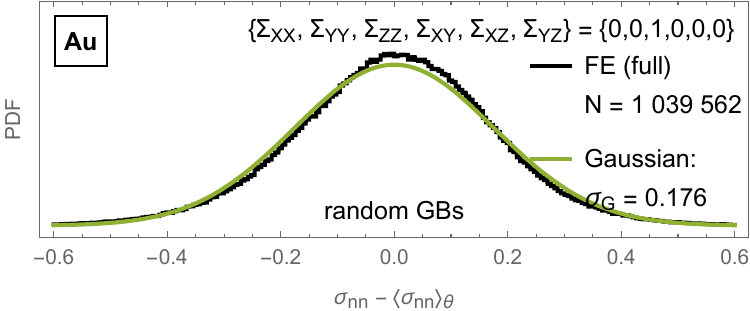}
	\includegraphics[width = 0.49\textwidth]{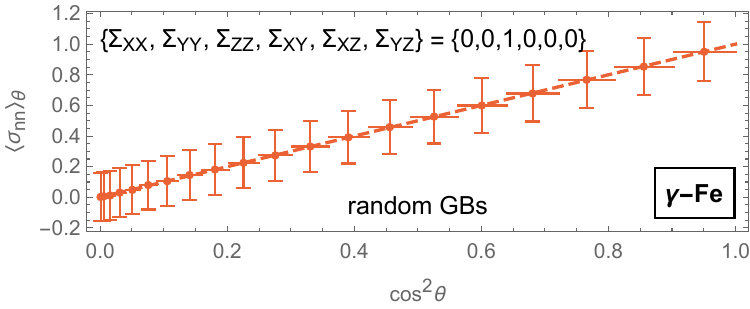}
	\includegraphics[width = 0.49\textwidth]{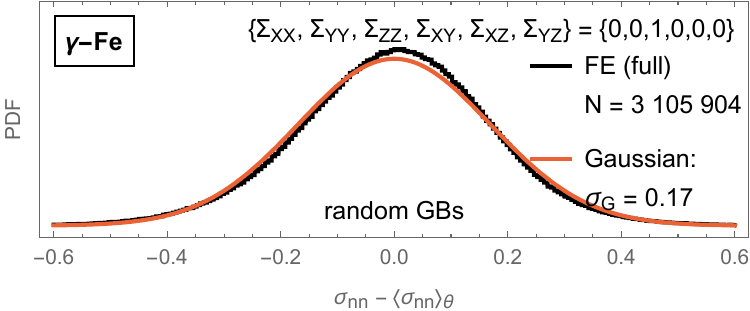}
	\includegraphics[width = 0.49\textwidth]{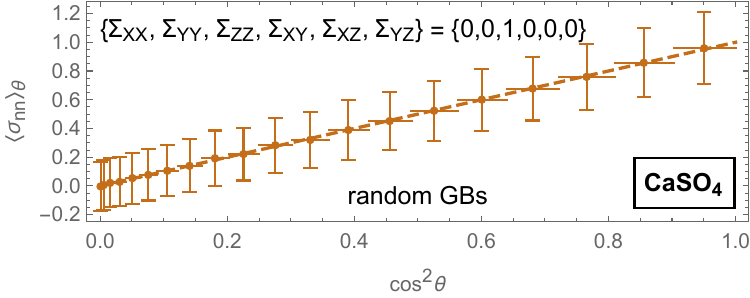}
	\includegraphics[width = 0.49\textwidth]{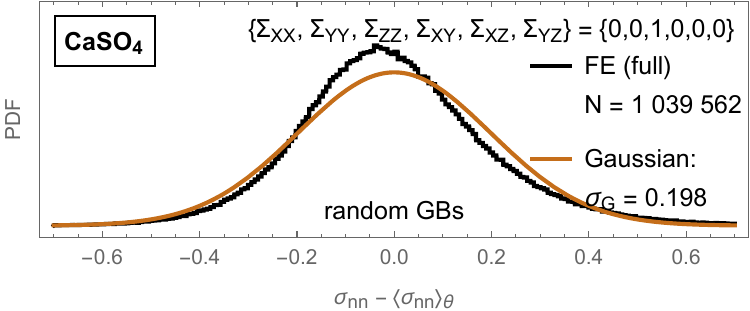}
	\includegraphics[width = 0.49\textwidth]{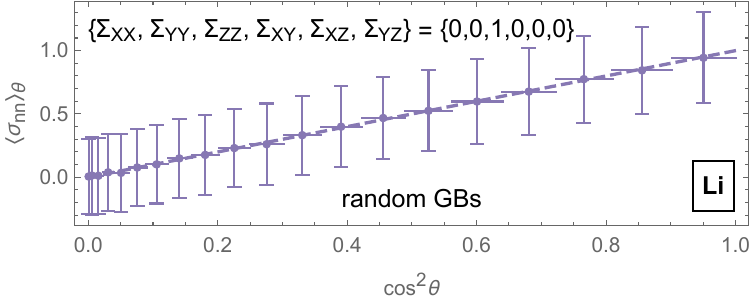}
	\includegraphics[width = 0.49\textwidth]{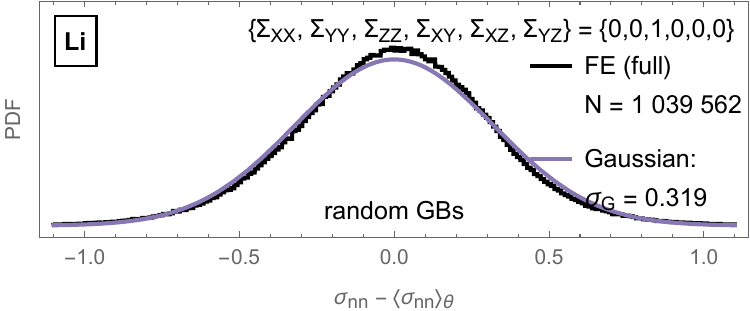}
	\caption{Left panels: The value of $\snn$ averaged over all (random) GBs in the aggregate with the same inclination $\theta$ relative to uniaxial external loading. Horizontal lines indicate the width of the bin $\cos\theta\pm 0.025$, while vertical bars correspond to the standard deviation of the $\snn$ distribution in each bin (which appears to be practically independent of $\theta$). The dashed line represents the value for a perfectly isotropic material ($\snn = \Sigma_{zz} = \cos^2\theta$). Right panels: The distribution of normal stresses around the values predicted by the pure isotropic model. Except for CaSO$_4$, it is very well-described by a Gaussian of width $\sigma_G$ (whose value was fitted rather than determined by the linear approximation shown in Fig.~\ref{fig:width_Gaussian_random}).}
	\label{fig:fluctuation_randomGBs}
\end{figure}

The reliability of the isotropic approximation for GB-normal stresses on random GBs is further illustrated in Fig.~\ref{fig:fluctuation_randomGBs}. As observed in its left panels, the average value of $\snn$ at GBs with the same inclination (denoted by $\langle\snn\rangle_{\theta}$) closely aligns with the prediction of the pure isotropic model. Additionally, the stress fluctuation around $\langle\snn\rangle_{\theta}$ is accurately captured by a Gaussian distribution, with a width that remains almost constant across different inclinations.%
\footnote{More precisely, $\sigma_G$ is a gently ascending linear function of $\cos^2\theta$.}
The validity of these observations holds true irrespective of whether we consider the reduced or full INS distributions. The only difference is that the Gaussian is wider in the latter case, as explained in Footnote~\ref{footnote:reduced_vs_full}.
For crystal lattices with non-cubic symmetry, the average value of $\snn$ at a chosen inclination is still predicted by the isotropic model, but the stresses no longer appear to be normally distributed around this value. Instead, their distribution appears slightly asymmetric.

\FloatBarrier

In conclusion, the presented isotropic model contains a single free parameter $k$ requiring fitting, and yet manages to successfully predict
the INS distributions on random GBs for any chosen material and uniform external loading.
However, this is still not sufficient for accurately estimating the likelihood of intergranular cracking initiation, as different GBs possess distinct thresholds for cracking.
To achieve that goal, one is forced to go beyond the isotropic approximation and consider the INS distributions on individual GB types.


\section{Bicrystal model\label{sec:bicrystal}}

Each GB type is characterized by a relationship between the GB plane and the orientations of the crystal lattices on either side of it. The GB orientation is specified by its normal vector, which contributes two degrees of freedom in each of the two crystallographic systems (being expressed in them as unit-length vectors $(a,b,c)$ and $(d,e,f)$, respectively), while a relative rotation about the GB normal between both crystallographic systems introduces another parameter, known as the twist angle ($\Delta\omega$). Alternatively, three degrees of freedom are required to relate the crystallographic systems of both adjacent grains, and an additional two degrees to specify the GB plane in one of them, again resulting in a total of five degrees of freedom to determine the GB type. 

To investigate the INS distribution across all GBs of a particular $[abc]$-$[def]$-$\Delta\omega$ type, one would require either a considerably large random aggregate (with crystallographic orientation of each grain and normal direction of every GB randomly assigned) to accumulate sufficient statistics, or a dedicated aggregate in which the number of GBs of that type is increased by design. 
We opted for the latter approach and thus created a distinct aggregate for each studied GB type. These aggregates share the same grain topology and FE-mesh structure, and differ only in the crystallographic orientations of the grains. However, due to topological constraints, only a limited subset of all GBs in them can be assigned a special character.

To enhance the statistical significance of our results, a selected GB type was imposed on the largest available GBs in the aggregate --- those with the largest area, which also typically contain the highest number of finite elements.  
In our case, featuring an aggregate with $4000$ grains and a total of $31255$ GBs, only $N = 1636$ GBs (representing approximately $17\%$ of the entire GB area and containing $159461$ finite elements) correspond to the designated GB type.
As a result, the accuracy of our numerical results is noticeably worse than in the isotropic approximation, where all GBs could be included in the analysis. Consequently, the FE distributions presented for individual GB types are less refined.

In this section, we begin by presenting the empirical findings from our previous numerical studies~\cite{ELSHAWISH2021104293,ELSHAWISH2023104940}. Then the (pure) bicrystal analytical model is introduced along with its general solution. Finally, the model predictions are assessed through a comparison with the FE results. Similar to the isotropic model, Gaussian modulation is introduced to improve the INS distributions.
However, unlike in the isotropic scenario, a single parameter ($\sigma_G$) is not sufficient to accurately describe the INS distribution across the GBs of a specific type. 
The reason behind this lies in the fact that the standard deviation of its pure bicrystal-model estimate can exceed the FE value if all grains are assumed to have average, aggregate-scale (bulk) elastic properties (despite such assumption manages to correctly reproduce the distribution's mean value).
Hence, we introduce two additional parameters, $\alpha$ and $\beta$, which are related to the stiffness of the GB. All three parameters ($\alpha$, $\beta$, and $\sigma_G$) are allowed to vary as functions of both the GB type and material properties.

While we have successfully found a simple phenomenological relation that links $\alpha$~and $\beta$ with $E_{12}$, this proves more difficult for $\sigma_G$ due to its sensitivity to uncertainties arising from finite aggregate size in numerical simulations. Consequently, expressing these parameters as functions of material properties, akin to the linear fit in the isotropic case, is even more challenging. Improving upon the proposed fits would thus require a substantial increase in the amount of numerical data. Nevertheless, our existing fits demonstrate a robust ability to predict INS distributions across a wide range of materials and GB types.


\subsection{Review of numerical simulation results\label{sec:FE}}

Below, the most important insights from our previous FE study~\cite{ELSHAWISH2021104293} of selected materials with \emph{cubic} crystal lattices are gathered:
\begin{enumerate}
	\item \textbf{\emph{twist-angle (in)dependence of $\text{PDF}(\sigma_{nn})$}:} \\
	The results for $\langle abc \rangle$-$twist$ GBs suggested that the first two statistical moments of INS distributions remain virtually unaffected by the \emph{twist angle} $\Delta\omega$ between both grains (although its value might still affect the corresponding GB strength). However, later research~\cite{ELSHAWISH2023104940} has revealed that while this observation holds for $\langle \sigma_{nn} \rangle$ of $[abc]$-$[def]$-$\Delta\omega$ GB types, it is no longer true for their $s(\sigma_{nn})$. Nonetheless, we consider this effect to be a higher-order correction.
	As a result, the value of $\Delta\omega$ can be assigned at random for each GB in numerical simulations, and $4$ degrees of freedom effectively suffice to specify a GB type in relation to the local GB stress. In our discussions, we colloquially refer to it as the $[abc]$-$[def]$ GB type.
	\item  \textbf{\emph{universal value of $\langle\sigma_{nn}\rangle$}:} \\
	INS distributions have a \emph{universal mean value} for every GB type and in any material with cubic lattice symmetry:
    \begin{align}
    	\langle\sigma_{nn}\rangle & = \frac{1}{3} \Tr{\mathbf{\Sigma}} \ . \label{eq:mean}
    \end{align}
    Consequently, the proportion of GBs of a particular type with stresses exceeding the threshold value is, in the first approximation, well characterized by the standard deviation of the corresponding INS distribution.
	\item \textbf{\emph{relevance of effective GB-stiffness parameter for $s(\sigma_{nn})$}:} \\
	The \emph{effective GB stiffness} $E_{12}$ has been identified as the sole parameter needed to relate the GB type to the \emph{standard deviation} of the corresponding INS distribution.
	An approximate phenomenological expression has been proposed
	\begin{align}
		\begin{split}
		s(\sigma_{nn}; E_{12}, A, \mathbf{\Sigma}) & := \sqrt{\langle (\sigma_{nn} - \langle\sigma_{nn}\rangle)^2 \rangle} \\ 
		& \approx A_1 (A) \arctan{\left (A_2 (A) \, E_{12}^{A_3 (A)}\right )} \, \Sigma_{\text{mis}} \ , \label{eq:std_dev}
		\end{split}
	\end{align}
    where $A_1$, $A_2$ and $A_3$ are material-specific%
    \footnote{In truth, the numerical simulations conducted in~\cite{ELSHAWISH2021104293} were limited to the uniaxial tensile loading. Consequently, the coefficients $A_k$ could in principle depend also on the external loading $\mathbf{\Sigma}$. However, a separate study~\cite{ELSHAWISH2023104940} showed that $\langle\sigma_{nn} \rangle$ and $s(\sigma_{nn})$ can be expressed as products of a loading-dependent part and a part depending on the GB type and material properties. Furthermore, this study identified the scaling factors as $\Tr\mathbf{\Sigma}$ and $\Sigma_{\text{mis}}$, respectively.}
    fitting coefficients, and
    \newpage 
    \begin{align}	
    	A & = \frac{2 (s_{11}-s_{12})}{s_{44}} \ , \label{eq:Zener} \\
    	E_{12} & = \frac{2 \xbar{E}^{-1}}{E_{abc}^{-1} + E_{def}^{-1}} \label{eq:E12}
    \end{align}
    represent the Zener elastic anisotropy index~\cite{zener}, and the effective GB stiffness, while
    \begin{align}
    	E_{hkl}^{-1} & = s_{11} - 2 s_0 \frac{(h k)^2+(h l)^2+(k l)^2}{(h^2+k^2+l^2)^2} \ , \label{eq:Ehkl} \\
    	s_0 & := s_{11} - s_{12} - \frac{s_{44}}{2} = \frac{s_{44}}{2} (A-1)
    \end{align}
    are the inverse Young's modulus of a grain along the $(h,k,l)$-direction, and a specific combination of the eigenbasis components ($s_{11}$, $s_{12}$, and $s_{44}$) of the compliance tensor in Voigt notation, which
    vanishes for isotropic materials (i.e., $s_0 = 0$ for $A=1$). 
    
    The parameter $\xbar{E}$ denotes the average (bulk) Young's modulus of the aggregate, given by
    \begin{align}
    	\xbar{E} & = \frac{9 K G}{3 K + G} \ , \label{eq:Eave}
    \end{align}
    where compression ($K$) and shear ($G$) elastic moduli can be estimated using a self-consistent approach~\cite{kroner58}. 
    For crystal grains with cubic lattice symmetry, assembled in a random (macroscopically isotropic) aggregate, these expressions simplify to the following forms~\cite{hershey}:
    \begin{align}
    	K & = \frac{1}{3 (s_{11}+2 s_{12})} \ , \label{eq:K}\\
    	G & = \left\{
    	\begin{array}{c}	
    		\!\! 8 \, (s_{11}\!+\!2 s_{12}) (s_{11}\!-\!s_{12}) \, s_{44} \, G^3 + (5 s_{11}\!+\!s_{12}) \, s_{44} \, G^2-(7 s_{11}\!+\!11 s_{12}) \, G = 1 \\  G>0
    	\end{array}
    	\right. \!\!. 
    	\label{eq:G}
    \end{align} 
\end{enumerate}

In summary, out of the five degrees of freedom specifying a $[abc]$-$[def]$-$\Delta\omega$ GB type, only one particular combination of $a$, $b$, $c$, $d$, $e$, and $f$ --- namely, the $E_{12}$ --- is needed to determine the local stress distribution.
Since $E_{12}$ can be considered essentially a ``bicrystal'' quantity (depending only on the properties of the two nearest neighbours), the above observations (based on ``realistic'' simulations that properly take into account all the grains in the aggregate, not just the closest two) suggest that there is no need to go beyond the second-order approximation (the bicrystal model) for the intended estimates of GB-stress distributions. 

\begin{figure}[htb]
	\centering
	\includegraphics[width=\textwidth]{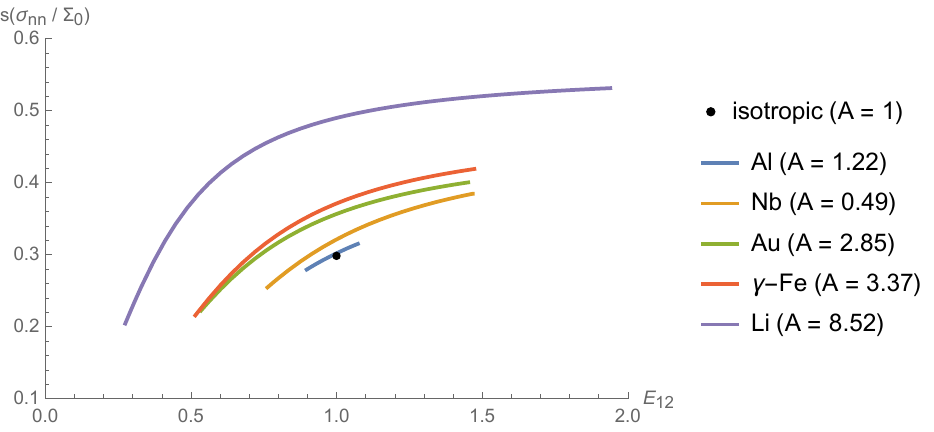}
	\caption{Standard deviation $s(\sigma_{nn}/\Sigma_0)$ as a function of GB stiffness $E_{12}$ for uniaxial external loading of magnitude $\Sigma_0$; showing its \emph{fit} specified in Eq.~\eqref{eq:std_dev}.
	Several (cubic) materials are presented, arranged in increasing order of anisotropy (specified as a deviation from the isotropic case, also provided for reference). The indicated values of Zener index $A$ uniquely determine the corresponding fitting coefficients $A_k$.
	The values on the vertical axis correspond to the \emph{full} INS distributions obtained from all finite elements within the aggregate that belong to GBs of a specific type. The same \emph{fitting} function (albeit with different coefficients $A_k$) could also be applied to the narrower \emph{reduced} INS distributions (see Fig.~\ref{fig:bicrystal_VS_FE}).}
	\label{fig:std_dev_FE}
\end{figure}

One of the primary objectives of this paper is thus to elucidate these observations
and to derive them within a simplified framework based on the bicrystal approximation. Especially relevant in this context is recognizing the effective stiffness of the bicrystal pair along the GB-normal direction ($E_{12}$) and the Zener anisotropy index ($A$) as key parameters. It's worth noting that material properties are contained not only in $A$ but also in $E_{12}$.
Its value implicitly depends on $s_{ij}$ and thus the choice of material; cf.~Eqs.~\eqref{eq:E12}--\eqref{eq:G}. The combined effect of material-dependent parameters $A$ and $E_{12}$ is illustrated in Fig.~\ref{fig:std_dev_FE}.

%


\subsection{The model\label{sec:bicrystal model}}

We wish to obtain a component of stress tensor along the GB-normal direction ($\sigma_{nn}$) for any chosen GB in the aggregate that is subjected to a macroscopic external loading $\mathbf{\Sigma}$. To achieve this, a generalized Hooke's law: 
\begin{align}
	\epsilon_{ij}^{(N)}& = \sum_{k,l = 1}^{3} \tilde{s}_{ijkl}^{(N)} \, \sigma_{kl}^{(N)} \ , \quad i,j=1,2,3 \ ,
	\label{eq:Hooke}
\end{align}
relating the components of local strain ($\epsilon_{ij}$), compliance ($\tilde{s}_{ijkl}$) and stress ($\sigma_{ij}$) tensors,	
or its inverse containing the elastic stiffness tensor ($\tilde{C}_{ijkl}$):
\begin{align}
	\sigma_{ij}^{(N)}& = \sum_{k,l = 1}^{3} \tilde{C}_{ijkl}^{(N)} \, \epsilon_{kl}^{(N)} \ , \quad i,j=1,2,3 \ , \label{eq:Hooke_alt}
\end{align}
needs to be solved for a pair of grains ($N = 1,2$) enclosing the investigated GB from both sides.
Depending on the position of the GB within the aggregate, the grains may be of various shapes and sizes, and their crystal lattices can posses arbitrary orientations in relation to external loading. The GB itself can also be oriented in any direction (uncorrelated with the crystallographic orientations of both surrounding grains). 

The simplest scenario considers the grains with cubic lattice symmetry. In such case, the components of the $rank$-$4$ compliance tensor $\tilde{s}_{ijkl}$ (or elasticity tensor $\tilde{C}_{ijkl}$) depend on just three material parameters ($s_{11}$, $s_{12}$ and $s_{44}$), and many of the relations simplify substantially. When expressed in a local (\emph{grain}) coordinate system $(n_1,n_2,n_3)$ aligned with the crystallographic axes of the underlying lattice, the tensor assumes the following form: 
\begin{align}
	\begin{split}
	s_{mmmm} & := s_{11} \ , \quad m=1,2,3 \ , \\
	s_{mnmn} = s_{mnnm} & := \frac{s_{44}}{4} \ , \quad m, n\neq m=1,2,3 \ , \\
	s_{mmnn} & := s_{12} \ , \quad m, n\neq m=1,2,3 \ , \\
	s_{mnop} & := 0 \ , \quad \text{otherwise} \ ,
	\end{split}
\end{align}
and similarly for the stiffness tensor:
\begin{align}
	\begin{split}
	C_{mmmm} & := C_{11} \ , \quad m=1,2,3 \ , \\
	C_{mnmn} = C_{mnnm} & := C_{44} \ , \quad m, n\neq m=1,2,3 \ , \\
	C_{mmnn} & := C_{12} \ , \quad m, n\neq m=1,2,3 \ , \\
	C_{mnop} & := 0 \ , \quad \text{otherwise} \ ,
	\end{split}
\end{align}
where the transformation rules between both
\begin{align}
	\begin{split}
	s_{11} & = \frac{C_{11}+C_{12}}{(C_{11}-C_{12}) (C_{11} + 2 C_{12})} \ , \\
	s_{12} & = -\frac{C_{12}}{(C_{11}-C_{12}) (C_{11} + 2 C_{12})} \ , \\
	s_{44} & = \frac{1}{C_{44}} \ , \\
	s_0 & = \frac{1}{C_{11}-C_{12}}-\frac{1}{2 C_{44}}
	\end{split} 
\end{align}
are symmetric to $s_{ij} \leftrightarrow C_{ij}$.

But for our purposes, it is more convenient to express all the quantities in a local (\emph{GB}) coordinate system $(x,y,z)$, in which the \mbox{$z$-axis} is aligned with the GB normal. The transformations 
\begin{align}
	\begin{split}
    \tilde{s}_{ijkl}^{(N)} & = \hspace{-.4cm} \sum_{m,n,o,p=1}^{3} \hspace{-.4cm} R_{im}^{(N)} R_{jn}^{(N)} R_{ko}^{(N)} R_{lp}^{(N)} \, s_{mnop} \\
    & = \left (\sum_{m=1}^{3} R_{im}^{(N)} R_{jm}^{(N)} R_{km}^{(N)} R_{lm}^{(N)} \right ) s_0 + (\delta_{ij}\delta_{kl}) \, s_{12} + (\delta_{ik}\delta_{jl} + \delta_{il}\delta_{jk}) \, \frac{s_{44}}{4} \ , 
    \end{split} \\
    \begin{split}
	\tilde{C}_{ijkl}^{(N)} & = \hspace{-.4cm} \sum_{m,n,o,p=1}^{3} \hspace{-.4cm} R_{im}^{(N)} R_{jn}^{(N)} R_{ko}^{(N)} R_{lp}^{(N)} \, C_{mnop} \\
	& = \left (\sum_{m=1}^{3} R_{im}^{(N)} R_{jm}^{(N)} R_{km}^{(N)} R_{lm}^{(N)} \right ) (C_{11}\!-\!C_{12}\!-\!2 \, C_{44}) + (\delta_{ij}\delta_{kl}) \, C_{12} + (\delta_{ik}\delta_{jl} \!+\! \delta_{il}\delta_{jk}) \, C_{44} \label{eq:Cijkl}
	\end{split}
\end{align}
involve the (orthogonal) rotation matrix
\begin{align}
\label{eq:Rcry}
\mathbf{R} & = \left(
\begin{array}{ccc}
	\phantom{-}\frac{h l \cos \omega }{\sqrt{h^2+k^2} \sqrt{h^2+k^2+l^2}} - \frac{k \sin \omega}{\sqrt{h^2+k^2}} & \phantom{-}\frac{k l \cos \omega}{\sqrt{h^2+k^2} \sqrt{h^2+k^2+l^2}} + \frac{h \sin \omega}{\sqrt{h^2+k^2}} & -\frac{\sqrt{h^2+k^2} \cos \omega
	}{\sqrt{h^2+k^2+l^2}} \\
	-\frac{h l \sin \omega}{\sqrt{h^2+k^2} \sqrt{h^2+k^2+l^2}} - \frac{k \cos \omega}{\sqrt{h^2+k^2}} & -\frac{k l \sin \omega}{\sqrt{h^2+k^2} \sqrt{h^2+k^2+l^2}} + \frac{h \cos \omega}{\sqrt{h^2+k^2}} & \phantom{-}\frac{\sqrt{h^2+k^2} \sin \omega
	}{\sqrt{h^2+k^2+l^2}} \\
	\frac{h}{\sqrt{h^2+k^2+l^2}} & \frac{k}{\sqrt{h^2+k^2+l^2}} & \frac{l}{\sqrt{h^2+k^2+l^2}}
\end{array}
\right) \ ,
\end{align}
with $(h,k,l)$ representing the direction of the GB normal expressed in the grain system $(n_1,n_2,n_3)$, and $\omega$ the angle of rotation (twist) about the GB normal. Two such matrices $\mathbf{R}^{(N)}$ are needed for every GB (one for each grain), with parameters $(a,b,c)$ and $\omega_1$ used for the first grain ($N=1$) and $(d,e,f)$ and $\omega_2$ for the second grain ($N=2$). Their numerical values determine the corresponding GB type $[abc]$-$[def]$-$\Delta\omega$.

The transformation of the macroscopic loading from the \emph{lab} coordinate system $(X,Y,Z)$ to the \emph{GB} coordinate system $(x,y,z)$ is carried out in the same manner as before, cf.~Eqs.~\eqref{eq:external_load}--\eqref{eq:Rlab}.
The expressions for the external stress tensor components in the GB coordinate system can be found in Appendix~\ref{app:loading in GB system}.

Since strain ($\epsilon_{ij}$) and stress ($\sigma_{ij}$) are both represented by symmetric tensors, Eq.~\eqref{eq:Hooke} yields $6$ constitutive equations for each grain, and thus $12$ altogether. To solve them, a matching number of \emph{boundary conditions} is needed. We will categorize them into $6$ ``inner'' and $6$ ``outer'' conditions, depending on whether they apply to the grain boundary or to the bicrystal pair. The inner conditions are:
\begin{enumerate}
	\item[] \emph{stress continuity across the GB}
	\begin{align}
		\sigma_{xz}^{(1)} = \sigma_{xz}^{(2)} \quad , \quad \sigma_{yz}^{(1)} = \sigma_{yz}^{(2)} \quad , \quad \sigma_{zz}^{(1)} = \sigma_{zz}^{(2)} \ , \label{eq:BC1}
	\end{align}
	\item[] \emph{strain compatibility across the GB}
	\begin{align}
		\epsilon_{xx}^{(1)} = \epsilon_{xx}^{(2)} \quad , \quad \epsilon_{xy}^{(1)} = \epsilon_{xy}^{(2)} \quad , \quad \epsilon_{yy}^{(1)} = \epsilon_{yy}^{(2)} \ , \label{eq:BC2}
	\end{align}
\end{enumerate}
while the outer conditions are specifying how the investigated pair of grains is deformed as a whole when a specific macroscopic external loading $\mathbf{\Sigma}$ is applied to the aggregate:
\begin{align}
	\begin{split}
		V_1 \, \epsilon_{xx}^{(1)} + V_2 \, \epsilon_{xx}^{(2)} = (V_1+V_2) \, \xbar{\epsilon}^{b}_{xx} \quad , \quad V_1 \, \epsilon_{yz}^{(1)} + V_2 \, \epsilon_{yz}^{(2)} = (V_1+V_2) \, \xbar{\epsilon}^{b}_{yz} \ , \\
		V_1 \, \epsilon_{yy}^{(1)} + V_2 \, \epsilon_{yy}^{(2)} = (V_1+V_2) \, \xbar{\epsilon}^{b}_{yy} \quad , \quad V_1 \, \epsilon_{xz}^{(1)} + V_2 \, \epsilon_{xz}^{(2)} = (V_1+V_2) \, \xbar{\epsilon}^{b}_{xz} \ , \\
		V_1 \, \epsilon_{zz}^{(1)} + V_2 \, \epsilon_{zz}^{(2)} = (V_1+V_2) \, \xbar{\epsilon}^{b}_{zz} \quad , \quad V_1 \, \epsilon_{xy}^{(1)} + V_2 \, \epsilon_{xy}^{(2)} = (V_1+V_2) \, \xbar{\epsilon}^{b}_{xy} \ . \label{eq:BC3}
	\end{split}
\end{align}
If the \emph{average strain of the bicrystal} $\xbar{\epsilon}^{b}_{ij}$ and the volumes of the corresponding grains, $V_1$ and $V_2$, were known for every pair of grains, the equations~\eqref{eq:Hooke} and~\eqref{eq:BC1}--\eqref{eq:BC3} would yield an exact solution for $\sigma_{nn} := \sigma_{zz}^{(1)} = \sigma_{zz}^{(2)}$ at each GB (neglecting that, in practice, strain and stress are not constant throughout the grains).
However, obtaining the correct values of $\xbar{\epsilon}^{b}_{ij}$ requires solving the equations for all the grains in the aggregate simultaneously, which is feasible only numerically, e.g., by relying on the finite element method.
Such approach is both computationally demanding and time-consuming due to the large number of grains involved. It is also impractical since it requires the knowledge of the exact configuration of all the grains in the aggregate.

Another way to address the problem is to make some reasonable assumptions about the deformation of the bicrystal pair (i.e., the outer boundary conditions), and try to solve the model analytically. Several factors contribute to the GB stress. Ranked in decreasing order of importance, the most relevant are: GB orientation, GB type, and the neighbourhood consisting of all the grains surrounding the bicrystal pair. As previously explained, in this study we adopted a perturbative approach, approximating the neighbourhood with a homogeneous and isotropic matrix material (into which the bicrystal pair is embedded).
This allows us to consider only a single pair of grains rather than the entire aggregate, and to derive a solution of the bicrystal-model equations for any GB (specified by its orientation and the corresponding GB type).

\begin{figure}[htb]
	\centering
	\includegraphics[width = \textwidth]{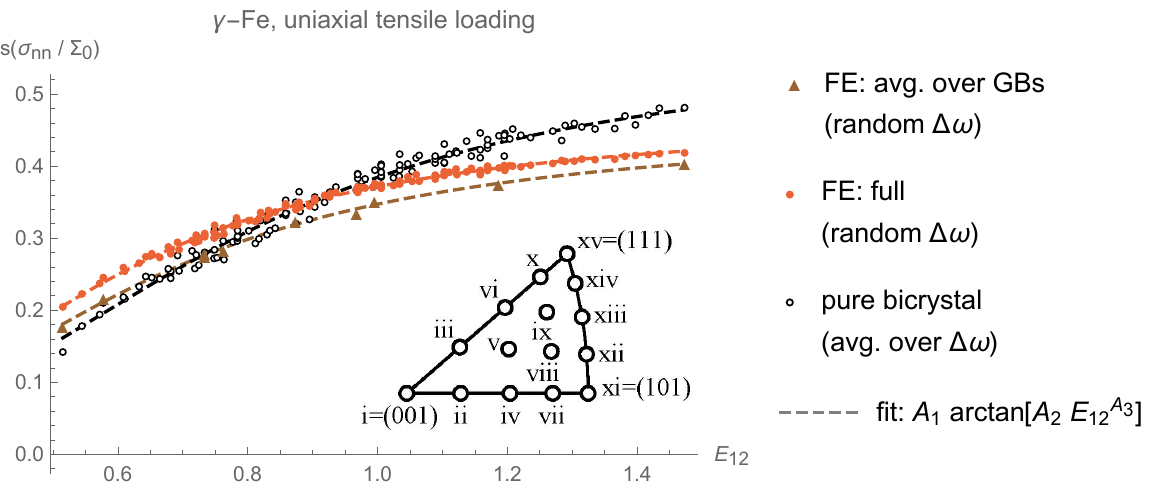}
	\caption{Standard deviation $s(\sigma_{nn}/\Sigma_0)$ as a function of GB stiffness $E_{12}$ for $\gamma$-Fe and uniaxial loading of strength $\Sigma_0$. Comparison is made between the results of FE simulation and the solution of a simplified bicrystal model. The $120$ dots in each case correspond to selecting GB normals $(a,b,c)$ and $(d,e,f)$ among the $15$ predefined directions from the standard stereographic triangle shown in the inset.
	In addition to the \emph{full} distribution, we also display the \emph{reduced} distribution (denoted by \emph{triangles}). The same fitting function~\eqref{eq:std_dev}, albeit with different coefficients $A_k$, was used in all three cases.}
	\label{fig:bicrystal_VS_FE}
\end{figure}

The simplest assumption that can be made about $\xbar{\epsilon}^{b}_{ij}$ is to use the \emph{average strain approximation}
\begin{align}
	\xbar{\epsilon}^{b}_{ij} = \xbar{\epsilon}_{ij} & = \frac{1+\xbar{\nu}}{\xbar{E}} \Sigma_{ij} - \frac{\xbar{\nu}}{\xbar{E}} (\Tr\mathbf{\Sigma}) \, \delta_{ij} \ , \label{eq:assumption}
\end{align}
where
\begin{align}
	\xbar{\nu} & = \frac{3 K - 2 G}{2 (3 K + G)} = \frac{1}{2} \left (1 - \xbar{E} \, (s_{11} + 2 s_{12}) \right ) \label{eq:nuave}
\end{align} 
is the average Poisson's ratio of the material (and last equality applies only to the cubic lattice symmetry).
The problem with this ansatz is that it produces an incorrect response --- the magnitude of induced stress $\sigma_{nn}$ is too large for very stiff GBs (as they should not deform as much as the bulk material) and too small for very soft GBs (which should deform more than the bulk). 
This can be observed in Fig.~\ref{fig:bicrystal_VS_FE}, showing a comparison between the results of FE simulation and the bicrystal model. Both exhibit similar trends for the ``width'' of INS distribution%
\footnote{On the other hand, their mean values are the same, namely $\langle \sigma_{nn} / \Sigma_0 \rangle = 1/3$ for every GB type.}
as a function of stiffness of the corresponding GB type, but having different slopes.
The realistic (FE) curve is less steep since for soft GBs (small $E_{12}$) both tensile and compressive strains are larger than the average strain of the aggregate ($\xbar{\epsilon}_{ij}$) and thus the magnitude of stress bigger than the bicrystal estimate based on Eq.~\eqref{eq:assumption}. The standard deviation is thus underestimated in the bicrystal model.
Conversely, for stiff GBs (large $E_{12}$), the bicrystal-model estimate surpasses the true standard deviation.

Different points in Fig.~\ref{fig:bicrystal_VS_FE} represent different $[abc]$-$[def]$ GB types obtained by selecting GB normals $(a,b,c)$ and $(d,e,f)$ from the $15$ directions within the standard stereographic triangle: $i=[001]$, $ii=[104]$, $iii=[114]$, $iv=[102]$, $v=[214]$, $vi=[112]$, $vii=[304]$, $viii=[314]$, $ix=[324]$, $x=[334]$, $xi=[101]$, $xii=[414]$, $xiii=[212]$, $xiv=[434]$ and $xv=[111]$.
Note that some of the GB types correspond to the same value of $E_{12}$ because the Young's moduli along different directions can be the same (specifically, $E_{hkl}^{-1} = s_{11} - s_0/2$ along the $[101]$, $[112]$ and $[314]$ directions).
For each GB type, the $\sigma_{nn}$ distribution (originating mainly from random orientations of GBs rather than from their different neighbourhoods) is obtained in FE simulation.
Its standard deviation (representing the ``width'' of the distribution) is shown on the vertical axis of Fig.~\ref{fig:bicrystal_VS_FE}.

To obtain $\sigma_{nn}$ in the bicrystal case, Eq.~\eqref{eq:Hooke} along with boundary conditions~\eqref{eq:BC1}--\eqref{eq:BC3}, and imposing
the approximation~\eqref{eq:assumption}, needs to be solved for a specific GB orientation and GB type. Here, this was done numerically, and thus material properties, loading conditions and the relative sizes of both grains had to be specified.
INS distribution for a chosen GB type is then produced by performing a
Monte Carlo (MC) sampling over the uniformly distributed parameters $\psi$, $\cos\theta$ and $\phi$ (to account for random orientations of GBs), $\omega_1$ and $\omega_2$ (hence averaging the GB types over the twist angle $\Delta\omega$), and potentially the grain sizes $V_1$ and $V_2$.%
\footnote{In principle, any distribution of grain sizes could be used; however, in this study, grains of equal size ($V_1=V_2$) are assumed for simplicity.}

The main shortcoming of the numerical approach outlined above is that obtaining the solution for $\sigma_{nn}$ requires specifying the values of all involved parameters (including selecting the material properties and loading conditions), which makes the results very difficult to generalize.
We will thus attempt to solve the equations analytically. To improve the agreement with FE results, we also need to go beyond the average strain approximation.
In the following, we will assume that $\xbar{\epsilon}^{b}_{ij}$ is of the form:
\begin{align}
	\xbar{\epsilon}^{b}_{ij} & = \alpha \, \Sigma_{ij} - \beta \, (\Tr\mathbf{\Sigma}) \, \delta_{ij} \ , \label{eq:assumption_general}
\end{align}
where the parameters $\alpha$ and $\beta$ are allowed to vary with GB type (and material properties), but they should remain independent of the GB's position within the aggregate or its orientation. In other words, $\alpha$ and $\beta$ are not functions of $\theta$, $\psi$, $\phi$, or the position vector $\vec{r}$.

The condition expressed in Eq.~\eqref{eq:assumption_general} can be interpreted as a form of Hooke's law. However, it is valid not at the local (grain) scale (where stresses are considered), but rather at the level of all the GBs of a chosen type, for which macroscopic isotropy can be safely assumed. Their collective Young's modulus and Poisson's ratio are then given by $E = (\alpha - \beta)^{-1}$ and $\nu = \beta/(\alpha - \beta)$, respectively.

Instead of using $\alpha$ and $\beta$, we could redefine Eq.~\eqref{eq:assumption_general} in terms of $\Delta\xbar{E} := E - \xbar{E}$ and $\Delta\xbar{\nu} := \nu - \xbar{\nu}$, representing the deviation of Young's modulus and Poisson's ratio from their aggregate values.
However, two reasons argue against this approach. Firstly, our choice is dictated by the simplicity of derived expressions, as it will later be shown that all the central moments of the pure bicrystal distribution depend solely on $\alpha$ (namely, being proportional to $\alpha^n$). Secondly, it proves easier to find a simple fitting function for $\alpha$ with respect to $E_{12}$ (and $A^u$) than it is for $\Delta\xbar{E}$ and $\Delta\xbar{\nu}$.

\subsection{Analytical solution}

\subsubsection{General case}

The general solution to a system of equations~\eqref{eq:Hooke_alt} and~\eqref{eq:BC1}--\eqref{eq:BC3}, valid for any lattice symmetry, material properties, GB type, and crystallographic orientations (all of them encoded in the local elasticity-tensor components $\tilde{C}_{ijkl}^{(N)}$), grain morphologies, and an arbitrary uniform external loading (determining the bicrystal-strain components $\xbar{\epsilon}^{b}_{ij}$) is
\begin{align}
	\sigma_{zz} & = \sum_{i,j=1}^{3} c_{ij} \, \xbar{\epsilon}^{b}_{ij} = c_{xx} \, \xbar{\epsilon}^{b}_{xx} + c_{yy} \, \xbar{\epsilon}^{b}_{yy} + c_{zz} \, \xbar{\epsilon}^{b}_{zz} + 2 \, (c_{xy} \, \xbar{\epsilon}^{b}_{xy} + c_{xz} \, \xbar{\epsilon}^{b}_{xz} + c_{yz} \, \xbar{\epsilon}^{b}_{yz}) \ , \label{eq:INS_c}
\end{align}
where
\begin{align}
	c_{ij} & := \frac{\tilde{C}^{(2)}_{ij33} \, \alpha^{(1)} \, V_2^3 + \beta^{(1,2)}_{ij} \, V_1 \, V_2^2 + \beta^{(2,1)}_{ij} \, V_1^2 \, V_2 + \tilde{C}^{(1)}_{ij33} \, \alpha^{(2)} \, V_1^3}{\alpha^{(1)} \, V_2^3 + \gamma^{(1,2)} \, V_1 \, V_2^2 + \gamma^{(2,1)} \, V_1^2 \, V_2 + \alpha^{(2)} \, V_1^3} \label{eq:c-coefficients}
\end{align}
for
\begin{align}
	\alpha^{(n)} & := \big (\hspace{-.5cm} \sum_{\substack{p=1 \\ (q<r; \, q,r\neq p)}}^{3} \hspace{-.5cm} \tilde{C}^{(n)}_{p3p3} \left (\tilde{C}^{(n)}_{q3r3} \right )^2 \big ) - \big ( 2 \, \tilde{C}^{(n)}_{p3q3} \, \tilde{C}^{(n)}_{p3r3} \, \tilde{C}^{(n)}_{q3r3} + \tilde{C}^{(n)}_{p3p3} \, \tilde{C}^{(n)}_{q3q3} \, \tilde{C}^{(n)}_{r3r3} \big )_{p<q<r} \label{eq:alpha_parameter}
	\\
	\begin{split}
	\beta^{(m,n)}_{ij} & := \hspace{-.5cm} \sum_{\substack{p,q=1 \\ (q\neq p; \, r\neq p,q)}}^{3} \hspace{-.5cm} \left ( \tilde{C}^{(m)}_{ijp3} \left ( \tilde{C}^{(m)}_{p3q3} \tilde{C}^{(m)}_{r3r3} - \tilde{C}^{(m)}_{p3r3} \tilde{C}^{(m)}_{q3r3} \right ) \tilde{C}^{(n)}_{q333} \right ) + \\
	& + \hspace{-.5cm} \sum_{\substack{p=1 \\ (q<r; \, q,r\neq p)}}^{3} \hspace{-.5cm} \big \{ \tilde{C}^{(m)}_{ijp3} \left ( \left ( \tilde{C}^{(m)}_{q3r3} \right )^2 - \tilde{C}^{(m)}_{q3q3} \tilde{C}^{(m)}_{r3r3} \right ) \tilde{C}^{(n)}_{p333} + \\ 
	& \hspace{1.3cm} + \left ( \tilde{C}^{(n)}_{ij33} \tilde{C}^{(n)}_{p3p3} - \tilde{C}^{(n)}_{ijp3} \tilde{C}^{(n)}_{p333} \right ) \, \left ( \left ( \tilde{C}^{(m)}_{q3r3} \right )^2 - \tilde{C}^{(m)}_{q3q3} \tilde{C}^{(m)}_{r3r3} \right ) + \\ 
	& \hspace{1.3cm} + \left ( \tilde{C}^{(n)}_{ij33} \tilde{C}^{(n)}_{p3qr} - \tilde{C}^{(n)}_{ijp3} \tilde{C}^{(n)}_{qr33} \right ) \, \left ( \tilde{C}^{(m)}_{p3qr} \tilde{C}^{(m)}_{3333} - \tilde{C}^{(m)}_{p333} \tilde{C}^{(m)}_{qr33} \right ) + \\
	& \hspace{1.3cm} + \left ( \tilde{C}^{(n)}_{ij33} \tilde{C}^{(n)}_{p333} - \tilde{C}^{(n)}_{ijp3} \tilde{C}^{(n)}_{3333} \right ) \, \left ( \tilde{C}^{(m)}_{p333} \tilde{C}^{(m)}_{qrqr} - \tilde{C}^{(m)}_{p3qr} \tilde{C}^{(m)}_{qr33} \right ) \big \} \ ,
	\end{split} \label{eq:beta_parameter}
	\\
	\gamma^{(m,n)} & := \hspace{-.5cm} \sum_{\substack{p=1 \\ (q<r; \, q,r\neq p)}}^{3} \hspace{-.5cm} \left \{ \tilde{C}^{(n)}_{p3p3} \left ( \left (\tilde{C}^{(m)}_{q3r3} \right )^2 - \tilde{C}^{(m)}_{q3q3} \, \tilde{C}^{(m)}_{r3r3} \right ) + 2 \, \tilde{C}^{(n)}_{q3r3} \left ( \tilde{C}^{(m)}_{q3r3} \, \tilde{C}^{(m)}_{p3p3} - \tilde{C}^{(m)}_{p3q3} \, \tilde{C}^{(m)}_{p3r3} \right ) \right \} \label{eq:gamma_parameter}
	\ .
\end{align}
As long as we are in a purely elastic regime and stress-strain fields can be taken as constant within the grains and on their boundaries (even though in reality they are not), this solution can be considered exact, producing the correct GB-normal stress at any chosen GB in the aggregate. 
That is because, at this point of the derivation, the precise values of bicrystal strains $\xbar{\epsilon}^{b}_{ij}$ have not yet been specified. 
In practice, our intention is to relate them to the external loading $\mathbf{\Sigma}$. 
Since the exact relation between $\xbar{\epsilon}^{b}_{ij}$ of a particular bicrystal pair and $\mathbf{\Sigma}$ is not known, we must rely on some sensible approximations. Invoking the assumption~\eqref{eq:assumption_general}, leads to the bicrystal model of a pair of grains surrounded by matrix material, which enables us to express $\snn$ in the form
\begin{align}
	\sigma_{zz} & = \sum_{i,j=1}^{3} a_{ij} \, \Sigma_{ij} = a_{xx} \, \Sigma_{xx} + a_{yy} \, \Sigma_{yy} + a_{zz} \, \Sigma_{zz} + 2 \, (a_{xy} \, \Sigma_{xy} + a_{xz} \, \Sigma_{xz} + a_{yz} \, \Sigma_{yz}) \ , \label{eq:INS_a}
\end{align}
that relates the induced stress to components of the external stress tensor through the linear coefficients
\begin{align} 
a_{ij} & = \alpha \, c_{ij}-\beta \, (c_{xx}+c_{yy}+c_{zz}) \, \delta_{ij} \ . \label{eq:c-a-connection}
\end{align}
For instance, the isotropic case described in the previous section is exactly of this kind, with a single non-zero coefficient \mbox{$a_{zz} = 1$}.

Note that both sets of coefficients ($c_{ij}$ and $a_{ij}$) are only functions of GB type, and as such, they do not depend on the GB orientation ($\theta$, $\psi$).%
\footnote{From Eqs.~\eqref{eq:INS_c} and~\eqref{eq:INS_a} it may appear as if $\sigma_{zz}$ depended on the choice of the local GB coordinate system (the direction of local axes $x$ and $y$ determined by angle $\phi$) through the corresponding strain and stress components ($\xbar{\epsilon}_{ij}$ and $\Sigma_{ij}$) and the angles $\omega_1$ and $\omega_2$ contained in coefficients $c_{ij}$ and $a_{ij}$.
But this dependence is only implicit and vanishes once we evaluate $\Sigma_{ij}$ (using the expressions provided in Appendix~\ref{app:loading in GB system}), since in that case $\sigma_{nn}$ becomes only a function of $\omega_1-\phi$ and $\omega_2-\phi$, which makes it independent of the orientation of the local (GB) coordinate system.
In addition, the INS distribution and all its statistical moments then manifestly exhibit the expected behaviour by depending solely on the parameter $\Delta\omega = \omega_2 - \omega_1$, which uniquely determines the corresponding GB type.}
Therefore, it is straightforward to compute the first two moments of the INS distribution as:
\begin{align}
	\langle \snn \rangle & = \frac{1}{8\pi^2} \int_0^{2 \pi}\int_0^{2 \pi}\int_{-1}^{1} \sigma_{zz} \, d\!\cos\theta \, d\psi \, d\phi \label{eq:mean_gen} \\ 
	& = (a_{xx} + a_{yy} + a_{zz}) \times \frac{1}{3} \Tr\mathbf{\Sigma} \nonumber \\ 
	& = (\alpha - 3 \, \beta) \, (c_{xx} + c_{yy} + c_{zz}) \times \frac{1}{3} \Tr\mathbf{\Sigma} \ , \nonumber \\
	s(\snn) & = \sqrt{\langle (\snn - \langle \snn \rangle)^2 \rangle} = \sqrt{\frac{1}{8\pi^2} \int_0^{2 \pi}\int_0^{2 \pi}\int_{-1}^{1} \left (\sigma_{zz} - \langle \sigma_{zz} \rangle \right )^2 \, d\!\cos\theta \, d\psi \, d\phi} \label{eq:std dev} \\
	& = \sqrt{\frac{1}{2} \, \big ((a_{xx} - a_{yy})^2 + (a_{yy} - a_{zz})^2 + (a_{zz} - a_{xx})^2 \big ) + 3 \, (a_{xy}^2 + a_{xz}^2 + a_{yz}^2)} \times \frac{2}{3\sqrt{5}} \Sigma_{\text{mis}} \nonumber \\ 
	& = |\alpha| \, \sqrt{\frac{1}{2} \, \big ((c_{xx} - c_{yy})^2 + (c_{yy} - c_{zz})^2 + (c_{zz} - c_{xx})^2 \big ) + 3 \, (c_{xy}^2 + c_{xz}^2 + c_{yz}^2)} \times \frac{2}{3\sqrt{5}} \Sigma_{\text{mis}} \nonumber \ .
\end{align}
We have thus shown that the first two statistical moments, in general (meaning for any lattice symmetry), scale with loading as $\Tr\mathbf{\Sigma}$ and $\Sigma_{\text{mis}}$, respectively. 
One might assume that the validity of this result extends even beyond the bicrystal-model approximation, as it arises from the linearity of Hooke's law~\eqref{eq:INS_a} and the assumption of no correlation between the orientations of GBs and the crystallographic orientations of grains (both uniformly distributed).
As such, it should then apply also to realistic cases (represented by numerical simulations). However, this would imply that for purely \emph{hydrostatic} loading (for which $\Sigma_{\text{mis}} = 0$), $s(\snn)$ should become zero, resulting in an infinitely narrow distribution, but this is only true for cubic symmetry. Hence, the above scaling relations are valid for arbitrary lattice symmetry only for purely \emph{deviatoric} loadings, whereas for cubic symmetry, they apply to any loading scenario. 

\subsubsection{Cubic symmetry}

In the case of cubic lattice symmetry, we can go a step further. By imposing the relations
\begin{align}
	\tilde{C}^{(n)}_{3333} & = -\tilde{C}^{(n)}_{1133} - \tilde{C}^{(n)}_{2233} + (C_{11}+2 \, C_{12}) \ , \\
	\tilde{C}^{(n)}_{1333} & = -\tilde{C}^{(n)}_{1311} - \tilde{C}^{(n)}_{1322} \ , \\
	\tilde{C}^{(n)}_{2333} & = -\tilde{C}^{(n)}_{2311} - \tilde{C}^{(n)}_{2322}
\end{align}
on Eq.~\eqref{eq:c-coefficients}, we obtain
\begin{align}
c_{xx} + c_{yy} + c_{zz} & = C_{11}+2 \, C_{12} = (s_{11}+2 \, s_{12})^{-1} \ , \label{eq:cubic_rule}
\end{align}
resulting in
\begin{align}
	\langle \snn \rangle & = (\alpha - 3 \, \beta) \, (C_{11}+2 \, C_{12}) \times \frac{1}{3} \Tr\mathbf{\Sigma} \ .
\end{align}
This introduces an additional relation between the parameters $\alpha$ and $\beta$: 
\begin{align}
	\beta & = \frac{1}{3} \, \left (\alpha - (s_{11} + 2 \, s_{12}) \right )  
	= \frac{1}{3} \, \left (\alpha - \frac{1-2\xbar{\nu}}{\xbar{E}} \right ) \ ,
	\label{eq:beta}
\end{align}
such that $\langle \snn \rangle = \frac{1}{3} \Tr\mathbf{\Sigma}$, as observed in the FE simulations.%
\footnote{This result can even be derived analytically~\cite{shawish2024extending} by assuming macroscopic isotropy of the aggregate (in the limit of infinitely many grains, the INS distribution for any GB type should become rotationally invariant), linearity of the elastic response, and the fact that for cubic lattice symmetry, the stress field within the aggregate under purely hydrostatic loading equals the external stress.}
While Eq.~\eqref{eq:beta} must be fulfilled for every GB type, the parameter $\alpha$ remains free and can be adjusted to produce a correct response $s(\snn)$.
For instance, in the average strain approximation~\eqref{eq:assumption} 
\begin{align}
	\begin{split}
		\alpha \to \xbar{\alpha} & := \frac{1+\xbar{\nu}}{\xbar{E}} = \frac{1}{2 \, G} \ , \\ 
		\beta \to \xbar{\beta} & := \frac{\xbar{\nu}}{\xbar{E}} = \frac{1}{3} \left (\frac{1}{2 \, G} - \frac{1}{3 \, K} \right ) = \frac{1}{3} \left (\frac{1}{2 \, G} - (s_{11}+2 \, s_{12}) \right ) \ , \label{eq:alpha_bar}
	\end{split}
\end{align} 
hence it fits perfectly into this category.%
\footnote{Interestingly, exactly the condition~\eqref{eq:assumption} is needed to correctly reproduce the isotropic limit ($A=1$) in the bicrystal model. Namely, when we insert $C_{44} = (C_{11} - C_{12})/2$ (and $V_1 = V_2$) into Eq.~\eqref{eq:c-coefficients}, the only non-zero coefficients in Eq.~\eqref{eq:INS_c} become $c_{xx} = c_{yy} = C_{12}$ and $c_{zz} = C_{11}$. Consequently, $a_{zz} = 1$, while the remaining $a_{ij}$ are zero (such that $\sigma_{zz} = \Sigma_{zz}$), precisely for $\alpha = 1/(2 \, G) = 1/(C_{11} - C_{12})$, and the value of $\beta$ in compliance with Eq.~\eqref{eq:beta}. The average strain approximation~\eqref{eq:assumption} should thus be regarded as the limit of condition~\eqref{eq:assumption_general} for isotropic materials. 
\label{footnote:isotropic}}
%

\subsubsection{Twist-angle dependence}

Drawing on the findings from FE simulations, we have concluded that the precise value of the twist angle should not significantly affect the INS distributions, and hence a random value of $\Delta\omega$ can be assigned to each GB.
Furthermore, the width $s(\snn)$ of the INS distribution is for any (cubic) material expected to primarily depend only on the effective stiffness $E_{12}$ of the corresponding GB type.

However, neither of these observations seems to hold true in the \emph{pure} bicrystal model. There, the INS distributions for individual GB types typically exhibit a dependence on $\Delta\omega$, and a considerable spread can exist between the minimal and maximal values of $s(\snn)$ with respect to $\Delta\omega$. This behavior could, in principle, be suppressed by using the $\Delta\omega$-dependent Gaussian fluctuations (namely, by varying their width $\sigma_G$).%
\footnote{In fact, as demonstrated in~\cite{ELSHAWISH2023104940}, the twist angle has only a minimal impact on the average value of stress at GBs with a given inclination, while the width of the corresponding stress distribution might significantly depend on it (though it does not depend on the inclination itself). As a result, the expected value of $\snn$ for any chosen $\Delta\omega$ can be accurately predicted by the pure bicrystal model for randomly assigned twist angles, while the true value of $\Delta\omega$ affects only the width of the Gaussian distribution $\sigma_G$.}

In addition to that, even the $\Delta\omega$-averaged values of $s(\snn)$ --- corresponding to randomly assigned $\Delta\omega$ --- can differ for the GB types that share the same value of $E_{12}$. 
Consequently, $\sigma_G$ for $\Delta\omega$-averaged INS distributions, in a given material, cannot be only a function of $E_{12}$. Nevertheless, in the subsequent analysis, we will proceed as if it were and thus disregard these relatively minor higher-order corrections.

To arrive at these conclusions and hence determine whether the bicrystal-model estimates of $\snn$ are independent of $\Delta\omega$, and whether the corresponding $s(\snn)$ is truly a function of (just) $E_{12}$ for cubic crystal lattices, we should take the following approach: First evaluate the
coefficients $\tilde{C}_{ijkl}^{(N)}$ for a general GB type $[abc]$-$[def]$-$\Delta\omega$, as defined in Eq.~\eqref{eq:Cijkl}, and then insert them into Eqs.~\eqref{eq:c-coefficients} and~\eqref{eq:INS_a}, using the relation~\eqref{eq:c-a-connection} (and assuming grains of uniform size).  
The difficulty is that this process leads to an exceedingly complex expression.

Hence, we have adopted here a different strategy by showcasing the issue through a careful selection of a few distinct GB types. 
When numerical values are assigned to $a$, $b$, $c$, $d$, $e$, and $f$, while keeping $\Delta\omega$ as a free parameter, the standard deviation of the corresponding INS distribution takes the form%
\footnote{Standard deviation of INS distribution has to be an \emph{even} function of $\Delta\omega$ because the value of $s(\sigma_{nn})$ should not change upon exchanging the labels $1$ and $2$ of both GB grains.}	 
\begin{align}
	s(\snn) & = |\alpha| \, \frac{\sqrt{\sum_{i=0}^k n_i \, \cos (i \, \Delta\omega)}}{\sum_{j=0}^m d_j \, \cos (j \, \Delta\omega)}  \times \frac{2}{3\sqrt{5}} \Sigma_{\text{mis}} \ ,
\end{align}
where $n_i$ and $d_j$ are polynomial functions of $C_{11}$, $C_{12}$, and $C_{44}$, with coefficients depending on the values of $a$, $b$, $c$, $d$, $e$, and $f$. The exact values of $m$ and $k$ (where $k \geq m$) also depend on the GB type. 
For a chosen material and a given GB type, we can thus plot $s(\snn)$ against $\Delta\omega$.
It is also convenient to normalize the values of standard deviation to uniaxial loading ($\Sigma_{\text{mis}} = 1$) and the average strain approximation ($\alpha = \xbar{\alpha} = 1/(2 \, G)$); cf.~Eq.~\eqref{eq:alpha_bar}.

\begin{figure}[htp]
	\centering
	\includegraphics[width=\textwidth]{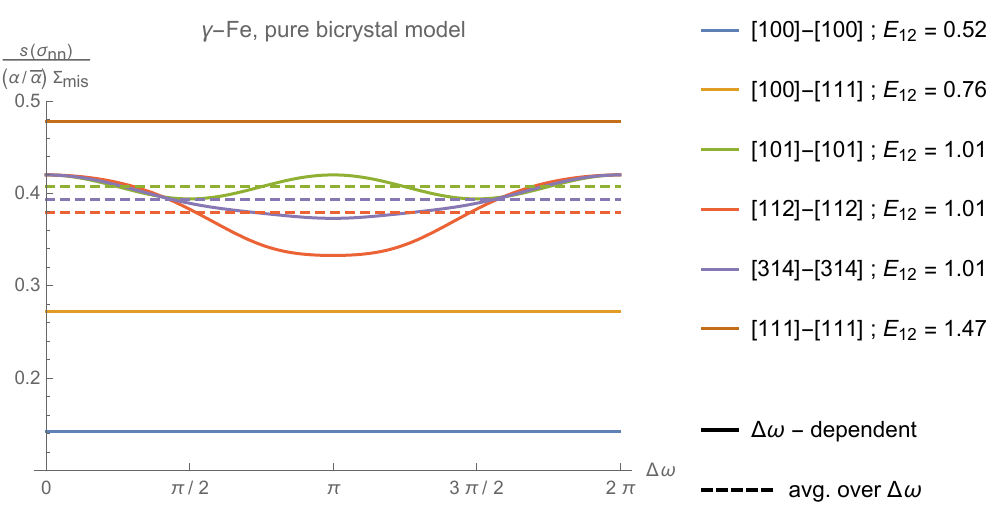}
	\caption{Standard deviation of $\sigma_{nn}$ distribution as a function of the twist-angle difference $\Delta\omega := \omega_2 - \omega_1$ for $\gamma$-Fe and six different GB types. For all of them maximal width of the corresponding INS distribution is reached at $\Delta\omega = 0$. Dashed lines represent the twist-angle averaged values.}
	\label{fig:twist_angle}
\end{figure}

\begin{figure}[htp]
	\centering
	\includegraphics[width=\textwidth]{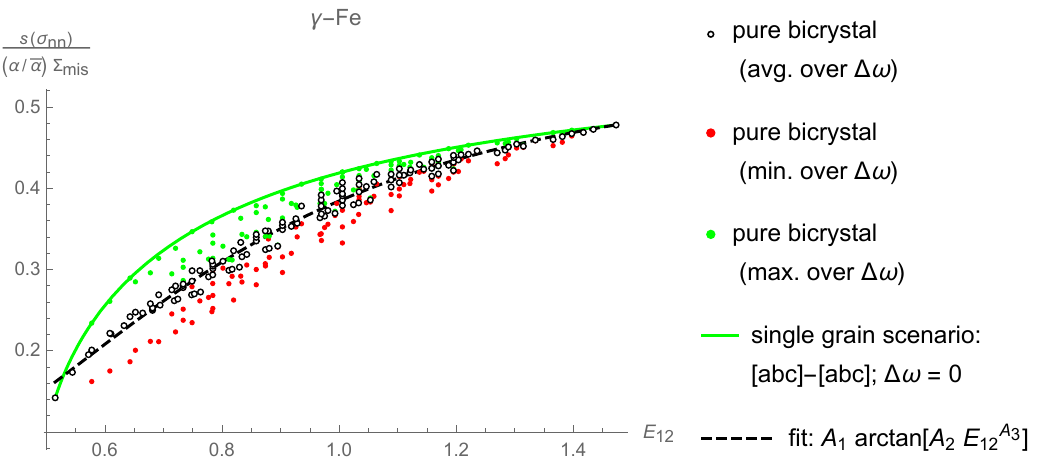}
	\caption{Maximal (\emph{green}), minimal (\emph{red}), and twist-angle averaged (\emph{black}) values of the standard deviation $s(\snn)$ for $120$ GB types corresponding to the selection of $(a,b,c)$ and $(d,e,f)$ from the $15$ preselected directions in the stereographic triangle, while keeping $\Delta\omega$ as a free parameter. The properties of $\gamma$-Fe have been chosen and inserted into the bicrystal solution, see Eqs.~\eqref{eq:std dev} and~\eqref{eq:c-coefficients}. The \emph{green curve} represents the single grain solution specified in Eq.~\eqref{eq:single-grain-limit},
	while the \emph{dashed black} line is the $A_1 \arctan{\left (A_2 \, E_{12}^{A_3}\right )}$ fit of $\Delta\omega$-averaged values. Normalization corresponds to uniaxial tensile loading ($\Sigma_{\text{mis}} = 1$) and the average strain approximation ($\alpha = \xbar{\alpha} = \frac{1}{2 \, G}$).}
	\label{fig:max_min}
\end{figure}

In Fig.~\ref{fig:twist_angle} the influence of the twist angle is illustrated for several GB types. 
Specific choices have been made based on different criteria.
Some GB types have been selected due to their independence from the twist angle $\Delta\omega$. This enables a direct comparison with the $\Delta\omega$-averaged outcomes presented in Fig.~\ref{fig:bicrystal_VS_FE}. Examples of such types include $[100]$-$[100]$, $[111]$-$[111]$, and $[100]$-$[111]$ configurations.%
\footnote{Since for $[100]$-$[100]$ and $[111]$-$[111]$ types, $\snn$ is independent of $\Delta\omega$, the bicrystal pair effectively behaves as a single crystal (as also evidenced by the end points of the \emph{green} curve in Fig.~\ref{fig:max_min} coinciding with the results for randomly assigned twist angles, denoted by \emph{black circles}). Consequently, the exact solution~\cite{f4a16d09-67e1-3bfb-9267-a5261a4d55c6,ac8407d4-2d87-3174-a1f8-941d54f90e52} could be applied to the strain field $\xbar{\epsilon}^{b}_{ij}$ of the bicrystal pair in these two cases.}
They correspond to the softest and stiffest GBs in the aggregate, as well as the GBs sandwiched between the softest and stiffest grains along the GB-normal direction, respectively.
The other three GB types, $[101]$-$[101]$, $[112]$-$[112]$, and $[314]$-$[314]$, are showcased because they all share a common value of $E_{12}$.  Despite that, their $\Delta\omega$-averaged values of $s(\snn)$ differ (albeit slightly, to the extent that this effect can be considered a next-order correction).

We can compute the value of $s(\snn)$ as a function of $\Delta\omega$ for any GB type $[abc]$-$[def]$ in a given material. 
Analysis of the resulting expressions reveals that the INS distributions are the widest when the twist-angle difference between the two grains is set to zero.%
\footnote{A similar conclusion was reached also in~\cite{ELSHAWISH2023104940}, where a simpler analytical model (the aforementioned buffer-grain model) was used for estimating the GB stresses.}
In Fig.~\ref{fig:max_min}, we show the maximal, minimal, and $\Delta\omega$-averaged values of $s(\snn)$ for $\gamma$-Fe and the range of GB types obtained by selecting GB normals $(a,b,c)$ and $(d,e,f)$ from a set of $15$ preselected directions within the stereographic triangle. 

The key observation is that the points associated with the maximal values of $[abc]$-$[abc]$-$\Delta\omega$ types (reached for $\Delta\omega = 0$) are forming a smooth curve. This curve represents an \emph{upper bound} for $s(\snn)$ as a function of $E_{12}$. The $[abc]$-$[abc]$-$\Delta\omega$ GB types with \mbox{$\Delta\omega = 0$} correspond precisely to the \emph{single grain} limit, which can be used for estimating the \emph{intragranular} normal stresses, i.e., to calculate the stress perpendicular to a chosen plane within a single grain (modelled in this framework as a region of constant stress).
In this limit, the relevant expressions simplify into:
\begin{align}
	s(\sigma_{nn}) & = \frac{2 \, |\alpha|}{(2 \, s_0+s_{44}) \, s_{44}} \, \sqrt{(6 \, s_{11} \, s_0 + 6 \, s_{11} \, s_{44} + s_{44}^2) - \frac{6 \, (s_0+s_{44}) \, \xbar{E}^{-1}}{E_{12}}}  \times \frac{2}{3\sqrt{5}} \Sigma_{\text{mis}} \ .
	\label{eq:single-grain-limit}
\end{align}

The most important aspect of the solution presented above is that $s(\snn)$ indeed becomes only a function of $E_{12}$. 
Hence, we succeeded in deriving the dependence of INS distribution's width (in fact, its upper bound) on the effective stiffness, at least for specific GB types, if not in the general case, and thereby explicitly proved the significance of the $E_{12}$ parameter, justifying its prior use. However, the functional dependence in Eq.~\eqref{eq:single-grain-limit} is not of the form $s(\sigma_{nn}) \approx A_1 \arctan{\left (A_2 \, E_{12}^{A_3}\right )} \, \Sigma_{\text{mis}}$, that successfully fits the $[abc]$-$[def]$ GB types averaged over $\Delta\omega$. Instead, it is described by a different relation, of the type
$s(\sigma_{nn}) = C_1 \sqrt{1 + C_2/E_{12}} \, \Sigma_{\text{mis}}$. Both sets of coefficients ($A_k$ and $C_k$) depend only on the material properties ($s_{11}$, $s_{12}$, and $s_{44}$).

For random twist angles, the expression~\eqref{eq:INS_a} predicting the GB-normal stress in the bicrystal model simplifies to (for any lattice symmetry):
\begin{align}
	\sigma_{zz}^{\Delta\omega} & = \xbar{a}_{xx,yy} \, (\Sigma_{xx} + \Sigma_{yy}) + \xbar{a}_{zz} \, \Sigma_{zz} \ , \label{eq:avg_twist}
\end{align}
where $\xbar{a}_{ij} = (2\, \pi)^{-2} \int_{0}^{2\pi}\int_{0}^{2\pi} a_{ij} \, d\omega_1 \, d\omega_2$, and $\xbar{a}_{xx} = \xbar{a}_{yy} := \xbar{a}_{xx,yy}$.
This result also provides justification for the buffer-grain model~\cite{ELSHAWISH2023104940}, which arrived at the expression of a similar form.
All the dependence on $\phi$ dropped out of it (while for uniaxial loading in the $Z$-direction, even the dependence on $\psi$ vanishes). Consequently,
\begin{align}
	\begin{split}
		s(\snn^{\Delta\omega}) & = |\xbar{a}_{xx,yy}-\xbar{a}_{zz}| \times \frac{2}{3\sqrt{5}} \Sigma_{\text{mis}} \\ 
		& = |\alpha| \, |\xbar{c}_{xx,yy}-\xbar{c}_{zz}| \times \frac{2}{3\sqrt{5}} \Sigma_{\text{mis}} \\
		& = \frac{|\alpha|}{2} \, |(C_{11}+2 \, C_{12})-3 \, \xbar{c}_{zz}| \times \frac{2}{3\sqrt{5}} \Sigma_{\text{mis}} \quad \text{(for cubic symmetry)} \ .
	\end{split}
\end{align}
%

\subsection{Comparison with FE simulations}

To compare the results of the bicrystal model with realistic stresses obtained in numerical simulations, we will examine the INS distributions generated by both approaches. We will focus on several representative GB types ($[100]$-$[100]$, $[102]$-$[334]$, and $[111]$-$[111]$) within a chosen material ($\gamma$-Fe).

Two scenarios will be considered: either the \emph{pure} bicrystal model embedded in a homogeneous and isotropic neighbourhood, or introducing Gaussian fluctuations to induced stresses, meant to mimic the effect of anisotropic grains surrounding the bicrystal pair. In the latter case, an additional free parameter, $\sigma_{G}$, is required alongside $\alpha$ and $\beta$.

\subsubsection{Pure bicrystal model\label{sec:pure bicrystal}}

As demonstrated in Fig.~\ref{fig:bicrystal_VS_FE}, the bicrystal model, when combined with the average strain approximation~\eqref{eq:assumption}, results
in a standard deviation curve that is overly steep.
To address this issue, we impose a less constraining condition on the bicrystal strain $\xbar{\epsilon}^{b}_{ij}$, introducing two additional parameters, $\alpha$ and $\beta$, see Eq.~\eqref{eq:assumption_general}.
By appropriately adjusting their values, we can reconcile the discrepancies between the $s(\snn)$ obtained from FE simulations and the model predictions. 
However, our goal extends beyond matching the standard deviation; we aim to fit the entire INS distribution, ensuring that all its moments agree with their predicted values.

For a chosen material ($C_{ij}$) and GB type ($a$, $b$, $c$, $d$, $e$, $f$), we can evaluate the first two moments of $\snn$ distribution in the bicrystal model, as described by Eqs.~\eqref{eq:mean_gen} and~\eqref{eq:std dev}. Then $\alpha$ and $\beta$ can be determined by averaging those expressions over $\Delta\omega$ and fitting them to the FE values. Specifically, we need to solve the equations
\begin{align}
	\mu_h^{(1)} & = \langle \snn \rangle_h = \mu_f^{(1)} \ , \label{eq:moment1_pure_bicrystal} \\
	\mu_h^{(2)} & = \langle (\snn - \langle \snn \rangle_h)^2 \rangle_h = \mu_f^{(2)} \label{eq:moment2_pure_bicrystal}
\end{align}
for the first two moments of FE distribution $h$ and bicrystal distribution $f$,
where
\begin{align}
	\mu_f^{(1)} & = \langle \snn \rangle_f \label{eq:moment1_bicrystal} \\
	& = (\alpha - 3 \, \beta) \, (c_{xx} + c_{yy} + c_{zz}) \times \frac{1}{3} \Tr\mathbf{\Sigma} \ ,  
	\nonumber \\
	\mu_f^{(2)} & = \langle (\snn - \langle \snn \rangle_f)^2 \rangle_f \label{eq:variance of bicrystal} \\
	& = \alpha^2 \left (\frac{1}{2} \, \big ((c_{xx} - c_{yy})^2 + (c_{yy} - c_{zz})^2 + (c_{zz} - c_{xx})^2 \big ) + 3 \, (c_{xy}^2 + c_{xz}^2 + c_{yz}^2) \right ) \times \frac{2}{15} \Tr(\mathbf{\Sigma}^2_{\text{dev}})
	\ . \nonumber
\end{align}

In principle, both parameters, $\alpha$ and $\beta$, could vary with GB type and material properties, potentially assuming different values for each $E_{12}$ (at a given $A^u$).
However, for cubic lattice symmetry, the relationship between $\alpha$ and $\beta$ is quite straightforward. In this limit, the requirement that the mean value should match $\tfrac{1}{3}\Tr\mathbf{\Sigma}$ for every GB type, effectively fixes the value of $\beta$, as indicated in Eq.~\eqref{eq:beta}. Consequently, we are left with just one adjustable parameter ($\alpha$). While it is possible to determine this parameter also through fitting to higher-order moments, the most precise approach typically involves using the standard deviation.

\begin{figure}[htb]
	\centering
	\includegraphics[width = 0.328\textwidth, valign=c]{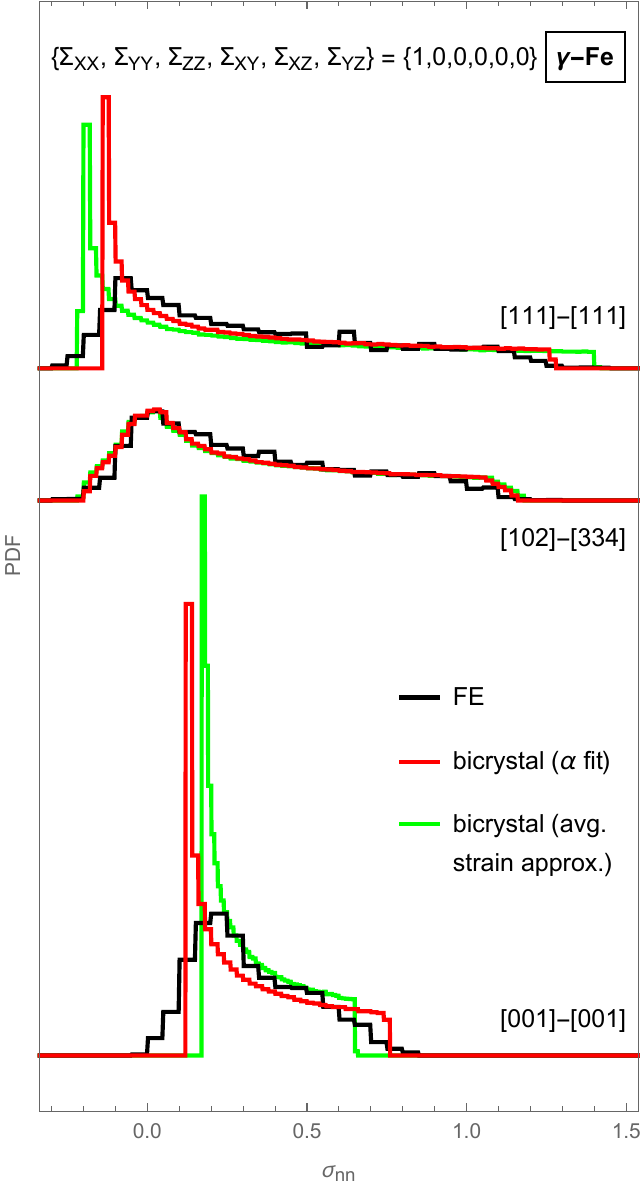}
	\includegraphics[width = 0.328\textwidth, valign=c]{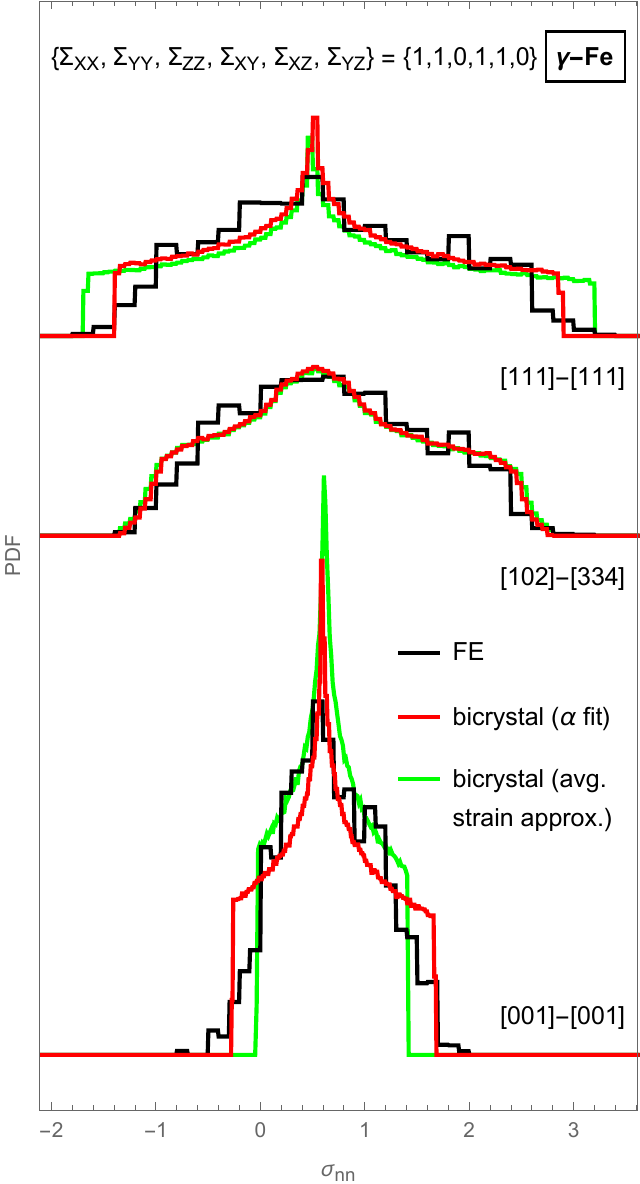}
	\includegraphics[width = 0.328\textwidth, valign=c]{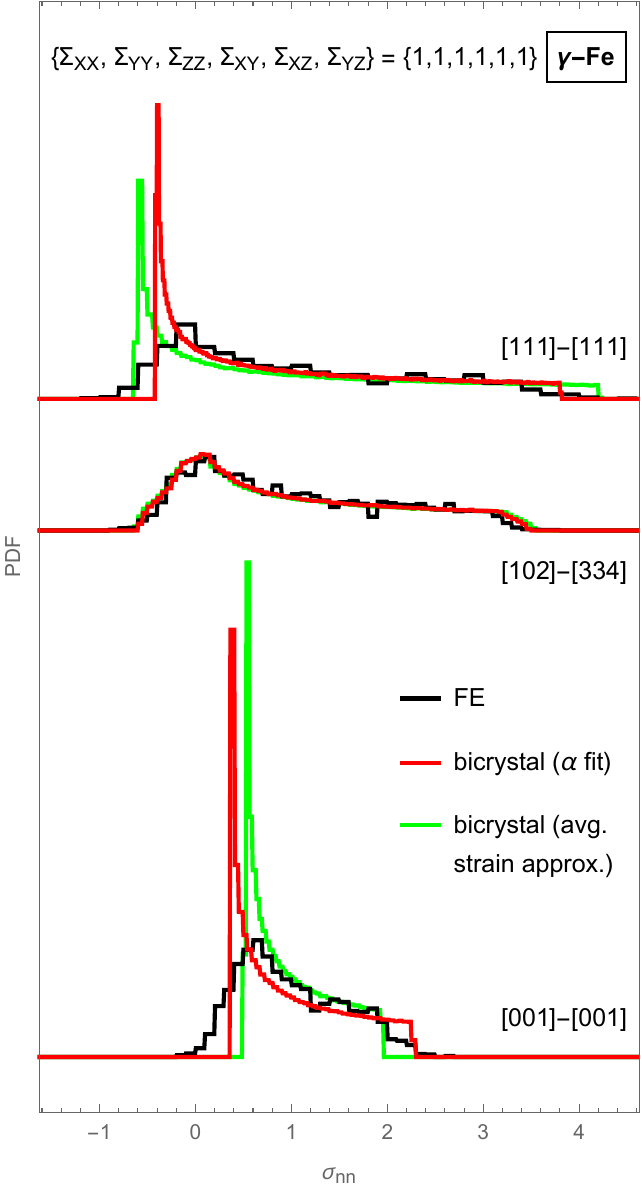}
	\caption{Comparison of (reduced) INS distributions from FE simulations (\emph{black}; $N = 1636$) with those from the (pure) bicrystal model. In the latter case, two variants are considered: 
	using the average strain approximation $\alpha = 1/(2\, G)$ (\emph{green}) or determining the values of $\alpha$ and $\beta$ through fitting specified in Eqs.~\eqref{eq:moment1_pure_bicrystal} and~\eqref{eq:moment2_pure_bicrystal} (\emph{red}). Results are presented for three external loadings and three different $\Delta\omega$-averaged GB types in $\gamma$-Fe. In each of these cases, the 	mean values of all three distributions are the same, while their standard deviations match only for the \emph{red} and \emph{black} curves.}
	\label{fig:distributions}
\end{figure}

Upon inserting the values of $\alpha$ and $\beta$ into $\snn$, we can assess how accurately we have replicated the actual INS distributions. Fig.~\ref{fig:distributions} features three distinct GB types: $[001]$-$[001]$, $[111]$-$[111]$, and $[102]$-$[334]$, representing the softest, stiffest, and an intermediate GB type. The choice of the latter has been motivated by the fact that its effective GB stiffness in $\gamma$-Fe is $E_{12} = 0.97$, which closely matches the stiffness of isotropic grains ($E_{12} = 1$) or the average stiffness of the aggregate (corresponding to random GBs, see Fig.~\ref{fig:random_distribution_GammaFe}). 

It is worth noting that the same $\alpha$ was used for all loadings in Fig.~\ref{fig:distributions}, hence confirming the validity of the scaling factors $\Tr\mathbf{\Sigma}$ and $\Sigma_{\text{mis}}$ for the first two moments, respectively. Nonetheless, it is also evident that $\alpha$ and $\beta$ alone are insufficient to accurately reproduce the FE distributions.

\FloatBarrier

\subsubsection{Gaussian modulation}

The approach taken in the previous section does not quite capture the genuine stress responses, particularly for very soft and very stiff GBs.
For these GB types, the predicted INS distributions manifestly disagree with the outcomes of the FE simulations, see Fig.~\ref{fig:distributions}. It is evident that the predictions of the bicrystal model exhibit a notably sharper and narrower profile, despite the same ``width'' of distributions in both cases (specifically, the parameters $\alpha$ and $\beta$ have been adjusted to ensure the agreement between the mean value and standard deviation of both distributions).

The naive \emph{two-parameter} fitting procedure thus needs to be extended to include the effect of the inhomogeneous neighbourhood of a bicrystal pair. In line with our perturbative framework (see also Eq.~\eqref{eq:perturbativity_1st_order}), this extension can be conceptualized as follows:
\begin{align}
	\small
	\begin{split} 
		\textbf{bicrystal model} + \textbf{Gaussian fluctuations} \approx & 
		\, \textbf{INS distributions on individual GB types} \ .
	\end{split}
	\normalsize
\end{align}
The approach again involves convolution with normally distributed random fluctuations.
Due to the additional parameter $\sigma_G$, we now have a \emph{three-parameter} fitting procedure: 
\begin{align}
	\mu_h^{(1)} & = \langle \snn \rangle_h = \mu_f^{(1)} \ , \label{eq:moment1} \\
	\mu_h^{(2)} & = \langle (\snn - \langle \snn \rangle_h)^2 \rangle_h = \mu_f^{(2)} + \sigma_G^2 \ , \label{eq:moment2} \\
	\mu_h^{(3)} & = \langle (\snn - \langle \snn \rangle_h)^3 \rangle_h = \mu_f^{(3)} \ , \label{eq:moment3}
\end{align}
providing the values of $\alpha$, $\beta$, and $\sigma_G$.
The first two statistical moments of the pure bicrystal distribution $f$ are provided in Eqs.~\eqref{eq:moment1_bicrystal} and~\eqref{eq:variance of bicrystal}, while the third is given by:
\begin{align}
	\mu_f^{(3)} & = \langle (\snn - \langle \snn \rangle_f)^3 \rangle_f \nonumber  \nonumber \\ 
	\begin{split}	
		& = \alpha^3 \, \frac{1}{2} \, \Big (2 \left (c_{xx}^3+c_{yy}^3+c_{zz}^3 \right ) + 12 \, c_{xx} \, c_{yy} \, c_{zz} + 54 \, c_{xy} \, c_{xz} \, c_{yz} + \\
		& + 9 \left (c_{xx} \, (c_{xy}^2+c_{xz}^2-2 \, c_{yz}^2) + c_{yy} \, (c_{xy}^2+c_{yz}^2-2 \, c_{xz}^2) + c_{zz} \, (c_{xz}^2+c_{yz}^2-2 \, c_{xy}^2) \right ) - \\
		& - 3 \left (c_{xx}^2 \, (c_{yy} + c_{zz}) + c_{yy}^2 \, (c_{xx} + c_{zz}) + c_{zz}^2 \, (c_{xx} + c_{yy}) \right ) \Big ) \times \frac{8}{105} \Tr(\mathbf{\Sigma}^3_{\text{dev}}) \ .
	\end{split} 
\end{align}
The required moments of the convolution $h$ are obtained from the $\snn$ distribution produced in FE simulation. In principle, we could use any three moments $\mu_h^{(n)}$ for the fitting, 
but from practical perspective it is best to take the lowest-order moments 
because they are most accurate, especially when dealing with imprecisely known FE distributions (computed on only finite aggregate).%
\footnote{A notable exception are the ``pathological'' cases in which the loading leads to some of the lowest three statistical moments $\mu_f^{(i)}$ becoming exactly zero due to $\Tr\mathbf{\Sigma}=0$, $\Tr(\mathbf{\Sigma}^2_{\text{dev}}) = 0$, or $\Tr(\mathbf{\Sigma}^3_{\text{dev}}) = 0$. However, since numerical simulations use aggregates of finite size, the corresponding moments $\mu_h^{(i)}$ do not precisely vanish in such instances. In these situations, it is preferable
to determine $\alpha$, $\beta$, and $\sigma_G$ using the lowest three \emph{non-vanishing} moments.}
\begin{figure}[!htb]
	%
	\includegraphics[width = 0.328\textwidth, valign=c]{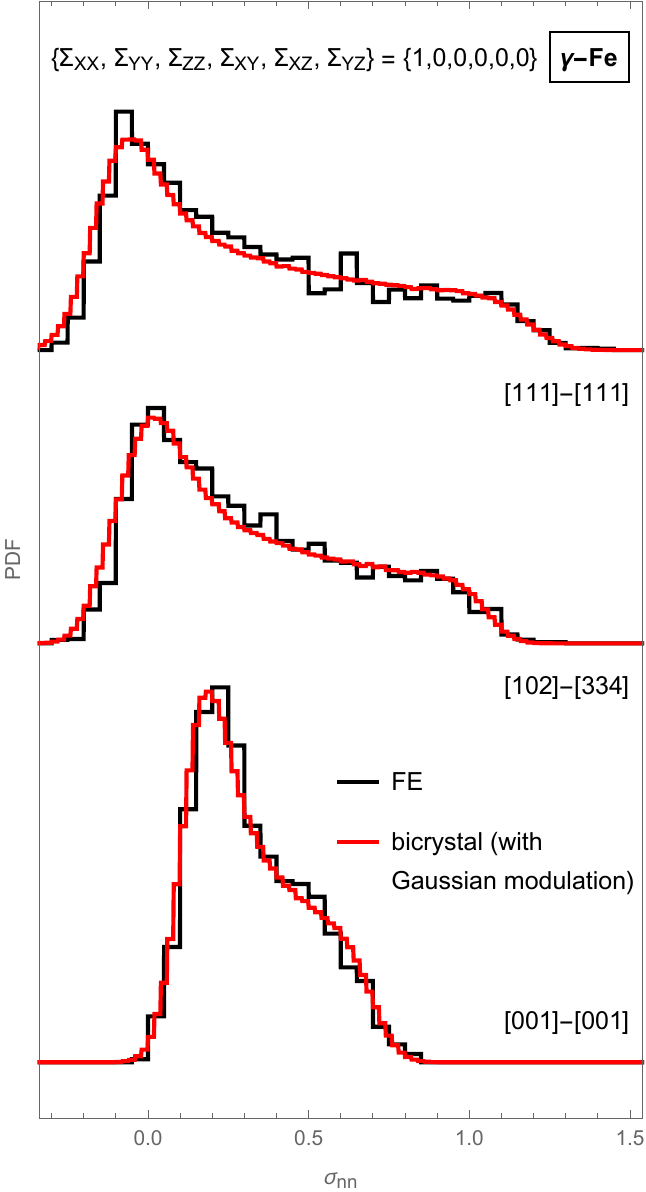}
	\includegraphics[width = 0.328\textwidth, valign=c]{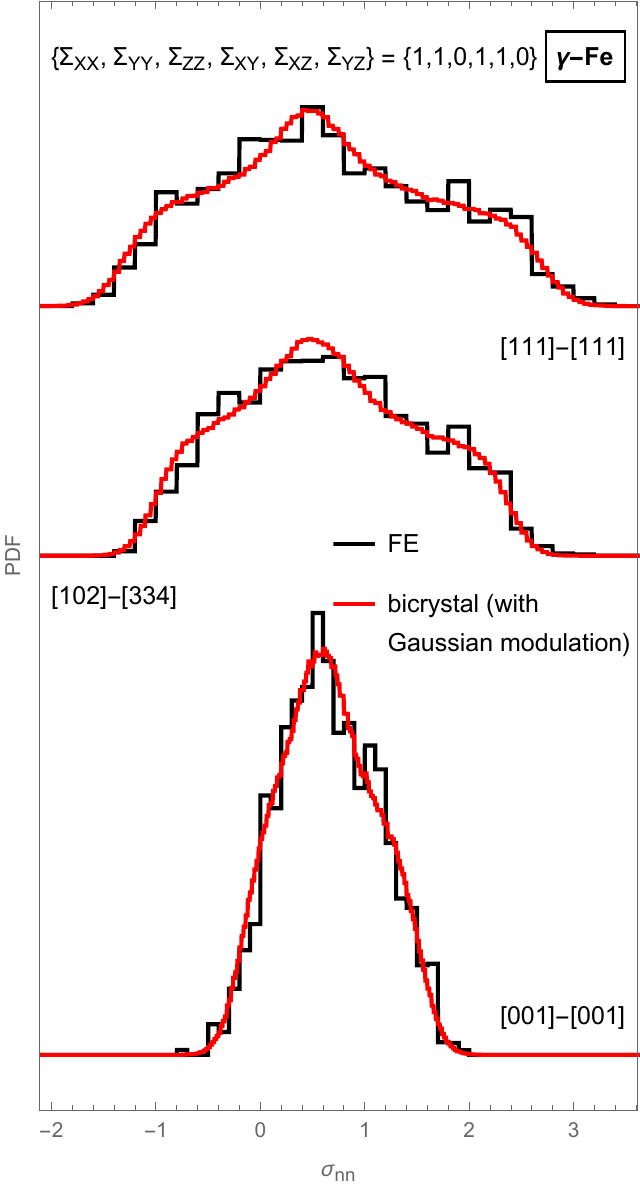}
	\includegraphics[width = 0.328\textwidth, valign=c]{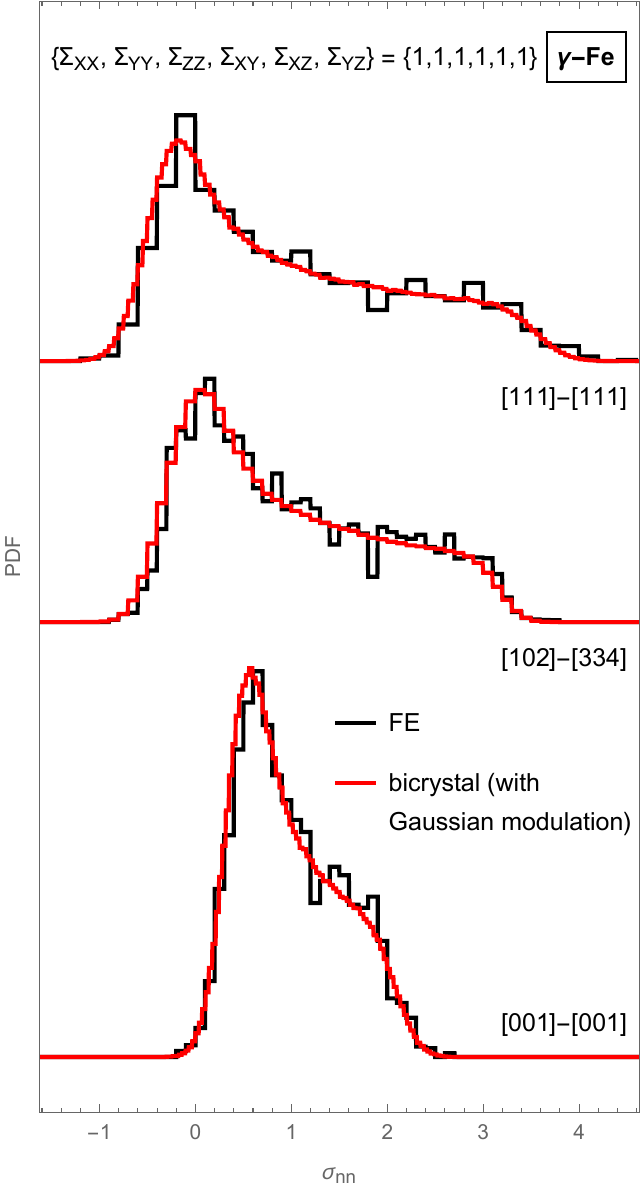}
	\caption{Same as in Fig.~\ref{fig:distributions}, but for the bicrystal model with Gaussian modulation. Parameters $\alpha$, $\beta$, and $\sigma_G/\Sigma_{\text{mis}}$ are determined from the lowest three statistical moments of FE distributions and depend only on the GB type, irrespective of the external loading. Thus, all three panels are generated using the same set of parameters.}
	\label{fig:distributions_conv}
\end{figure}

With the fitted values of $\alpha$, $\beta$, and $\sigma_G$, we can now compare the convoluted bicrystal-model distributions with the FE results, as shown in Fig.~\ref{fig:distributions_conv}.
The agreement appears significantly improved compared to Fig.~\ref{fig:distributions}, which did not include the Gaussian fluctuations.
However, the accuracy of the proposed method is constrained by the inherent imprecision of the FE results due to limited statistics. In particular, the FE simulations are based on only $1636$ GBs of a selected GB type, which is far from sufficient to produce a uniform distribution of GB orientations. 
In contrast, the bicrystal case employs $10^{6}$ random GBs, resulting in a much smoother distribution.
For each specific GB type, the same values of $\alpha$, $\beta$, and $\sigma_G/\Sigma_{\text{mis}}$ have been used in all the panels of Fig.~\ref{fig:distributions_conv}.

\begin{figure}[!htb]
	\centering
	\includegraphics[width = 0.5\textwidth, valign=c]{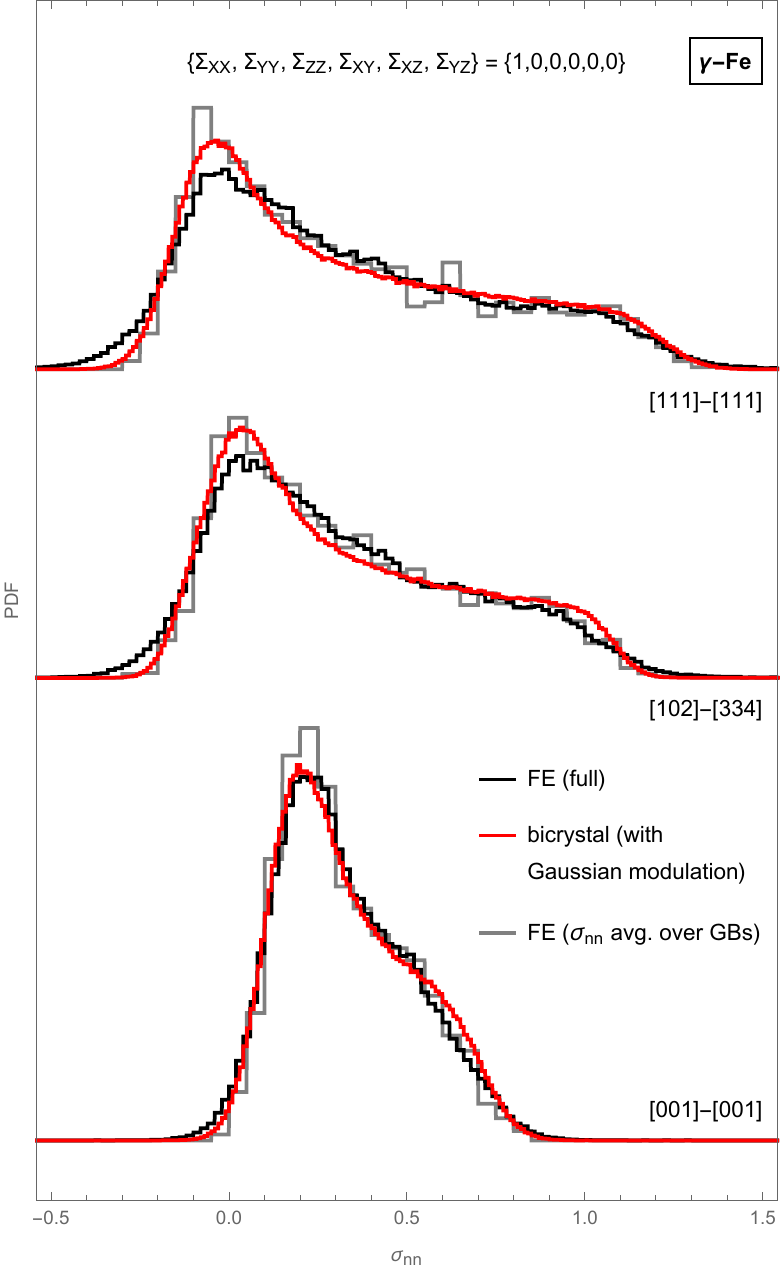}
	\caption{A comparison between the modelled (\emph{red}) and realistic (\emph{black}; $N = 159461$) \emph{full} INS distributions for $\gamma$-Fe, three different GB types (with random $\Delta\omega$), and uniaxial tensile loading. The values of $\alpha$, $\beta$, and $\sigma_G$ have been determined from the lowest three statistical moments of the (full) FE distributions.
	Additionally, the \emph{reduced} INS distributions (\emph{gray}; $N = 1636$) are also displayed.}
	\label{fig:distributions_conv_NOavg}
\end{figure}

At last, in Fig.~\ref{fig:distributions_conv_NOavg} we present a comparison between the bicrystal-model predictions and the \emph{full} INS distributions, corresponding to normal stresses across all $159461$ finite elements associated with the GBs of a selected type. 
These distributions are wider than the \emph{reduced} versions in Fig.~\ref{fig:distributions_conv}, but given the substantial number of elements compared to the number of GBs, they also offer increased accuracy and detail.

%

\subsection{Fits across GB types and material properties\label{sec:phenomenological fits}}

Up to this point, the procedure outlined in the previous section was completely general and applicable to any lattice symmetry. On the downside, it requires numerical simulations to determine for a given material and GB type the values of parameters $\alpha$, $\beta$, and $\sigma_G$, extracting them from the first three statistical moments of the FE distribution. 

\looseness-1
Given the absence of a mechanism to determine these parameters from first principles, our approach in this section is to fit them empirically. 
With the help of these approximate fits, one can almost instantaneously (and with minimal resources) estimate the INS distribution for any material and GB type,
circumventing the need for extensive numerical simulations.
 
Here, we will focus only on materials with \emph{cubic} lattices. In this particular scenario, the procedure simplifies in several significant aspects. First of all, just two parameters ($\alpha$ and $\sigma_G$) require fitting in this case, as $\beta$ is already fully determined by Eq.~\eqref{eq:beta} due to the fact that, in cubic polycrystallines, $\langle\snn\rangle = \tfrac{1}{3}\,\Tr{\mathbf{\Sigma}}$ for any GB type.%
\footnote{\looseness-1 As pointed out later, for non-cubic materials, we can approximate the parameter $\beta$ using Eqs.~\eqref{eq:n_approx} and~\eqref{eq:uni_general}.}
Secondly, for materials with cubic symmetry, $\alpha$ and $\sigma_G$ are solely functions of $E_{12}$,%
\footnote{In truth, this expectation cannot be entirely accurate, as bicrystal-model results exhibit much greater scatter across different GB types of the same stiffness compared to the FE results (see Fig.~\ref{fig:bicrystal_VS_FE}). Consequently, $\alpha$, $\beta$, and $\sigma_G$ cannot be solely functions of $E_{12}$, even for a selected material with cubic lattice symmetry; they need to be GB-type specific.}
while for lower symmetries of the crystal lattice, they depend on additional parameters (such as $\nu_{12}$, see~\cite{ELSHAWISH2023104940}). 
Thirdly, for non-cubic lattices, $\sigma_G$ does not precisely scale with $\Sigma_{\text{mis}}$, as evidenced by the material's response to hydrostatic loading. 
And lastly, it also remains unclear to what extent we can approximate the fluctuation $\Delta\snn$ with a Gaussian distribution in that case.	

\looseness-1
Systematically exploring the parameter space of various cubic materials and GB types makes it possible to express (or rather fit) the parameters $\alpha$, $\beta$, and $\sigma_G/\Sigma_{\text{mis}}$ as functions of $A^u$ and $E_{12}$. For any material with cubic lattice symmetry, the parameter $\alpha$ roughly follows (neglecting the scatter in the bicrystal case) a decreasing exponential trend with respect to $E_{12}$:
\begin{align}
	\frac{\alpha}{\xbar{\alpha}} & = b_1 \, E_{12}^{-b_2} \ . \label{eq:alpha_fit}
\end{align}
Here, $\xbar{\alpha} = (2 \, G)^{-1}$ represents the value of $\alpha$ in 
the isotropic limit (cf.~Footnote~\ref{footnote:isotropic}), while $b_1$ and $b_2$ denote some material-specific parameters.
The width of Gaussian modulation ($\sigma_G$) required to match the full FE stress distribution is, to a good approximation, an increasing linear function of $E_{12}$:
\begin{align}
	\frac{\sigma_G}{\Sigma_{\text{mis}}} & = d_1 \, E_{12} + d_2 \ . \label{eq:sigma_fit}
\end{align}

\begin{figure}[!htb]
	\centering
	\includegraphics[width = 0.49\textwidth]{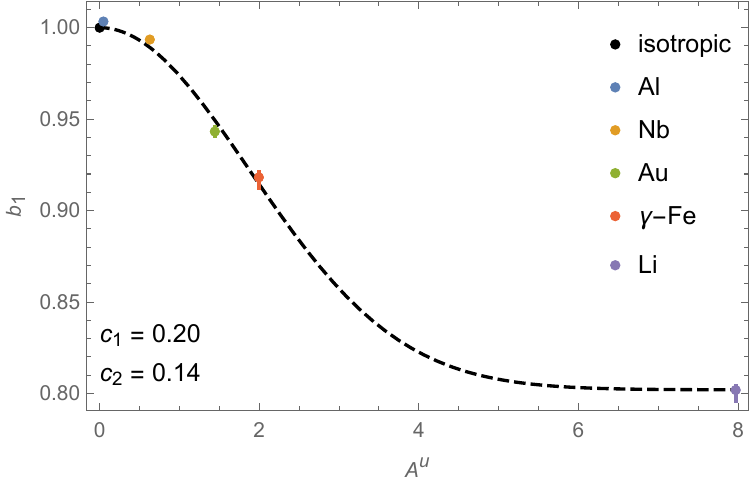}
	\includegraphics[width = 0.49\textwidth]{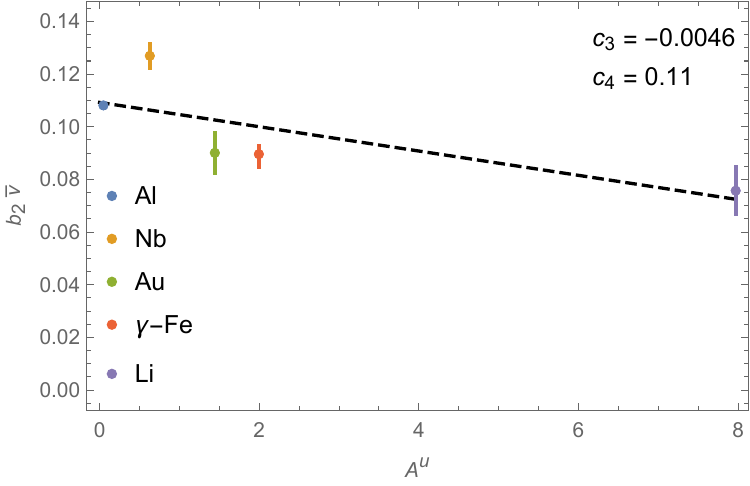}
	\includegraphics[width = 0.49\textwidth]{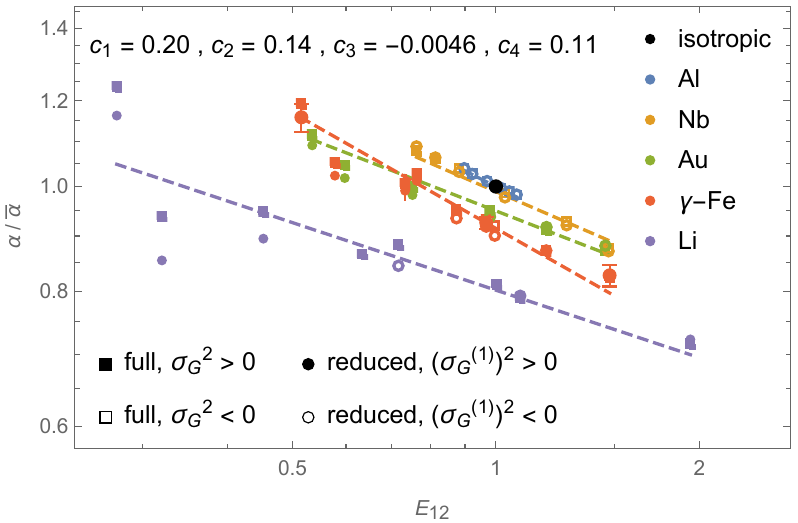}
	\includegraphics[width = 0.49\textwidth]{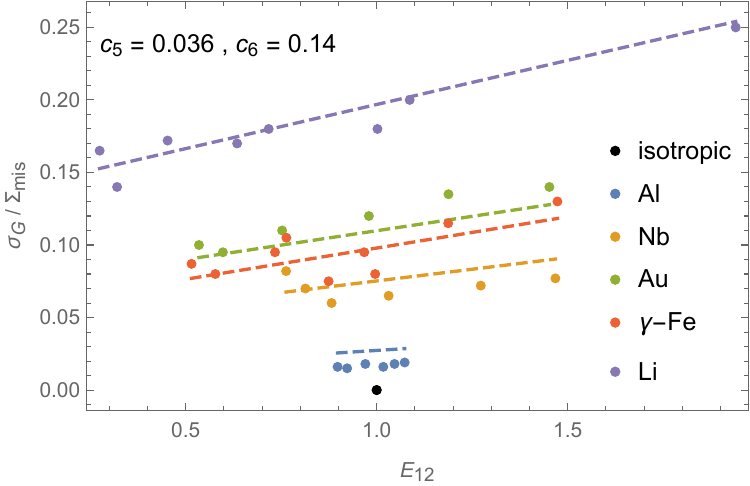}
	\caption{\looseness-1 \emph{Bottom left}: The parameter $\alpha$ was determined for various GB types and cubic materials by matching the third moments of the pure bicrystal model with either full (\emph{squares}) or reduced (\emph{circles}) FE distributions. Due to the imprecision of simulation results, the second moment of the pure bicrystal distribution can exceed that from FE (\emph{empty} markers). Error bars indicate the estimated variation of FE distribution, either due to different loading configurations or different Voronoi aggregates. The \emph{dashed line} represents the postulated $\alpha(E_{12})$ relation~\eqref{eq:alpha_fit}, based on fitting $b_1$ (\emph{top left}) and $b_2$ (\emph{top right}) to $A^u$ through expressions~\eqref{eq:fitting_parameters_1} and~\eqref{eq:fitting_parameters_2}, respectively, where the error bars are due to slightly different values obtained from full and reduced distributions. Using Eq.~\eqref{eq:alpha_fit} to assign the value to $\alpha$, the width of required Gaussian fluctuation $(\sigma_G)$ is obtained by matching the second moment of the full FE distribution (\emph{bottom right}). \emph{Dashed lines} represent the fit~\eqref{eq:sigma_fit} with $d_1$ and $d_2$ determined from Eqs.~\eqref{eq:fitting_parameters_3} and~\eqref{eq:fitting_parameters_4}, respectively. The values from Table~\ref{tab:fitting_parameters} have been used in all the \emph{dashed} fits.}
	\label{fig:fitting_parameters}
\end{figure}

For practical use, we provide here phenomenological fits of parameters $b_1$, $b_2$, $d_1$, and $d_2$ to the values of $A^u$, see Fig.~\ref{fig:fitting_parameters}. The following fitting functions have been selected for their simplicity%
\footnote{The proposed fitting functions are also perfectly consistent with the isotropic case, as $b_1$ (representing the value of $\alpha / \xbar{\alpha}$ at $E_{12} = 1$) indeed becomes $1$ for $A^u = 0$, while the width of the Gaussian ($\sigma_G$) in this case reduces to zero ($d_1 = -d_2$).}
and small number of fitting parameters (gathered in Table~\ref{tab:fitting_parameters}):
\begin{align}
	b_1 & = c_1 \, \exp{\left (-c_2 \, (A^u)^2\right )} + (1-c_1)  \ , \label{eq:fitting_parameters_1} \\
	b_2 & = \left (c_3 \, A^u + c_4\right ) \, \xbar{\nu}^{-1} \ , \label{eq:fitting_parameters_2} \\
	d_1 & = c_5 \, (A^u)^{1/4} \ , \label{eq:fitting_parameters_3} \\
	d_2 & = c_6 \, (A^u)^{1/2} \, \xbar{\nu} \ . \label{eq:fitting_parameters_4}
\end{align}	
\vspace*{-6mm}
\begin{table}[h]
	\caption{\label{tab:fitting_parameters}
		Optimal values of fitting parameters (for polycrystalline materials with cubic lattice symmetry), as were defined in Eqs.~\eqref{eq:fitting_parameters_1}--\eqref{eq:fitting_parameters_4}.}
	\centering
		\begin{tabular}{*{7}{c}}
			\toprule
			$c_1$ & $c_2$ & $c_3$ & $c_4$ & & $c_5$ & $c_6$ \\
			\midrule
			0.20 & 0.14 & -0.0046 & 0.11 & & 0.036 & 0.14 \\
			\bottomrule
	\end{tabular}
\end{table}

\begin{figure}[!htb]
	\centering
	\includegraphics[width = 0.5\textwidth, valign=c]{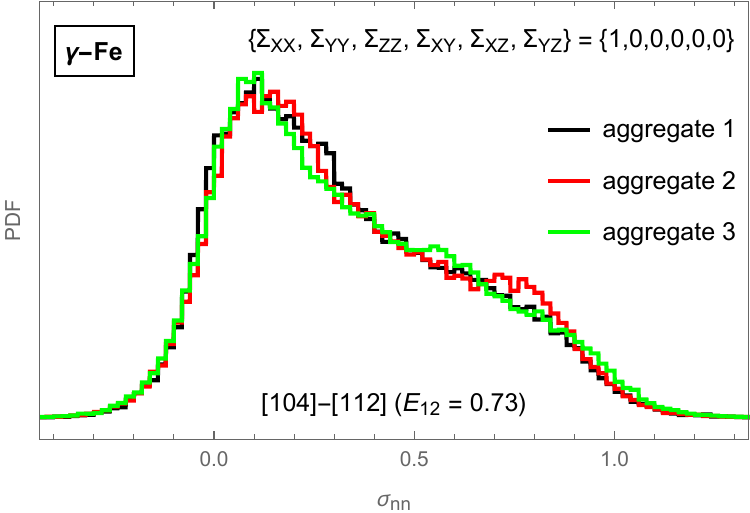}
	\caption{A full INS distribution for the $[104]$-$[112]$ GB type (with randomly assigned twist angles $\Delta\omega$) in $\gamma$-Fe under uniaxial loading. Comparison is shown between the FE simulation results obtained from three different Voronoi aggregates with $4000$ grains.}
	\label{fig:3_aggregates}
\end{figure}

\begin{figure}[!htb]
	%
	\includegraphics[width = 0.328 \textwidth, valign=c]{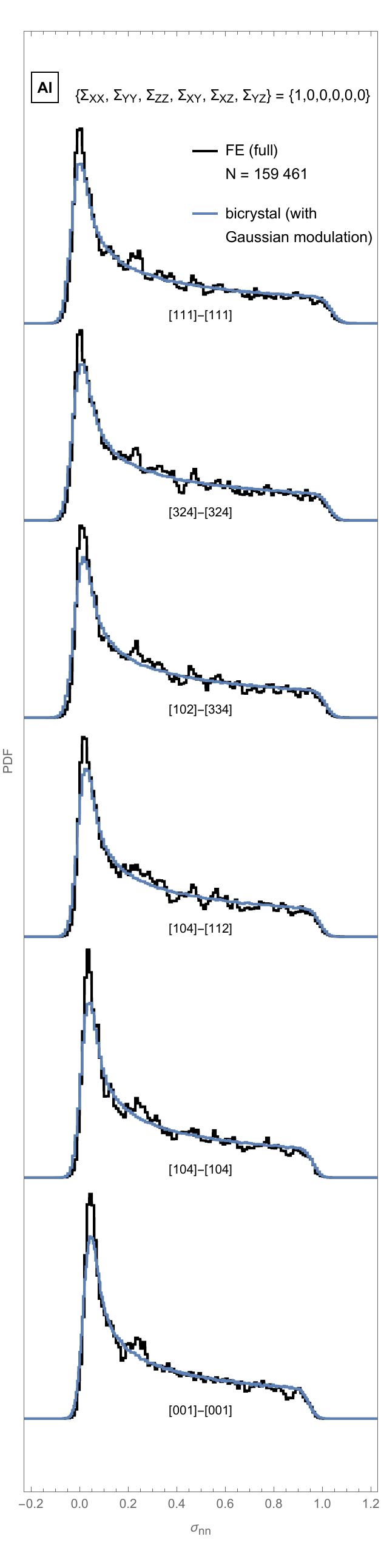} 
	\includegraphics[width = 0.328 \textwidth, valign=c]{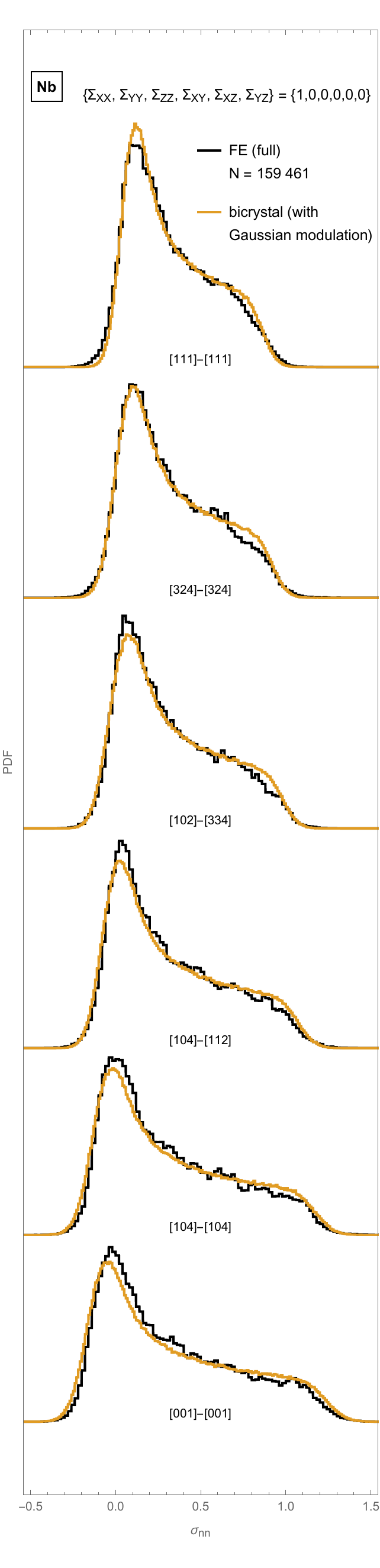}
	\includegraphics[width = 0.328 \textwidth, valign=c]{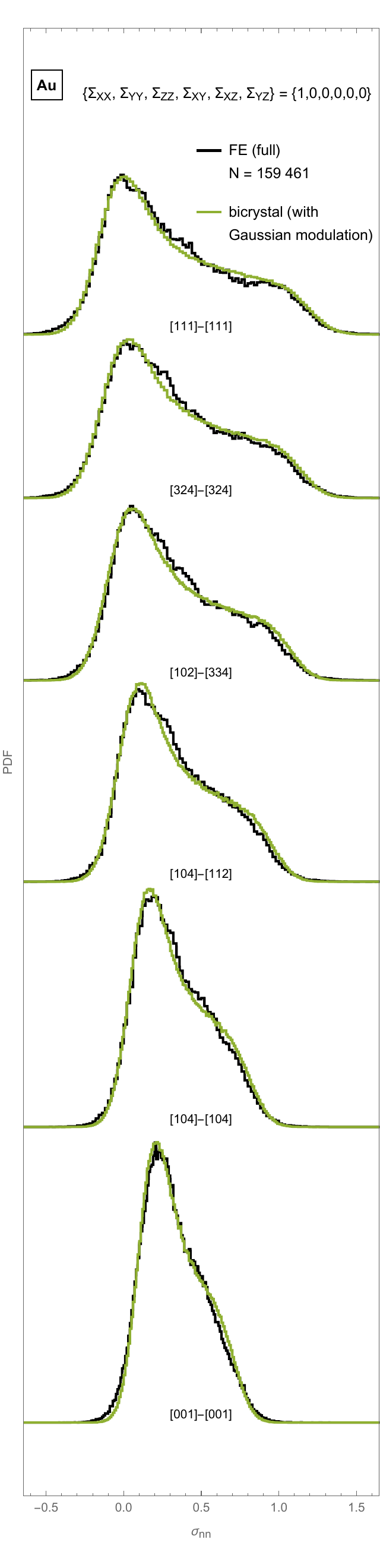}
	\caption{A comparison between the modelled (\emph{colored}) and FE (\emph{black}; $N = 159461$) \emph{full} INS distributions for several GB types (with random $\Delta\omega$) in Al, Nb, and Au under uniaxial tensile loading. The values of $\alpha$, $\beta$, and $\sigma_G$ have been determined from fits; cf.~Eqs.~\eqref{eq:beta}, \eqref{eq:alpha_fit}--\eqref{eq:fitting_parameters_4}, and Table~\ref{tab:fitting_parameters}.}
	\label{fig:distributions_materials}
\end{figure}

\begin{figure}[!htb]
	%
	\includegraphics[width = 0.328 \textwidth, valign=c]{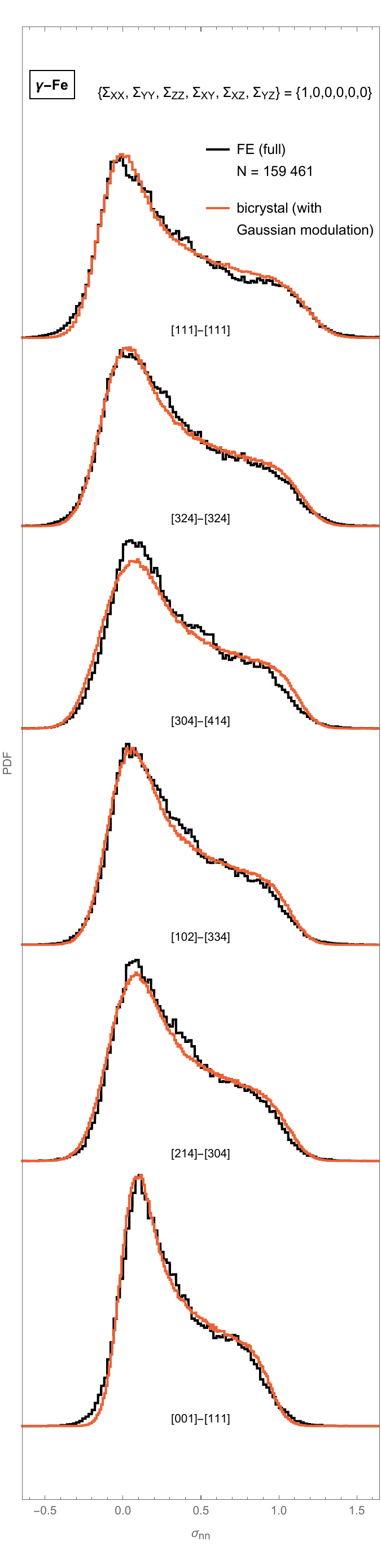}
	\includegraphics[width = 0.328 \textwidth, valign=c]{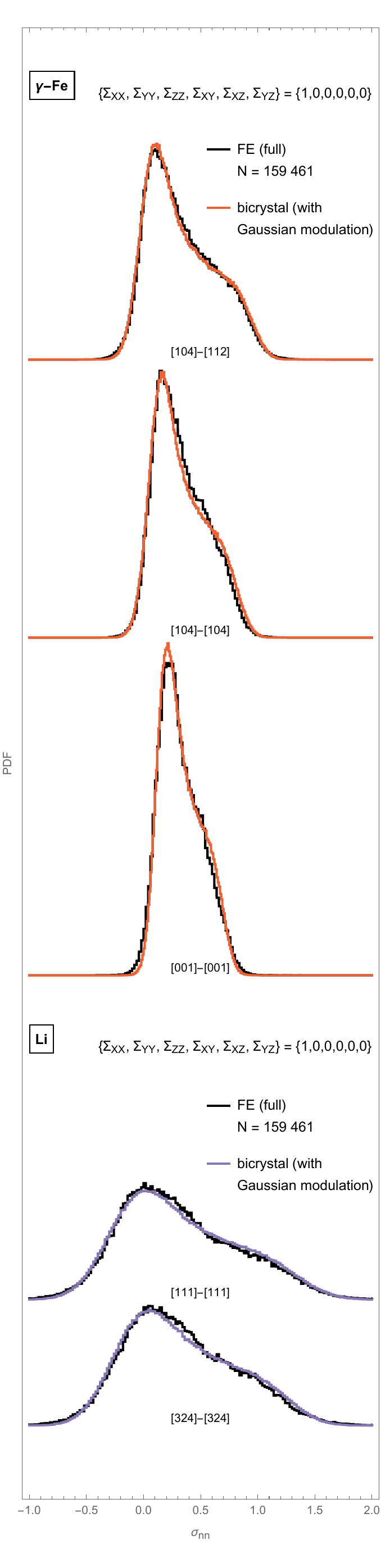}
	\includegraphics[width = 0.328 \textwidth, valign=c]{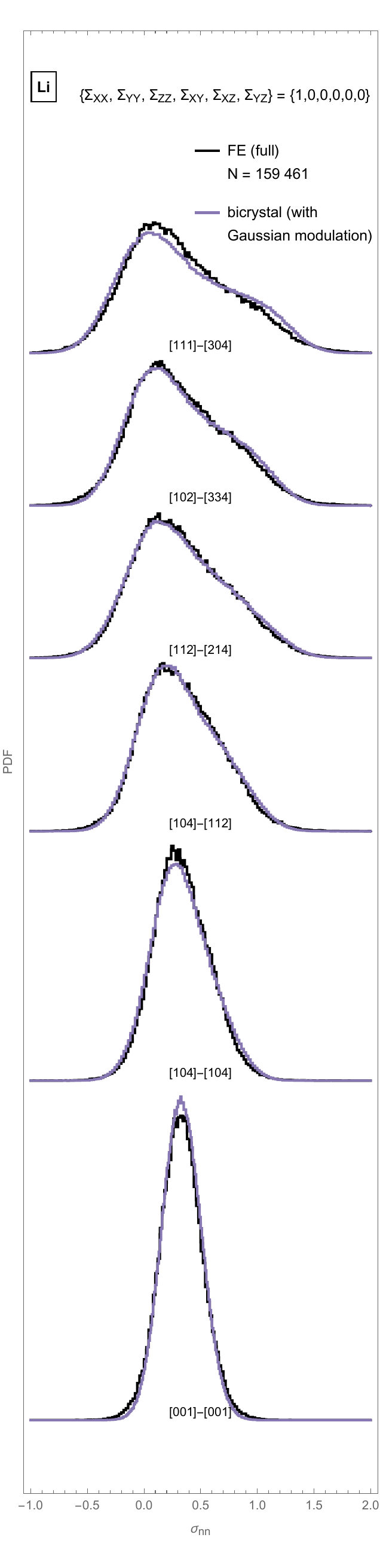}
	\caption{Same as in Fig.~\ref{fig:distributions_materials}, but for $\gamma$-Fe and Li. The INS distribution for the $[104]$-$[112]$ GB type in $\gamma$-Fe is compiled from distributions on three different aggregates (with $4000$ grains each) and contains $478213$ finite elements. It is thus the smoothest and most accurate, see also Fig.~\ref{fig:3_aggregates} for a comparison between the three distributions in individual aggregates.}
	\label{fig:distributions_materials_2}
\end{figure}

For demonstration purposes, a comparison between INS distributions produced by FE simulations and the bicrystal model using the above fits is shown in Figs.~\ref{fig:distributions_materials} and~\ref{fig:distributions_materials_2} for five different materials and several randomly selected GB types. A good agreement can be observed, despite a rather large uncertainty in the FE distribution obtained on an aggregate with only $4000$ grains. To assess its typical fluctuation, refer to Fig.~\ref{fig:3_aggregates}, depicting the full INS distribution on a specific GB type using three different aggregates.

\FloatBarrier

\nopagebreak

The $\alpha(E_{12},A^{u})$ and $\sigma_G(E_{12},A^{u})$ fits provided here should primarily be regarded only as a proof of concept, as the origin and exact form of functional  dependency remain largely unknown at present. The main obstacle to their refinement represents the lack of accuracy in the obtained results, as fitted parameters are subject to substantial uncertainties due to limited statistical data. In practical terms, we would need either a larger aggregate or a substantial number of them for each GB type and material under consideration. 
Accumulating a sufficient number of data points in the $E_{12}$--$A^u$ plane should presumably lead to the identification of a more appropriate fitting functions.


\FloatBarrier


\subsection{Alternative approach: the analysis of uniaxial loading\label{sec:uniaxial loading}}

The drawback of the aforementioned procedure is its critical dependence on the accurate extraction of the first three statistical moments from numerical distributions obtained on a finite (and relatively small) aggregate.

As an alternative approach, we can use the bicrystal model to examine the INS distribution under \emph{uniaxial loading} conditions. As will be demonstrated below, this distribution (for any chosen GB type) resembles the isotropic distribution.
We can then extend it to apply to a selected uniform loading scenario. This extension can be achieved through the techniques outlined in~\cite{shawish2024extending}, or by first generalizing the GB stresses according to Eq.~\eqref{eq:uniaxial_generalized} and then generating their distribution anew. Finally, the resulting \emph{pure} bicrystal-model INS distribution should be convolved with a normal distribution (fluctuations).

\begin{figure}[htb]
	\centering
	\includegraphics[width = 0.49\textwidth]{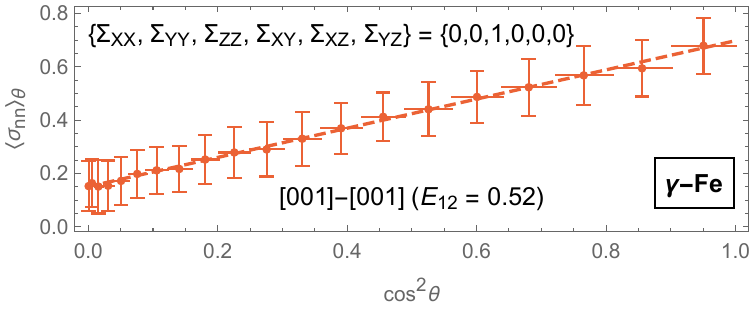}
	\includegraphics[width = 0.49\textwidth]{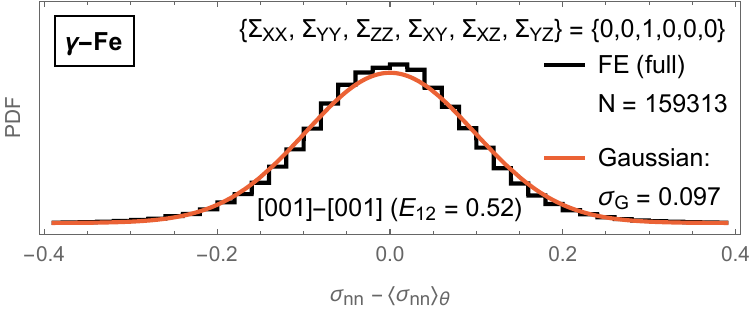}
	\includegraphics[width = 0.49\textwidth]{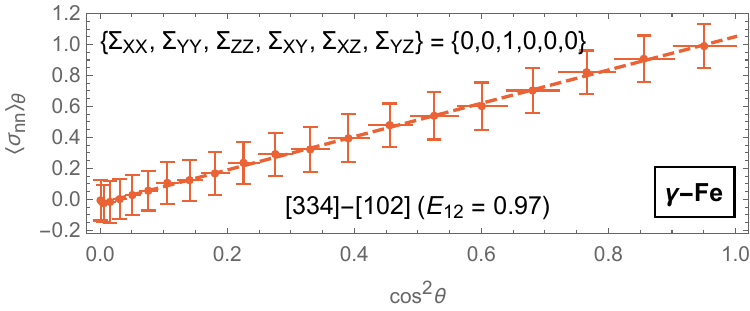}
	\includegraphics[width = 0.49\textwidth]{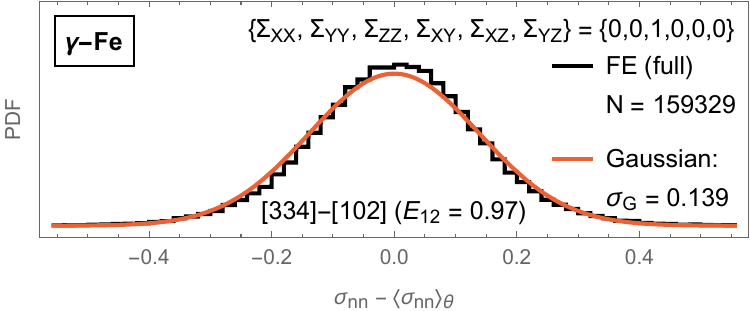}
	\includegraphics[width = 0.49\textwidth]{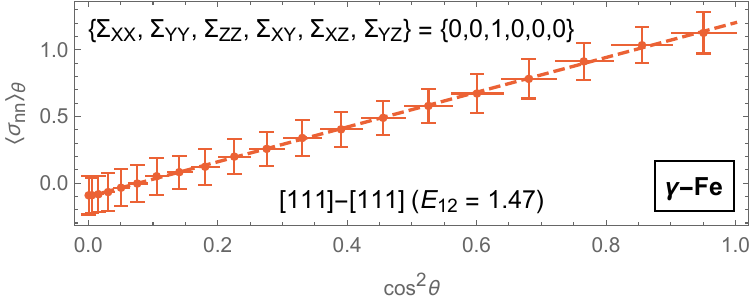}
	\includegraphics[width = 0.49\textwidth]{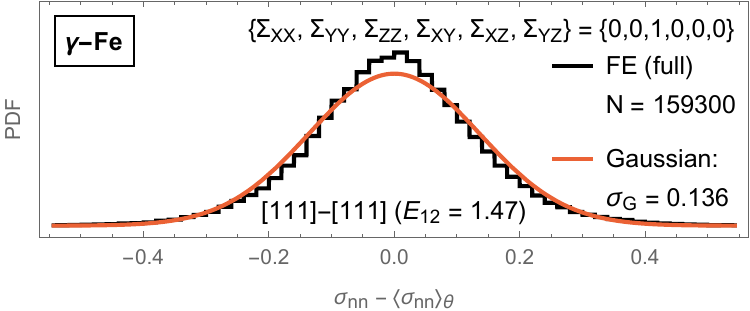}
	\includegraphics[width = 0.49\textwidth]{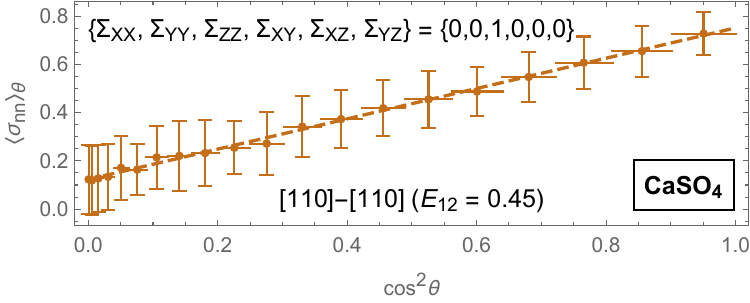}
	\includegraphics[width = 0.49\textwidth]{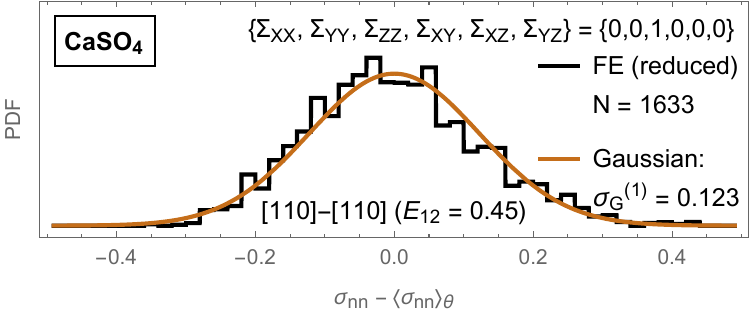}
	\includegraphics[width = 0.49\textwidth]{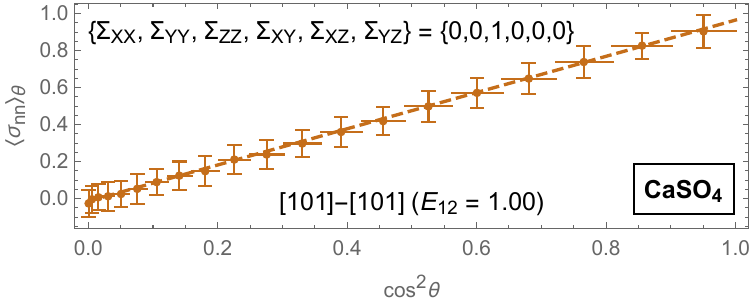}
	\includegraphics[width = 0.49\textwidth]{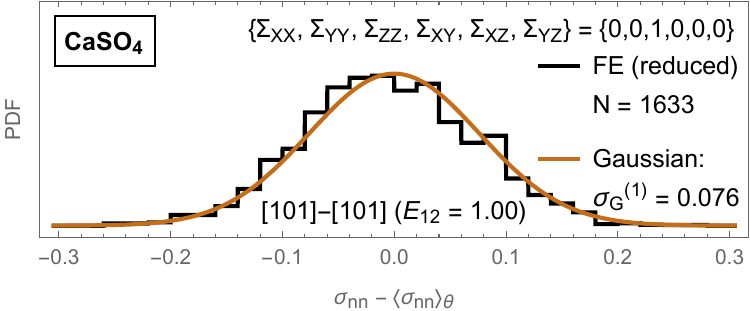}
	\includegraphics[width = 0.49\textwidth]{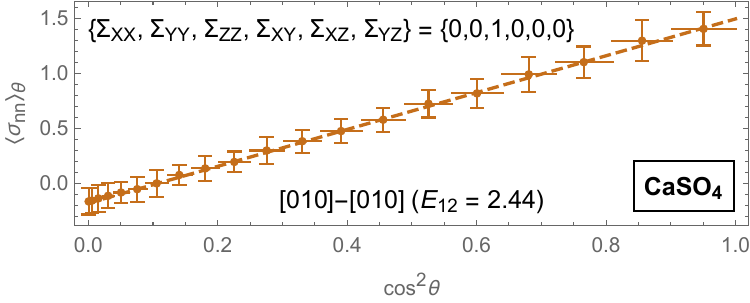}
	\includegraphics[width = 0.49\textwidth]{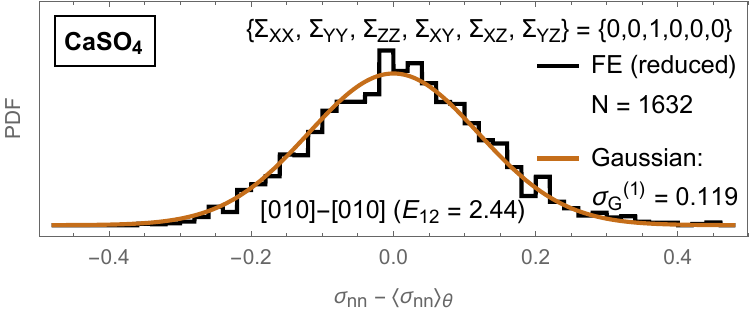}
	\caption{Left panels: The average value of $\snn$ across all GBs in the aggregate that are of a given type and have the same inclination $\theta$ with respect to uniaxial external loading. Horizontal lines indicate the width of the bin $\cos\theta\pm 0.025$, while vertical bars correspond to the standard deviation of the $\snn$ distribution in each bin. The \emph{dashed line} is a linear fit~\eqref{eq:linear} with $n$ defined in Eq.~\eqref{eq:n_approx}. Right panels: The distribution of normal stresses around their expected values (represented by the dashed line on the left) is again well-fitted by a Gaussian. Results are shown for three GB types and two selected materials.}
	\label{fig:fluctuation_specialGBs}
\end{figure}

In the bicrystal model, all GBs with the same orientation $\hat{n}$ correspond to equal $\snn$, whereas in a realistic aggregate, GB stresses can differ 
due to variations in the local neighbourhood of each GB. 
The distribution of these stresses exhibits a Gaussian profile.
For uniaxial loading (of strength $\Sigma_0$), the average normal stress $\langle\snn\rangle_{\theta}$ on GBs with inclination $\theta$ relative to the loading direction is given by:%
\footnote{The corresponding INS distribution is of a similar form to the isotropic distribution, only shifted by $n$ and stretched by $k$ along the horizontal axis, and compressed by $k$ along the vertical axis: $\text{PDF}(\tilde{\sigma}_{nn}) = \left (2 \sqrt{k \, (\tilde{\sigma}_{nn}-n)}\right )^{-1}$ in the range of $\tilde{\sigma}_{nn} := \tfrac{\sigma_{nn}}{\Sigma_0} \in [n,n+k]$. Its mean value is $\langle\snn\rangle = (k/3 + n) \, \Sigma_0$, and its standard deviation $s(\snn) = 2/(3 \, \sqrt{5}) \, k \, \Sigma_0$. Consequently, the overall width of the true INS distribution should be $\mu_h^{(2)} = 4/45 \, k^2 \, \Sigma_0^2 + \sigma_G^2$.\label{footnote:uniaxial}}
\begin{align}
	\frac{\langle\snn\rangle_{\theta}}{\Sigma_0} & = k \, \cos^2\theta + n \ , \label{eq:linear}
\end{align}
where $k$ and $n$ are some material and GB-type dependent coefficients.
This relation seems to hold universally, i.e., for any lattice symmetry, as illustrated in Fig.~\ref{fig:fluctuation_specialGBs}.
The approximate value of the $y$-\emph{intercept} ($n$) in Eq.~\eqref{eq:linear} was determined in~\cite{ELSHAWISH2023104940} to be:
\begin{align}
	n & \approx -\frac{1}{3} \left (k - 
	\frac{2 \, (1 + \nu_{12} - \xbar{\nu})}{E_{12}^{-1} + 
		1} \right ) \ , \label{eq:n_approx}
\end{align}
where the effective GB stiffness ($E_{12}$) is in general computed as:
\begin{align}
	E_{12} = 2 \, \xbar{E}^{-1} \left (\tilde{s}_{3333}^{(1)} + \tilde{s}_{3333}^{(2)} \right )^{-1} \ ,
\end{align}
and 
the effective GB Poisson's ratio ($\nu_{12}$) as:
\begin{align}
	\nu_{12} & = -\frac{1}{4} \, \xbar{E} \left (\tilde{s}_{1133}^{(1)} + \tilde{s}_{1133}^{(2)} + \tilde{s}_{2233}^{(1)} + \tilde{s}_{2233}^{(2)} \right ) \ .
\end{align}
In crystals with cubic symmetry, the latter reduces to $\nu_{12} = \xbar{\nu} + \tfrac{1}{2} (E_{12}^{-1} - 1)$, thereby becoming a function of $E_{12}$. This results in $n^{\text{cubic}} = -\tfrac{1}{3} \, (k-1)$, yielding $\langle\snn\rangle = \Sigma_0/3$, as confirmed by numerical simulations.

\begin{figure}[htb]
	\centering
	\includegraphics[width = 0.49\textwidth]{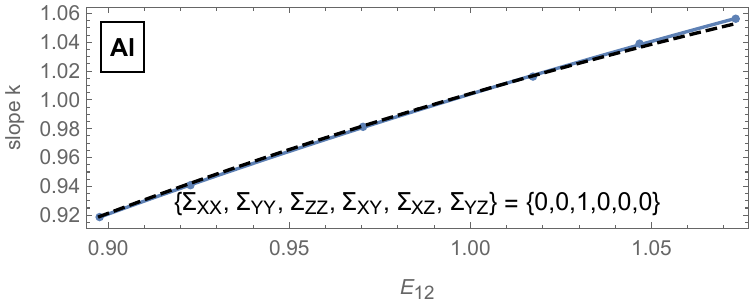}
	\includegraphics[width = 0.49\textwidth]{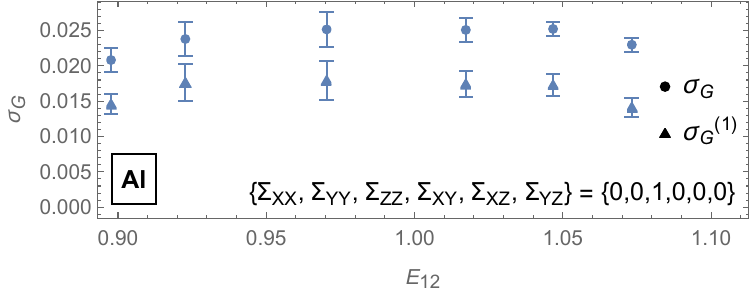}
	\includegraphics[width = 0.49\textwidth]{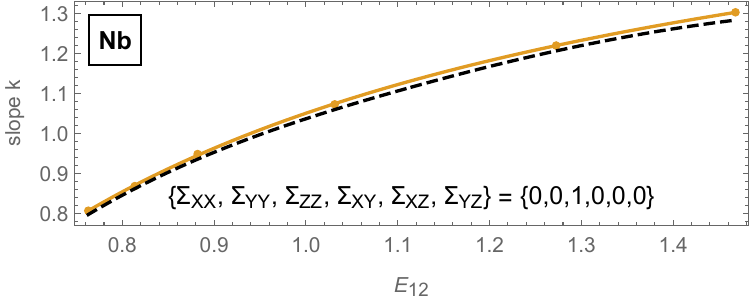}
	\includegraphics[width = 0.49\textwidth]{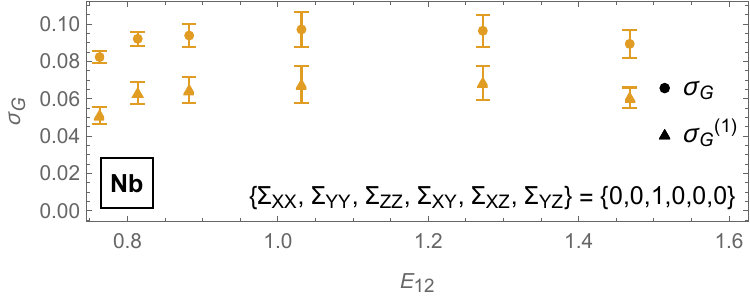}
	\includegraphics[width = 0.49\textwidth]{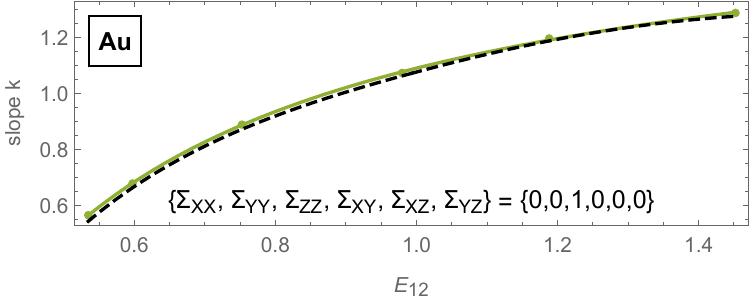}
	\includegraphics[width = 0.49\textwidth]{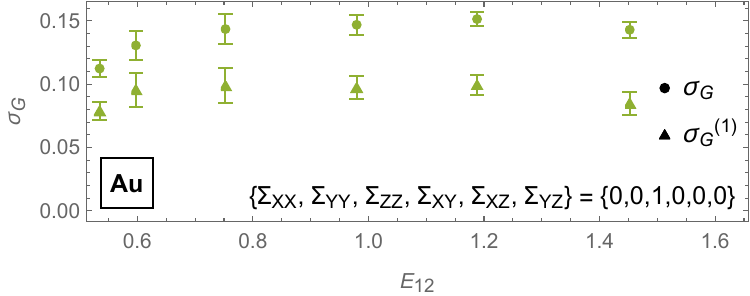}
	\includegraphics[width = 0.49\textwidth]{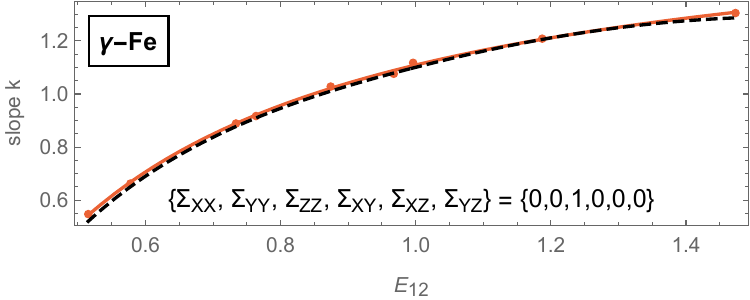}
	\includegraphics[width = 0.49\textwidth]{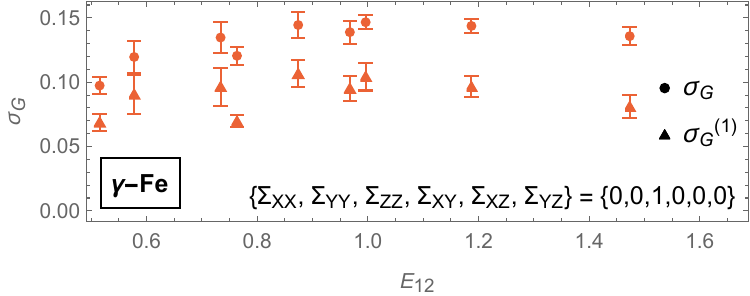}
	\includegraphics[width = 0.49\textwidth]{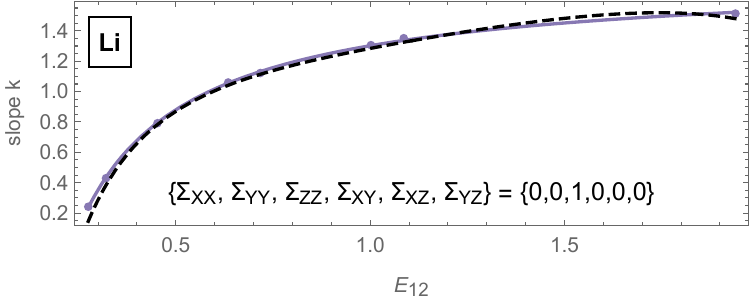}
	\includegraphics[width = 0.49\textwidth]{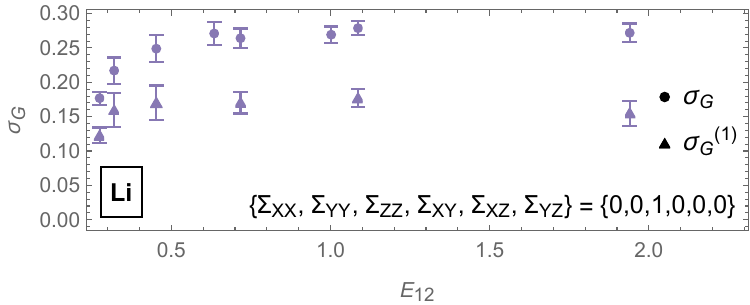}
	\includegraphics[width = 0.49\textwidth]{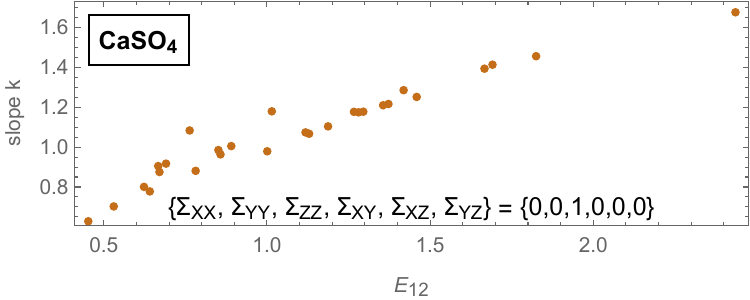}
	\includegraphics[width = 0.49\textwidth]{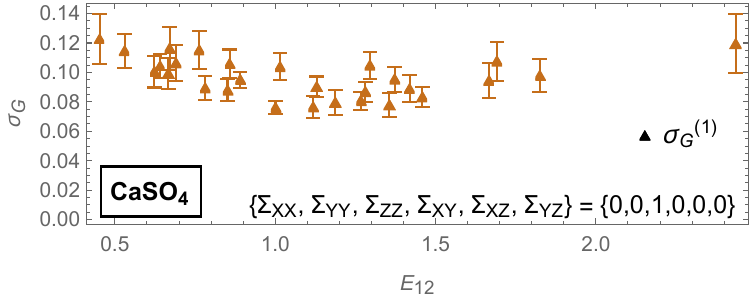}
	\caption{Left panels: The \emph{slope} ($k$) as a function of effective GB stiffness ($E_{12}$) for various materials. For lattices with cubic symmetry, it is well-fitted by Eq.~\eqref{eq:k_cubic} (\emph{colored curves}), while the fitting function~\eqref{eq:k_cubic_approx} proposed in the context of the buffer-grain model~\protect\cite{ELSHAWISH2023104940} (\emph{black dashed curves}) also provides a good fit. For non-cubic materials, the results are more scattered (indicating a dependence beyond just $E_{12}$). Right panels: The width of the Gaussian (and its uncertainty due to variation over different GB inclinations) obtained from fitting the \emph{full} ($\sigma_G$) and \emph{reduced} ($\sigma_G^{(1)}$) INS distributions.}
	\label{fig:slope}
\end{figure}

For cubic lattices, the \emph{slope} ($k$) is well-fitted by the following ansatz (see Fig.~\ref{fig:slope}):
\begin{align}
	k^{\text{cubic}} & = a_1 \, E_{12}^{-a_2} + a_3 \ , \label{eq:k_cubic}
\end{align}
where $a_1$, $a_2$, and $a_3$ are material-dependent coefficients, presumably functions of only $A^u$ or $A$.
A slightly different approximation was suggested in the context of the ($L_n = 2$) buffer-grain model~\cite{ELSHAWISH2023104940}:
\begin{align}
	k^{\text{cubic}} & \approx \frac{3}{1+ \left (E_{12} + f_1 - \left\vert E_{12}- \frac{E_{12}^{\text{min}}+E_{12}^{\text{max}}}{2}\right\vert^{f_2}\right )^{-1}} - \frac{1}{2} \ , \label{eq:k_cubic_approx}
\end{align}
with
\begin{align}
	f_1 & = 0.08 \, (A^u)^{0.85} \ , \\
	f_2 & = \frac{\log{f_1}}{\log{\frac{E_{12}^{\text{max}}-E_{12}^{\text{min}}}{2}}} \ ,
\end{align}
where $E_{12}^{\text{min}} = \xbar{E}^{-1}/s_{11}$, and $E_{12}^{\text{max}} = \xbar{E}^{-1}/(s_{11} - \tfrac{2}{3} \, s_0)$ for $A\geq 1$ (and vice versa for $A<1$).

For a general uniform loading, cubic lattice symmetry, and a random twist angle,
the average GB-normal stress across all GBs with the same inclination $\hat{n}$ is:
\begin{align}
	\langle\snn\rangle_{\hat{n}} & = \tfrac{1}{3} \, (2 \, k + 1) \, \Sigma_{zz} - \tfrac{1}{3} \, (k-1) \, (\Sigma_{xx} + \Sigma_{yy}) \ , \label{eq:uniaxial_generalized}
\end{align}
(with the components $\Sigma_{ij}$ given in Appendix~\ref{app:loading in GB system}), while in the bicrystal model (under identical conditions), it is given by Eq.~\eqref{eq:avg_twist}.
Under uniaxial loading along the $Z$-direction, the latter further reduces to the linear trend~\eqref{eq:linear} observed in FE simulations: 
\begin{align}
	\frac{\sigma_{zz}^{\Delta\omega}}{\Sigma_0} & = \alpha \, (\xbar{c}_{zz} - \xbar{c}_{xx,yy}) \cos^2\theta + \left (\alpha \, \xbar{c}_{xx,yy} - \beta \, (2 \, \xbar{c}_{xx,yy}+\xbar{c}_{zz}) \right ) \ . \label{eq:uni_general}
\end{align}
By taking into account Eqs.~\eqref{eq:cubic_rule} and~\eqref{eq:beta}, it can be shown that for cubic crystal lattices, $n$ is indeed equal to $n^{\text{cubic}} = -\tfrac{1}{3} (k-1)$:
\begin{align}
	\frac{\sigma_{zz}^{\Delta\omega}}{\Sigma_0} & = \tfrac{1}{2} \, \alpha \left (3 \, \xbar{c}_{zz} - (C_{11} + 2 \, C_{12}) \right ) \cos^2\theta - \tfrac{1}{3} \left (\tfrac{1}{2} \, \alpha  (3 \, \xbar{c}_{zz} - (C_{11} + 2 \, C_{12})) - 1 \right ) \ .
\end{align}

\begin{figure}[!htb]
	\centering
	\includegraphics[width = 0.49\textwidth]{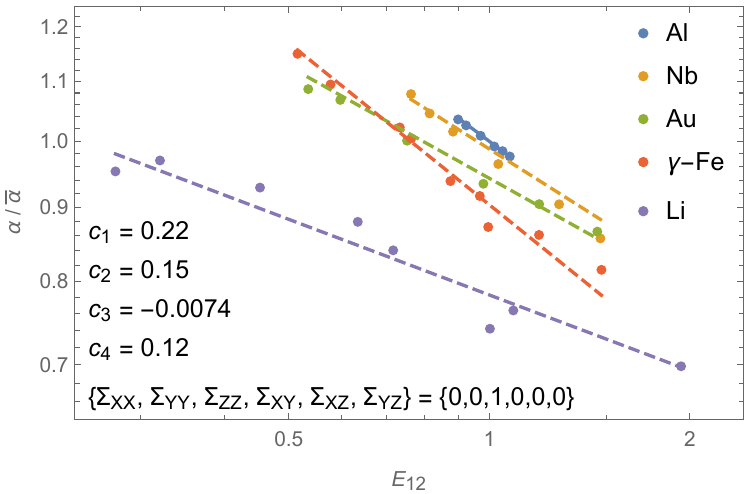} 
	\includegraphics[width = 0.49\textwidth]{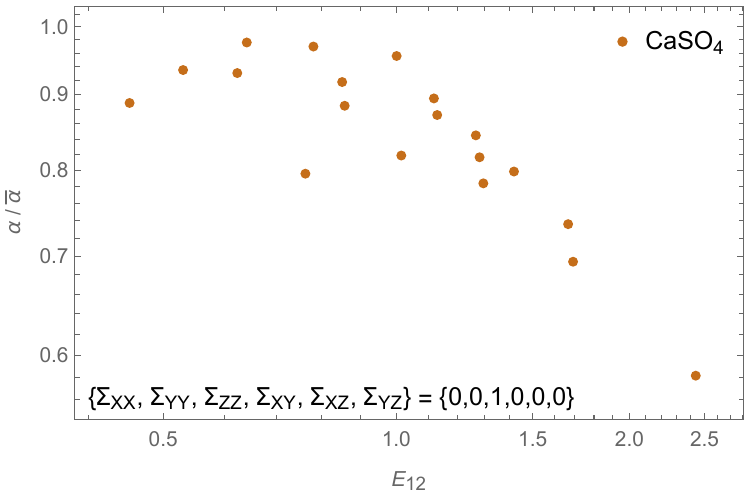}
	\caption{The parameter $\alpha$ plotted against effective GB stiffness $E_{12}$ for various materials. For lattices with cubic symmetry (\emph{left panel}), the dependence is well-fitted by Eq.~\eqref{eq:alpha_fit} (\emph{dashed curves}), with $b_1$ and $b_2$ related to material's anisotropy $A^u$ through Eqs.~\eqref{eq:fitting_parameters_1} and~\eqref{eq:fitting_parameters_2}, respectively, and the values of fitting parameters listed in the inset.
	For non-cubic materials (\emph{right panel}), the results are more scattered (suggesting a dependence beyond $E_{12}$). The value of $\alpha$ is normalized to its average strain approximation, $\xbar{\alpha} := (1+\xbar{\nu})/\xbar{E}$ (which for cubic symmetry simplifies to $\xbar{\alpha} \to \frac{1}{2 G}$), see Eq.~\eqref{eq:assumption}.}
	\label{fig:alpha_alt}
\end{figure}

\begin{figure}[!htb]
	\centering
	\includegraphics[width = 0.49\textwidth]{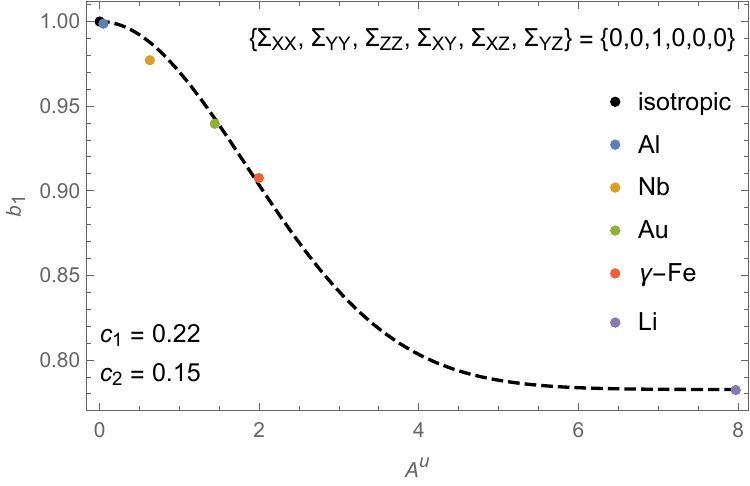}
	\includegraphics[width = 0.49\textwidth]{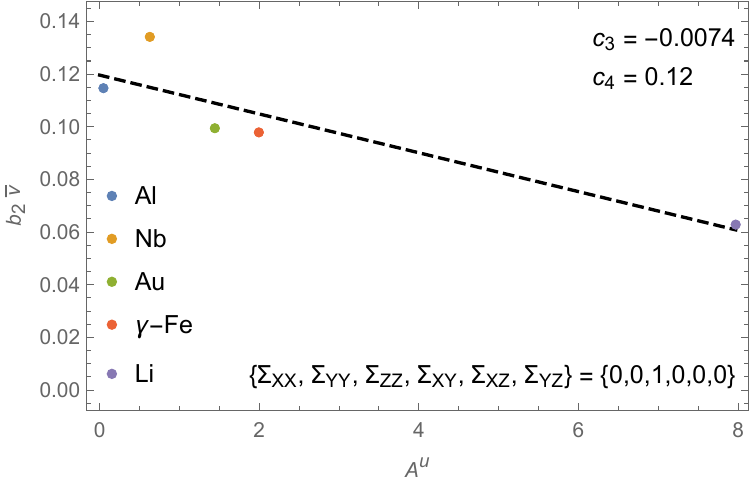}
	\caption{The parameters $b_1$ (\emph{left panel}) and $b_2$ (\emph{right panel}) plotted against the universal anisotropy index $A^u$. The dependencies are effectively captured by Eqs.~\eqref{eq:fitting_parameters_1} and~\eqref{eq:fitting_parameters_2}, respectively (shown as \emph{dashed curves}).}
	\label{fig:b1_b2_alt}
\end{figure}

From either of these two expressions, the slope $k$ and the $y$-intercept $n$ can be readily extracted. Comparing these values to those obtained from FE simulations allows us to determine the values of $\alpha$ and $\beta$.
As shown in Fig.~\ref{fig:alpha_alt}, for cubic lattices they can be effectively fitted by Eqs.~\eqref{eq:alpha_fit} and~\eqref{eq:beta},
where the coefficients $b_1$ and $b_2$ in Fig.~\ref{fig:b1_b2_alt} exhibit similar trends as those observed in the previous section, following the relationships in Eqs.~\eqref{eq:fitting_parameters_1} and~\eqref{eq:fitting_parameters_2}. The fitting coefficients ($c_1 = 0.22$, $c_2 = 0.15$, $c_3 = -0.0074$, and $c_4 = 0.12$) possess only marginally different values from those listed in Table~\ref{tab:fitting_parameters}.


\FloatBarrier

\section{Discussion}

The perturbative model presented in this paper displays several limitations and weaknesses. It is inherently probabilistic, primarily suited for estimating INS distributions rather than determining the stresses on specific GBs. Consequently, the model can help us assess only the statistical likelihood for microcrack initiation, but it does not allow us to pinpoint their exact locations. 

Furthermore, the model's validity is confined to the elastic regime since it does not account for plastic deformations, despite their crucial role in the formation of stress concentrations responsible for initiating the cracks.
Another drawback of the model is the assumption of constant stress and strain throughout the grains and along their GBs. When making a direct comparison with realistic distributions, this corresponds to averaging the GB stresses over the respective GB areas. And while we typically assume that the stresses on each GB follow a normal distribution, this remains unconfirmed at this point.
Lastly, it should also be mentioned that although our analytical solution was developed for a general case, many of the derived expressions and results provided here are valid only for cubic crystal lattices and grains of equal size ($V_1 = V_2$).

In principle, the entire ``machinery'' of the bicrystal model, particularly Eqs.~\eqref{eq:c-coefficients}--\eqref{eq:gamma_parameter}, was developed only to determine the parameters $\alpha$, $\beta$, and $\sigma_G$. 
However, once their fits are obtained, we can either numerically solve the constitutive equations for a chosen GB orientation and GB type, 
or use their analytical solution~\eqref{eq:INS_a} for the coefficients defined in Eqs.~\eqref{eq:c-a-connection} and~\eqref{eq:c-coefficients}. After producing the distribution across different GB orientations (and twist angles), we must convolve it with a Gaussian to obtain the actual (realistic) INS distribution.

Below are the detailed steps for producing the INS distribution for a chosen GB type in a selected material with cubic lattice symmetry under a given uniform external loading:
\begin{enumerate}
	\item For a chosen material ($A^u$), GB type ($E_{12}$), and uniform external loading ($\mathbf{\Sigma}$), compute the values of $\alpha$, and $\beta$ from fits in Eqs.~\eqref{eq:alpha_fit}, \eqref{eq:fitting_parameters_1}, \eqref{eq:fitting_parameters_2}, and~\eqref{eq:beta}. This allows us to produce $\snn$ (in the pure bicrystal model) by numerically solving Eqs.~\eqref{eq:Hooke} and~\eqref{eq:BC1}--\eqref{eq:BC3} along with condition~\eqref{eq:assumption_general} for any given GB orientation ($\psi$, $\cos\theta$, $\phi$) and twist angles ($\omega_1$ and $\omega_2$).
	\item By taking a uniform distribution of these angles, we can produce the INS distribution in the pure bicrystal model (e.g., by Monte Carlo sampling).
	\item Convolve the resulting distribution with the Gaussian distribution of width $\sigma_G$, obtained from fits in Eqs.~\eqref{eq:sigma_fit}
	, \eqref{eq:fitting_parameters_3}, and~\eqref{eq:fitting_parameters_4}.
\end{enumerate}

While the presented bicrystal model is, in a sense, ``exact'', the proposed phenomenological fits of $\alpha$ and $\sigma_G$ as functions of $E_{12}$, and especially of $A^u$, are not. They are only approximate and are meant only as a proof of concept. 
Beside the fact that there is no motivation for the chosen fitting functions beyond their simplicity and small number of parameters, also the fitted values are subject to significant uncertainties due to the smallness of the used aggregate, containing just $1636$ GBs of a chosen GB type. This effect becomes apparent when considering INS distributions under equivalent (i.e., rotated) loadings (such as $\{\Sigma_{XX},\Sigma_{YY},\Sigma_{ZZ},\Sigma_{XY},\Sigma_{XZ},\Sigma_{YZ}\} = \{1,0,0,0,0,0\}$ and $\{1,1,1,1,1,1\}$ in Fig.~\ref{fig:distributions_conv}), which should generate identical distributions for an infinitely large aggregate, whereas for the aggregate used in practice, the moments can vary by as much as a few percent. The variation is significant enough that it can even mean the difference between the existence and non-existence of a solution~\eqref{eq:moment1}--\eqref{eq:moment3} with real-valued $\alpha$, $\beta$, and $\sigma_G$.%
\footnote{What typically occurs is that the parameter $\alpha$, obtained from the third central moment of the pure bicrystal distribution, sometimes tends to be overestimated, causing the second moment of the pure bicrystal distribution to exceed that of the FE distribution (resulting in $\sigma_G^2 < 0$). Interestingly, this most often occurs for the least anisotropic materials.}
Those are, in fact, needed to be able to produce the fits in the first place.

Although the proposed fits perform reasonably well with a handful of fitting parameters, more accurate fits (both in terms of more suitable fitting functions and refined values of fitting coefficients) could likely be found. For instance, using a non-trivial distribution of grain sizes in the bicrystal model ($V_1\neq V_2$) or considering additional variables besides $E_{12}$ and $A^u$ could enhance the accuracy of the fits. This would require more extensive numerical data, including larger aggregates, a broader range of materials, GB types, and loadings.

Other potential sources of discrepancies with realistic distributions include:
the use of the bicrystal-strain condition~\eqref{eq:assumption_general} with GB-orientation-independent parameters $\alpha$ and $\beta$ in the pure bicrystal model. However, the actual relevance of this assumption is best demonstrated by the fact that the same values of $\alpha$ and $\beta$ apply to both the reduced (i.e., averaged over the GB) and the full INS distribution, with only the width of Gaussian modulation (the value of $\sigma_G$) changing. 

Another possible issue arises from the assumption that $\alpha$, $\beta$, and $\sigma_G$ are (only) functions of $E_{12}$ and $A^u$, and that the fluctuation ($\Delta\snn$) is Gaussian. And while $\alpha$ and $\beta$ are loading-independent, it is still not entirely clear how $\sigma_G$ correctly scales with loading. It is assumed that for cubic crystal lattices it is proportional to $\Sigma_{\text{mis}}$, but for non-cubic lattices, it remains an open question.

\section{Conclusions}

In the first part of the paper, we have demonstrated that we can successfully predict the INS distribution on random GBs for any polycrystalline material under uniform loading conditions. A single parameter was needed that connected the width of Gaussian fluctuations with the material's anisotropy. This parameter was determined from numerical fits. 

In the second part of the paper, we have shown how to generate INS distributions on particular GB types. This can again be done for any material and loading configuration, but for simplicity reasons, we restricted our analysis 
to lattices with cubic symmetry. In this case, instead of three parameters ($\alpha$, $\beta$, and $\sigma_G$), only two suffice to determine the INS distribution for a chosen GB type within a given material since $\alpha$ and $\beta$ are not independent of each other (in large part because the average Young's modulus and the average Poisson's ratio of the material are not linearly independent quantities in cubic crystals). 

A simple functional dependence between $\alpha$ and $E_{12}$ has been proposed, with material-specific coefficients determined for several different materials, and correlated with their $A^u$. Similarly for $\sigma_G$. The existing fits involve just \emph{six} parameters, the numerical values of which need to be determined only once.
Improvement of these fits could make the bicrystal model extremely useful for quick and reliable predictions of INS distributions for any chosen GB type, material, and uniform external loading conditions. As such, it can serve as a valuable tool for estimating the probability of intergranular cracking initiation or for determining the GB strengths of selected GB types.
A statistical knowledge of stresses (INS), resulting from a microscopic description (bicrystal model), can thus be used to address a macroscopic phenomenon (intergranular cracking initiation).


\section*{Acknowledgments}

The authors gratefully acknowledge financial support provided by Slovenian Research and Innovation Agency (grant P2-0026).

\newpage


\appendix

\section{Elastic properties of materials\label{app:mat}}

In Table~\ref{tab:elastic_constants} elastic constants of single crystals (monocrystals), together with their aggregate properties, are listed for several representative materials used in this study. Chosen polycrystallines have either \emph{orthorhombic} or \emph{cubic} lattice symmetry. The materials are ordered according to their universal elastic anisotropy index $A^u$~\cite{ranganathan}
defined as
\begin{align}
	A^u & = 5 \frac{G^V}{G^R} + \frac{K^V}{K^R} - 6 \ ,
\end{align}
where $G$ and $K$ are their \emph{shear} and \emph{compression} elastic moduli in Voigt ($V$) and Reuss ($R$) limits, respectively, which for macroscopically isotropic (untextured) polycrystallines assume the following forms~\cite{tromans}
\begin{align}
	K^V & = \tfrac{1}{9} \big (C_{11}+C_{22}+C_{33}+2 \, (C_{12}+C_{13}+C_{23})\big ) \ , \\
	G^V & = \tfrac{1}{15} \big (C_{11}+C_{22}+C_{33}+3 \, (C_{44}+C_{55}+C_{66}) - (C_{12}+C_{13}+C_{23})\big ) \ , \\
	K^R & = \big (s_{11}+s_{22}+s_{33}+2 \, (s_{12}+s_{13}+s_{23})\big )^{-1} \ , \\
	G^R & = \left (\tfrac{1}{15} \big (4 \, (s_{11}+s_{22}+s_{33})+3 \, (s_{44}+s_{55}+s_{66}) - 4 \, (s_{12}+s_{13}+s_{23})\big ) \right )^{-1} \ .
\end{align}
For cubic lattice symmetry, $C_{11}=C_{22}=C_{33}$, $C_{12}=C_{13}=C_{23}$ and  $C_{44}=C_{55}=C_{66}$, hence the universal anisotropy index reduces to 
\begin{align}
A^u & = \frac{3 \, (C_{11}-C_{12}-2 \, C_{44})^2}{5 \, (C_{11}-C_{12}) \, C_{44}} = \frac{6 \, (A-1)^2}{5 A} \ , 
\end{align}
where
\begin{align}
A = \frac{2 \, C_{44}}{C_{11}-C_{12}} 
\end{align}
represents the Zener elastic anisotropy index~\cite{zener} of the material. It is defined only for cubic lattices, with $A=1$ (and $A^u = 0$) denoting a perfectly isotropic crystal.

\begin{table}[h]
	\caption{\label{tab:elastic_constants}
		Elasticity (stiffness) tensor components $C_{ij}$ (in Voigt notation) of single crystals with orthorhombic ($\dagger$)~\cite{simmonswang} or cubic ($\ast$)~\cite{bower} lattice symmetry, and their aggregate properties in polycrystallines (with $\xbar{E}$ and $\xbar{\nu}$ denoting the average Young's modulus and Poisson's ratio of a macroscopic aggregate, which are for cubic materials computed according to Eqs.~\eqref{eq:Eave} and~\eqref{eq:nuave}, respectively). $C_{ij}$ and $\xbar{E}$ are given in GPa.}
	\resizebox{\textwidth}{!}{%
		\begin{tabular}{*{14}{c}}
			\toprule
			Crystal & $C_{11}$ & $C_{22}$ & $C_{33}$ & $C_{12}$ & $C_{13}$ & $C_{23}$ & $C_{44}$ & $C_{55}$ & $C_{66}$ & $\xbar{E}$ & $\xbar{\nu}$ & $A^u$ & $A$ \\
			\midrule
			$^{\ast}$Al          & 107.3 & & & 60.9  & & & 28.3  & & & 70.41 & 0.346 & 0.05 & 1.22 \\
			$^{\ast}$Nb          & 240.2 & & & 125.6 & & & 28.2  & & & 104.9 & 0.393 & 0.63 & 0.49 \\
			$^{\ast}$Au          & 192.9 & & & 163.8 & & & 41.5  & & & 79.40  & 0.424 & 1.44 & 2.85 \\
			$^{\ast}\gamma$-Fe & 197.5 & & & 125.0 & & & 122.0 & & & 195.2 & 0.282 & 2.00 & 3.37 \\
			$^{\dagger}$CaSO$_4$    & 93.82 & 185.5 & 111.8 & 16.51 & 15.20 & 31.73 & 32.47 & 26.53 &	9.26 & 71.77 & 0.282 & 2.78 & / \\
			$^{\ast}$Li          & 13.5  & & & 11.44 & & & 8.78  & & & 10.94 & 0.350 & 7.97 & 8.52 \\
			\bottomrule
	\end{tabular}}
\end{table}
%

\section{Most general uniform external loading tensor $\mathbf{\Sigma}^{\text{lab}}$ expressed in GB coordinate system\label{app:loading in GB system}}

%
\begin{align}
	\begin{split}
	\Sigma_{xx} & = \left ( \sin^2\theta \, \cos^2\phi \right ) \, \Sigma_{ZZ} + \left (\cos\theta \, \cos\psi \, \cos\phi - \sin\psi \, \sin\phi \right )^2 \, \Sigma_{XX} + \\
	& + \left (\cos\theta \, \sin\psi \, \cos\phi + \cos\psi \, \sin\phi \right )^2 \, \Sigma_{YY} + \\
	& + \left ( \cos^2\theta \, \sin 2\psi \, \cos^2\phi + \cos\theta \, \cos 2\psi \, \sin 2\phi - \sin 2\psi \, \sin^2\phi \right ) \, \Sigma_{XY} + \\
	& + \sin\theta \left ( \sin\psi \, \sin 2\phi - 2 \, \cos\theta \, \cos\psi \, \cos^2\phi \right ) \, \Sigma_{XZ} - \\
	& - 2 \sin\theta \, \cos\phi \left (\cos\theta \, \sin\psi \, \cos\phi + \cos\psi \, \sin\phi \right ) \, \Sigma_{YZ} \ ,
    \end{split} \\
	\begin{split}
	\Sigma_{yy} & = \left ( \sin^2\theta \, \sin^2\phi \right ) \, \Sigma_{ZZ} + \left (\cos\theta \, \cos\psi \, \sin\phi + \sin\psi \, \cos\phi \right )^2 \, \Sigma_{XX} + \\
	& + \left ( \cos\psi \, \cos\phi - \cos\theta \, \sin\psi \, \sin\phi \right )^2  \, \Sigma_{YY} + \\
	& + \left ( \cos^2\theta \, \sin 2\psi \, \sin^2\phi - \cos\theta \, \cos 2\psi \, \sin 2\phi - \sin 2\psi \, \cos^2\phi \right ) \, \Sigma_{XY} - \\
	& - 2 \sin\theta \, \sin\phi \left (\cos\theta \, \cos\psi \, \sin\phi + \sin\psi \, \cos\phi \right ) \, \Sigma_{XZ} + \\
	& + \sin\theta \left ( \cos\psi \, \sin 2\phi - 2 \, \cos\theta \, \sin\psi \, \sin^2\phi \right ) \, \Sigma_{YZ} \ ,
	\end{split} \\
    \begin{split}
    \Sigma_{zz} & = \left ( \cos^2\theta \right ) \, \Sigma_{ZZ} + \left ( \sin^2\theta \, \cos^2\psi \right ) \, \Sigma_{XX} + \left ( \sin^2\theta \, \sin^2\psi \right ) \, \Sigma_{YY} + \\
    & + \left ( \sin^2\theta \, \sin 2\psi \right ) \, \Sigma_{XY} + \left ( \sin 2\theta \, \cos\psi \right ) \, \Sigma_{XZ} + \left ( \sin 2\theta \, \sin\psi \right ) \, \Sigma_{YZ} \ ,
    \end{split} \\
    \begin{split}
    \Sigma_{xy} & = - \left ( \sin^2\theta \, \sin\phi \, \cos\phi \right ) \, \Sigma_{ZZ} - \\
    & - \frac{1}{2} \left ( \cos^2\theta \, \cos^2\psi \, \sin 2\phi + \cos\theta \, \sin 2\psi \, \cos 2\phi - \sin^2\psi \, \sin 2\phi \right ) \, \Sigma_{XX} - \\
    & - \frac{1}{2} \left ( \cos^2\theta \, \sin^2\psi \, \sin 2\phi - \cos\theta \, \sin 2\psi \, \cos 2\phi - \cos^2\psi \, \sin 2\phi \right ) \, \Sigma_{YY} - \\
    & - \frac{1}{2} \left (\left ( \cos^2\theta + 1 \right ) \sin 2\psi \, \sin 2\phi - 2 \, \cos\theta \, \cos 2\psi \, \cos 2\phi \right ) \, \Sigma_{XY} + \\
    & + \sin\theta \left (\cos\theta \, \cos\psi \, \sin 2\phi + \sin\psi \, \cos 2\phi \right ) \, \Sigma_{XZ} + \\
    & + \sin\theta \left ( \cos\theta \, \sin\psi \, \sin 2\phi - \cos\psi \, \cos 2\phi \right ) \, \Sigma_{YZ} \ ,
    \end{split} \\
    \begin{split}
    \Sigma_{xz} & = - \left ( \sin\theta \, \cos\theta \, \cos\phi \right ) \, \Sigma_{ZZ} + \sin\theta \, \cos\psi \left ( \cos\theta \, \cos\psi \, \cos\phi - \sin\psi \, \sin\phi \right ) \, \Sigma_{XX} + \\
    & + \sin\theta \, \sin\psi \left ( \cos\theta \, \sin\psi \, \cos\phi + \cos\psi \, \sin\phi \right ) \, \Sigma_{YY} + \\
    & + \sin\theta \left ( \cos\theta \, \sin 2\psi \, \cos\phi + \cos 2\psi \, \sin\phi \right ) \, \Sigma_{XY} + \\
    & + \left ( \cos 2\theta \, \cos\psi \, \cos\phi - \cos\theta \, \sin\psi \, \sin\phi \right ) \, \Sigma_{XZ} + \\
    & + \left ( \cos\theta \, \cos\psi \, \sin\phi + \cos 2\theta \, \sin\psi \, \cos\phi \right ) \, \Sigma_{YZ} \ ,
    \end{split} \\
    \begin{split}
    \Sigma_{yz} & = \left ( \sin\theta \, \cos\theta \, \sin\phi \right ) \, \Sigma_{ZZ} - \sin\theta \, \cos\psi \left ( \cos\theta \, \cos\psi \, \sin\phi + \sin\psi \, \cos\phi \right ) \, \Sigma_{XX} - \\
    & - \sin\theta \, \sin\psi \left ( \cos\theta \, \sin\psi \, \sin\phi - \cos\psi \, \cos\phi \right ) \, \Sigma_{YY} - \\
    & - \sin\theta \left ( \cos\theta \, \sin 2\psi \, \sin\phi - \cos 2\psi \, \cos\phi \right ) \, \Sigma_{XY} - \\
    & - \left ( \cos 2\theta \, \cos\psi \, \sin\phi + \cos\theta \, \sin\psi \, \cos\phi \right ) \, \Sigma_{XZ} + \\
    & + \left ( \cos\theta \, \cos\psi \, \cos\phi - \cos 2\theta \, \sin\psi \, \sin\phi \right ) \, \Sigma_{YZ}
    \end{split}
\end{align}
%


\bibliography{references}{} 

\end{document}